\documentstyle[12pt,epsf]{article}

\newcommand{\beq}{\begin{equation}}
\newcommand{\eeq}{\end{equation}}
\newcommand{\bea}{\begin{eqnarray}}
\newcommand{\eea}{\end{eqnarray}}
\def\nn{\nonumber}
\newcommand{\e}{\mbox{e}}
\newcommand{\tr}{\mbox{tr}}
\newcommand{\eqn}[1]{(\ref{#1})}
\def\semiprod{{\supset\!\!\!\!\!\!\!\times~}} 
\newcommand{\quater}{{\bb H}} 
\newcommand{\quaters}{{\bbs H}} 
\newcommand{\complex}{{\bb C}} 
\newcommand{\complexs}{{\bbs C}} 
\newcommand{\zed}{{\bb Z}} 
\newcommand{\real}{{\bb R}} 
\newcommand{\reals}{{\bbs R}} 
\newcommand{\zeds}{{\bbs Z}} 
\newcommand{\rat}{{\bb Q}} 
\newcommand{\id}{{\bb I}} 
\newcommand{\klein}{{(-1)^{F_{\rm L}}}} 
\newcommand{\cliff}{C\!\ell} 
\newcommand{\ccliff}{{\bb C}\!\ell} 
\newcommand{\dirac}{D\!\!\!\!/\,} 
\newcommand{\point}{{\rm pt}} 
\newcommand{\ch}{{\rm ch}} 
\newcommand{\rk}{{\rm rk}} 
\newcommand{\ind}{{\rm index}} 
\newcommand{\Ind}{{\rm Index}} 
\newcommand{\Grass}{{\rm Gr}} 
\newcommand{\NS}{{\rm NS}} 
\newcommand{\RR}{{\rm R}} 
\def\K{{\rm K}}
\def\KO{{\rm KO}}
\def\KU{{\rm KU}}
\def\KSp{{\rm KSp}}
\def\KR{{\rm KR}}
\def\KSC{{\rm KSC}}
\def\wt{\widetilde}

\newcommand{\newsection}{
\setcounter{equation}{0}
\section}

\font\mybb=msbm10 at 12pt
\def\bb#1{\hbox{\mybb#1}}
\font\mybbs=msbm10 at 9pt
\def\bbs#1{\hbox{\mybbs#1}}

\font\mybb=msbm10 at 12pt
\font\mybbs=msbm10 at 9pt

\makeatletter
\newdimen\normalarrayskip              
\newdimen\minarrayskip                 
\normalarrayskip\baselineskip
\minarrayskip\jot
\newif\ifold             \oldtrue            \def\new{\oldfalse}

\makeatother

\newlength{\extraspace}
\newlength{\extraspaces}
\begin{document}

\addtolength{\baselineskip}{.8mm}

\thispagestyle{empty}

\begin{flushright}
\baselineskip=12pt
NBI-HE-99-17\\
hep-th/9907140\\
\hfill{  }\\July 1999
\end{flushright}
\vspace{.25cm}

\begin{center}

\baselineskip=24pt

{\LARGE{CONSTRUCTING D-BRANES\\ FROM K-THEORY}}\\[10mm]

\baselineskip=12pt

{\bf Kasper Olsen}$^{a,}$\footnote{Address after September 1, 1999: Lyman
Laboratory
of Physics, Harvard University, Cambridge, MA 02138, U.S.A.}
and {\bf Richard J. Szabo}$^{a,b}$
\\[5mm]

$^a$ {\it The Niels Bohr Institute\\ Blegdamsvej 17, DK-2100 Copenhagen \O,
Denmark\\[2mm]
$^b$ {\it Department of Physics and Astronomy, University of British Columbia\\
6224 Agricultural Road, Vancouver, B.C. V6T 1Z1, Canada}\\[2mm] {\tt kolsen ,
szabo @nbi.dk}}
\\[15mm]

{\sc Abstract}

\begin{center}
\begin{minipage}{15cm}

A detailed review of recent developments in the topological classification of
D-branes
in superstring theory is presented. Beginning with a thorough, self-contained
introduction to the
techniques and applications of topological K-theory, the relationships
between the classic constructions of K-theory and the recent realizations of
D-branes
as tachyonic solitons, coming from bound states of higher dimensional systems
of unstable branes, are
described. It is shown how
the K-theory formalism naturally reproduces the known spectra of BPS and
non-BPS D-branes,
and how it can be systematically used to predict the existence of new states.
The emphasis
is placed on the new interpretations of D-branes as conventional topological
solitons in
other brane worldvolumes, how the mathematical formalism can be used to deduce
the
gauge field content on both supersymmetric and non-BPS branes, and also how
K-theory predicts
new relationships between the various superstring theories and their D-brane
spectra. The implementations of duality symmetries as natural isomorphisms of
K-groups are
discussed. The relationship with the standard cohomological classification
is presented and used to derive an explicit formula for D-brane charges. Some
string
theoretical constructions of the K-theory predictions are also briefly
described.

\end{minipage}
\end{center}

\end{center}

\baselineskip=18pt

\noindent
\vfill
\newpage
\pagestyle{plain}
\setcounter{page}{1}

\tableofcontents
\newpage

\newsection{Introduction and Overview}

The second superstring revolution (see \cite{Mrev} for reviews)
came with the realization that all five consistent
superstring theories in ten dimensions (Type I, Type IIA/B, Heterotic
$SO(32)$/$E_8
\times E_8$) along with 11-dimensional supergravity are merely different
perturbation
expansions of a single 11-dimensional quantum theory called M-Theory
\cite{wittenM}. The evidence for this is provided by the various
non-perturbative
duality relations that connect the different corners of the moduli space of
M-Theory
corresponding to the various string theories. The classic examples are the
self-duality of the Type IIB superstring \cite{schwarz} and the duality between
the
Type I and $SO(32)$ heterotic strings \cite{wittenM,polwitten}.

A new impetus into the duality conjectures came with the realization that
certain
nonperturbative degrees of freedom, known as Dirichlet $p$-branes (or
D$p$-branes
for short),
are charged with respect to the $p+1$-form gauge potentials of the
closed string Ramond-Ramond (RR) sector of Type II superstring theory
\cite{polchinski}.
D$p$-branes are supersymmetric extended objects which form $p+1$-dimensional
hypersurfaces in spacetime
on which the endpoints of open strings can attach (with Dirichlet boundary
conditions).
They can be thought of as
topological defects in spacetime which give explicit realizations of string
solitons
\cite{duffrev}. The crucial observation \cite{polchinski} was that D-branes
have
precisely the correct properties to fill out duality multiplets whose other
elements
are fundamental string states and ordinary field theoretic solitons. D-branes
have
thereby provided a more complete and detailed dynamical picture of string
duality.
They have also provided surprising new insights into the quantum mechanics
of black holes and into the nature of spacetime at very short distance scales.

The important property of D-branes is that they are examples of BPS states,
which
may be characterized by the property that their mass is completely determined
by their
charge with respect to some gauge field. They form ultra-short multiplets of
the
supersymmetry algebra of the string theory, and are thereby stable and
protected from
quantum radiative corrections. Their properties can therefore be analysed
perturbatively at weak
coupling in a given theory and then extrapolated to strong coupling where they
can
be reinterpreted as non-perturbative configurations of the dual theory. For
some
time it was thought that this supersymmetry property, which protects the
D-brane
configurations via non-renormalization theorems, was crucial to
ensure their stability and provide the appropriate non-perturbative tests of
the
duality conjectures.

However, this picture of D-branes has drastically changed in the last year and
a
half. It may be observed \cite{senbps}
that the spectrum of a superstring theory can
contain states which do not have the BPS property, but which are nevertheless
stable
because they are the lightest states of the theory which carry a given set of
conserved quantum numbers which prevent them from decaying.
Such stable non-BPS states can be studied using standard
string perturbation theory and their properties determined at weak coupling. It
has been
realized recently \cite{sentachyon}--\cite{bg} that when these states are
extrapolated
to strong coupling, the resulting non-perturbative configuration behaves in all
respects like an ordinary D-brane (see \cite{senrev} for recent
reviews). This provides a highly non-trivial check
of the non-perturbative duality conjectures beyond the level of BPS
configurations.
For instance, this idea can be applied to heterotic-Type I duality at a non-BPS
level
\cite{seninst,senparticle}.
The $SO(32)$ heterotic string contains states which are not supersymmetric,
but are stable because
they are the lightest states that carry the quantum numbers of the spinor
representation of the $SO(32)$ gauge group. It turns out that the corresponding
non-perturbative stable configuration which is a spinor of $SO(32)$ is the
object
that comes from the bound state of a Type I D-string and anti-D-string (wrapped
on
a circle and with a $\zed_2$-valued Wilson line in the worldvolume). The
D-string
pair becomes tightly bound, forming a solitonic
kink which behaves exactly as a D-particle but
which carries a non-additive charge taking values in $\zed_2$ that prevents one
from
building stacks of non-BPS D-branes.

Generally, this new perspective for understanding D-branes and their conserved
charges treats the branes as topological defects in the worldvolumes of
higher dimensional unstable systems of branes (such as brane-antibrane pairs).
Such systems are unstable because their spectrum contains a tachyonic state
that is
not removed by the usual GSO projection.
However, it is unclear whether these modes are
incurable instabilities in the system or if they play a more subtle role in the
dynamics. A better understanding of the string theory tachyon has been recently
achieved
\cite{senbps}--\cite{seninst},\cite{gbs}--\cite{tachrecent},
with the new belief that the tachyonic mode of an open string stretching
between a D-brane
and an anti-D-brane (or connecting an unstable brane to itself) is a
Higgs-type excitation which develops a stable vacuum expectation value, and the
unstable state decays into a stable state. Configurations of unstable D-branes
can
sometimes carry lower dimensional D-brane charges, so that when the tachyon
field
rolls down to the minimum of its potential and the state decays, it leaves
behind
a state which differs from the vacuum configuration by a lower-dimensional
D-brane charge.
The resulting stable state thereby contains topological defects that correspond
to
stable D-branes.

In addition to producing new D-brane configurations, the bound state
construction of
branes through the process of tachyon condensation can be achieved for the
known spectrum
of supersymmetric branes. This leads to various new connections between
different types
of D-branes which are known as ``descent
relations" \cite{sendescent},\cite{witten}--\cite{os}. These relations form a
remarkable web of
mappings between BPS and non-BPS branes that provides various different ways of
thinking about the origins of D-branes, and they could lead to a better
understanding of
the dynamics of different D-branes and their roles in string theory and in
M-Theory.
The situation in the case of Type II
superstring theory is depicted in fig. 1 \cite{sendescent,senrev}. If we
consider,
say, a D$p$-brane anti-D$p$-brane (or D$\overline{p}$-brane for short) bound
state pair
of Type IIB string theory ($p$ odd), then its open string spectrum contains a
tachyonic
excitation whose ground state corresponds to the supersymmetric vacuum
configuration. However, one can consider instead a tachyonic kink solution on
the brane-antibrane pair which describes a non-BPS D$(p-1)$-brane of the IIB
theory. This system also contains a tachyonic excitation in its worldvolume
field theory, so that one can consider a tachyonic kink solution on the
D$(p-1)$-brane which results in a BPS D$(p-2)$-brane of IIB.

\begin{figure}[htb]
\epsfxsize=5in
\bigskip
\centerline{\epsffile{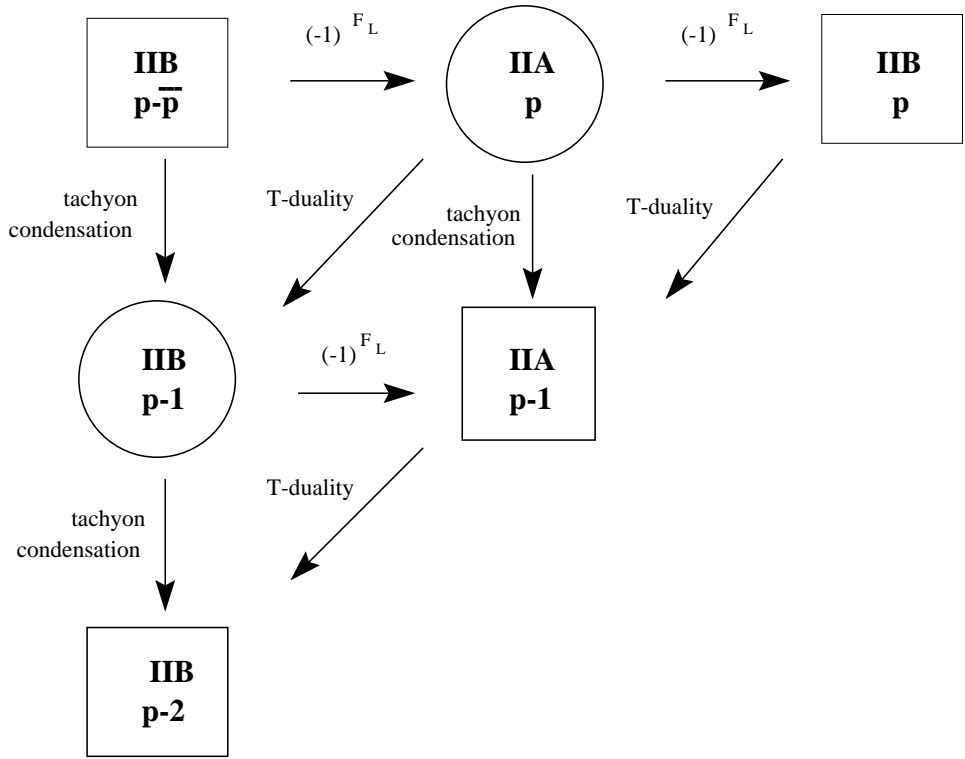}}
\caption{\baselineskip=12pt {\it The relationships between different D-branes
in Type II superstring theory. The squares represent stable supersymmetric BPS
branes or a combination of such a brane with its antibrane, while the circles
depict unstable non-BPS configurations. The horizontal arrows represent the
result of quotienting the theory by the operator $\klein$, the vertical arrows
the effect of constructing a tachyonic kink solution in the brane worldvolume
field theory, and the diagonal arrows the usual $T$-duality transformations.}}
\bigskip
\label{descentfig}\end{figure}

Another set of relations comes from modding out the $p$-$\overline{p}$ brane
pair by the operator $\klein$ which acts as $-1$ on all the Ramond sector
states in the left-moving part of the fundamental string worldsheet, and leaves
all other sectors unchanged. In particular, it exchanges a D-brane with its
antibrane, so
that a brane-antibrane pair is invariant under $\klein$ and it makes sense to
take the quotient of this configuration.
A careful study of the open string spectrum
reveals that the result is a non-supersymmetric D$p$-brane
of IIA, and that a further quotient by $\klein$ yields a supersymmetric
$p$-brane of IIB \cite{sendescent,daspark}. When combined with the usual
$T$-duality transformations between the Type IIB and IIA theories
\cite{Tduality}, we find that
any $p$-brane configuration in Type II superstring theory may be obtained from
any higher dimensional brane configuration. In particular, all branes of the
Type II theories descend from a bound state of D9-D$\overline{9}$ pairs.
Thus all possible stable D-branes appear as topological defects in the
worldvolume
tachyonic Higgs field on the spacetime filling D9-branes, so that the spacetime
filling
brane system provides a universal medium in which all stable D-brane charges
are carried by conventional topological solitons.

The standard coupling in Type II superstring theory
of a BPS D$p$-brane to a closed string $p+1$-form RR potential $C^{(p+1)}$ is
described by the action \cite{polchinski}
\beq
S_{(p)}=\mu_{(p)}\int\limits_{{\cal M}_p}C^{(p+1)}
\label{DpRR}\eeq
where $\mu_{(p)}$ is the $p+1$-form charge of the $p$-brane. In addition, the
topological
charge on the worldvolume
manifold ${\cal M}_p$ of a D$p$-brane couples to the spacetime RR fields
through generalized Wess-Zumino type actions \cite{li,douglasinst} (here we
work in
string units with $2\pi\alpha'=1$ and suppress the dependence on the
Neveu-Schwarz
two-form field $B$ as well as on correction terms due to non-vanishing
manifold curvature):
\beq
S_{\rm WZ}^{(p)}=\mu_{(p)}\int\limits_{{\cal
M}_p}\tr\left(\e^F\right)\wedge\sum_{p'}
C^{(p'+1)}
\label{WZaction}\eeq
where $F$ is the field strength of some gauge field which lives on
${\cal M}_p$. The nature of the gauge fields depends on the configurations of
D-branes. When
$N$ branes are brought infinitesimally close to one another, their generic
$U(1)^N$ gauge
symmetry is enhanced to $G=U(N)$ \cite{wittenp}. This introduces the
possibility of
embedding supersymmetric gauge theories of various dimensions into string
theory (see
\cite{gaugerev} for a review). The coupling \eqn{WZaction} also allows an
alternative interpretation of the topological charge as the RR charge due to
the
presence of lower dimensional branes in the worldvolume of higher dimensional
branes
\cite{douglasinst,strom}. This enables the topological
classification of RR charge in terms of worldvolume defects \cite{semz} in much
the same
spirit as that described above.

In fact, the new understanding of the tachyon in an unstable brane
configuration as a
Higgs type excitation in the spectrum of open string states leads to a
topological
classification of the resulting brane charges when D-branes are
viewed as the tachyonic solitons. Generally, the topological charges of these
objects are determined by the homotopy groups of a homogeneous space $G/H$,
where
$G$ is a compact Lie group and $H$ is a closed subgroup of $G$. The fibration
\beq
H~{\buildrel i\over\hookrightarrow}~G~{\buildrel\pi\over\longrightarrow}~G/H
\eeq
with $i$
the inclusion and $\pi$ the canonical projection, induces a long exact sequence
of homotopy groups,
\beq
\dots~\longrightarrow~\pi_{n-1}(H)~{\buildrel{i^*}\over\longrightarrow}~
\pi_{n-1}(G)~{\buildrel{\pi^*}\over\longrightarrow}~\pi_{n-1}(G/H)~{\buildrel
{\partial^*}\over\longrightarrow}~\pi_{n-2}(H)~\longrightarrow~\dots \ .
\label{longexacthtpy}\eeq
In the present case, $G$ is the worldvolume gauge group
of a given configuration of branes and the tachyon scalar field $T$ is a Higgs
field for the breaking of the gauge symmetry down to the subgroup $H$. The
tachyonic soliton must be accompanied by a worldvolume gauge field $A$ of
corresponding topological charge in the unbroken subgroup of the gauge group,
in order that
the energy per unit worldvolume of the induced lower dimensional brane be
finite.
It can be argued \cite{senbps,seninst} that the brane worldvolume field theory
admits
finite energy, static soliton
solutions which have asymptotic pure gauge configurations at infinity,
\beq
T\simeq T_v\,U~~~~~~,~~~~~~A\simeq i\,U^{-1}\,dU
\label{puregauge}\eeq
where $T_v$ is a constant, and $U$ is a $G/H$ valued function
corresponding to the identity map (of a given winding number) from the
asymptotic boundary of the worldvolume soliton to the group manifold of the
space $G/H$
of vacua. This leads to topologically distinct sectors in the space of all
field
configurations, and the charges which distinguish these sectors take values in
the
appropriate homotopy group of the vacuum manifold.
Precisely, if the induced brane configuration has codimension $n$ in the higher
dimensional worldvolume, then the corresponding soliton carries topological
charge taking values in $\pi_{n-1}(G/H)$. This homotopy group may be computed
using
the exact sequence \eqn{longexacthtpy} \cite{semz}
(for instance, if the induced boundary homomorphism
$\partial^*$ is trivial mapping, so that $\ker\partial^*=\pi_{n-1}(G/H)$, then
$\pi_{n-1}(G/H)=\pi_{n-1}(G)/\pi_{n-1}(H)$).

The coupling \eqn{DpRR} would seem to imply that, since the massless RR fields
$C^{(p+1)}$
are differential forms, the RR charges of D-branes are determined by cohomology
classes,
i.e. by integrating the $C^{(p+1)}$ over suitable cycles of the spacetime
manifold $X$.
However, the new interpretation of D-brane charge as a topological charge
actually
suggests a different characterization (at least when all spacetime dimensions
are much
larger than the string scale so that no new stringy phenomena occur).
Let us consider Type IIB superstring theory, and
go back to the realization of RR charge in terms of a configuration of $N$
9-branes and
$M$ $\overline{9}$-branes. Type II theories have no gauge group, so in order to
cancel
the tadpole anomaly there must be the same number of 9-branes and
$\overline{9}$-branes,
$N=M$. The 9-branes and $\overline{9}$-branes fill out the spacetime manifold
$X$. The system
of $N$ 9-branes carries a $U(N)$ gauge bundle $E$ and the system of $N$
$\overline{9}$-branes
carries a $U(N)$ gauge bundle $F$. The system of $9-\overline{9}$-branes can
therefore be
labelled by a pair of $U(N)$ vector bundles $(E,F)$ which characterize the
gauge field
configurations corresponding to the soliton.

We would now like to determine when two such pairs of bundles $(E,F)$ and
$(E',F')$ should
be considered equivalent. They should be regarded the same if they determine
the same
topological class of the soliton. It turns out that the basic equivalence
relation is
brane-antibrane creation and annihilation. The $9-\overline{9}$ and
$\overline{9}-9$
open strings have the opposite GSO projection, so that the massless vector
fields are
projected out and the tachyonic mode survives \cite{gbs}. As we have discussed
above,
it is conjectured that the
instability associated with the tachyon represents a flow toward annihilation
of the
brane-antibrane pair, i.e. by giving the tachyon field a suitable expectation
value one
can return to the vacuum state without this pair \cite{senbps}--\cite{seninst}.
Thus if we add an equal number $M$ of 9-branes and $\overline{9}$-branes with
the same
$U(M)$ gauge bundle $H$ on them, then the tachyon field associated with the
open strings
stretched between the 9-branes and the $\overline{9}$-branes is a section of
a trivial bundle, and hence it can condense to the minimum of its potential
everywhere
on the $9-\overline{9}$ worldvolume. We suppose that any such collection of
brane-antibrane
pairs can be created and annihilated, so that the configuration is equivalent
to the
vacuum which carries no D-brane charges
(this is much like the situation in ordinary quantum field theory). We conclude
that adding such pairs has no effect on the topological class of the soliton,
i.e.
the pair $(E,F)$ can be smoothly deformed to the pair $(E\oplus H,F\oplus H)$
for any such
bundle $H$. Thus in terms of the conserved D-brane charges, a property of the
system that
is invariant under smooth deformations, we conclude that RR-charge is
classified
topologically by
specifying a pair of $U(N)$ vector bundles $(E,F)$ subject to the equivalence
relation
\beq
(E,F)\sim(E\oplus H,F\oplus H)
\eeq
for any $U(M)$ vector bundle $H$. In a manner of speaking (that will soon be
made
precise), the D-brane charge is determined by the ``difference" between the
Chan-Paton
gauge bundles on the 9-branes and anti-9-branes.

The mathematical conditions described above define the so-called K-theory group
$\K(X)$
of the spacetime $X$. This proposal that D-brane charge takes values in the
K-theory of
spacetime was made initially in \cite{witten}, and then
extended in \cite{horava,os},\cite{garcia}--\cite{vancea}. However, the
solitonic
description of D-brane states discussed above was not the first evidence that
RR-charge
should be understood in terms of K-theory rather than cohomology. The strongest
prior
proposal \cite{minmoore} had been the observation (extending earlier
calculations
in \cite{li,douglasinst,strom,cheung})
that when a D-brane wraps a submanifold $Y$ of spacetime,
its RR-charge depends on the geometry of $Y$, of its normal bundle and on the
gauge
fields on $Y$ in a manner which suggests that D-brane charges take values in
$\K(X)$.
Other earlier hints at a connection with K-theory may be found in
\cite{horavaeq,quivers}.

The arguments presented above for spacetime filling branes clearly show that
when a
D-brane wraps a submanifold $Y$ of the spacetime $X$, its charges are
classified by the group
$\K(Y)$. One of the profound observations of \cite{witten} is that there is a
standard
K-theory construction, called the Thom isomorphism, which embeds $\K(Y)$ into
$\K(X)$ and
is equivalent to the bound state construction of D-branes described above, and
hence
to the representation of all branes in terms of 9-branes and antibranes. In
this way,
one gets a complete classification of D-brane charges in terms of the topology
of the
underlying spacetime manifold. The main feature of K-theory which parallels the
above
soliton constructions is its intimate relationship with homotopy theory.
Moreover, another standard K-theory construction, known
as the Atiyah-Bott-Shapiro construction, can be used to obtain explicit forms
for the
classical gauge field configurations which live on a given D-brane. These
remarkable facts
have been used to reproduce the construction of the Type I non-BPS D-particle
discovered
in \cite{seninst}, and to predict
the existence of new D-branes in the spectrum of the Type I theory and
also other superstring theories (the homotopic soliton construction of the Type
I D-string
was first carried out in \cite{dufflu}). Indeed, the K-groups of a spacetime
can be much more general than
the corresponding cohomology groups. In many instances the K-groups can have
torsion
while the cohomology groups are torsion free, lending a natural explanation to
the fact
some D-branes (such as the Type I 0-brane) carry torsion charges
\cite{minmoore}.
The recent string theoretical construction of these new Type I objects
\cite{frau} illustrates the
strong predictions that can follow from the K-theory formalism.
In addition, for the spectrum of supersymmetric D-branes (where the RR charge
is integer
valued) there is a mapping, known as the Chern homomorphism, onto cohomology,
thereby making
contact with the expectations which follow from the coupling \eqn{DpRR} to the
spacetime
RR fields.

\subsection{Outline}

In this paper we will review the mathematical formalism of topological K-theory
and its use as
a systematic tool in the topological classification of D-branes in superstring
theory.
As K-theory now turns out to be at the forefront of mathematical physics as far
as its
applications to string theory are concerned and, while cohomology and
differential
geometry are already well-known to most
theoretical physicists, K-theory may seem rather obscure, we have
attempted to merge the mathematics with the physics in such a way that the
naturality
of K-theory as a classification tool is evident. The main purpose will be to
collect all
the relevant mathematical material in one place
in a way that should be accessible to a rather general audience of string
theorists and
mathematicians. The level of this review is geared at string theorists with a
relatively
good background in algebraic topology and differential geometry (at the level
of the
books \cite{spanier} and the review article \cite{eguchi}),
and at mathematicians with a rudimentary background in string theory (at the
level of the
books \cite{joesbook}). More references and background will be given as we
proceed.

Before giving a quick outline of the structure of this paper, let us briefly
indicate
the omissions in our presentation, which can also be taken as directions for
further
research. Throughout this review we will consider only superstring
compactifications for
which the curvature of the Neveu-Schwarz $B$-field is cohomologically
trivial. The problems with
incorporating this two-form field are discussed in \cite{witten}, and at
present it is
not fully understood what the appropriate K-theory should be in these
instances. Some
steps in this direction have appeared recently in \cite{vancea,yi,kalkkinen}.
Related to this problem is how to correctly incorporate $S$-duality into
K-theoretic
terms, and in particular the description of the self-duality of Type IIB
superstring
theory. The analysis of \cite{yi} is a first step in this direction. Another
related aspect
is making contact with the correct construction for M-Theory. The description
of M-branes
has been discussed in \cite{vancea,yi} and the appropriate relations in Matrix
Theory
in \cite{ahh}. Using the approach of \cite{vancea}, which is based on {\it
algebraic}
K-theory, there may be an intimate connection with the gauge bundles for Matrix
Theory
compactifications used in \cite{ncg} based on noncommutative geometry. These
are all
problems that do not as of yet have a natural description in terms of K-theory.
It is
hoped that the exposition of this paper, in addition to providing the reader
with the
necessary tools to pursue the subject further, could provoke some detailed
investigations of such matters. Finally, we note that the analysis given in the
following
is meant to serve only as a topological classification of the spectra of branes
in
the various string theories. The second step, which is omitted in this review,
is to
actually carry out string
theoretical constructions of the D-branes predicted by the K-theory
formalism and hence describe their dynamics, especially for the new non-BPS
states.
This is addressed in \cite{senbps}--\cite{senrev},\cite{witten,gukov,bgh,frau}.
Indeed,
the K-theory classification of D-branes has revealed many interesting new
effects and
constructions in string theory. We shall only briefly touch upon such matters
here, in order
to keep the presentation as self-contained as possible.

The structure
of this review is as follows. In section 2 we present a thorough,
self-contained
introduction to the ideas and fundamental constructions of topological
K-theory. This
section deals with the mathematical highlights of the formalism that will
follow in
subsequent sections. Many tools for computing K-groups are described which are
useful in
particular to the various superstring applications that we shall discuss. They
will in
addition turn out to give many unexpected connections between the various
different
superstring theories. Here the reader is
assumed to have a good background in algebraic topology and the theory of fiber
bundles.
In section 3 we begin the classification of D-branes using K-theory, starting
with the simplest
case of Type IIB superstring theory. We start by giving a quick description of
the relevant
physics of the brane-antibrane pair. For more details, the reader is refered to
the
original papers \cite{senbps}--\cite{bg} and the recent review articles
\cite{senrev}. We
then describe the bound state construction and how it naturally implies the
pertinent
connection to K-theory, following \cite{witten} for the most part. In section 4
we carry
out the analogous constructions for Type IIA superstring theory. The relevant
K-theory
for Type IIA D-branes was suggested in \cite{witten} and developed in detail in
\cite{horava}. In section 5 we then move on to Type I superstring theory, in
which new
non-BPS D-branes are predicted, again following \cite{witten} for a large part.
In section 6 we turn our attention to orbifold and orientifold superstring
theories. Orbifolds
were dealt with in \cite{garcia} while orientifolds were described in
\cite{gukov,os}. The
various $T$-dual orientifolds of the Type I theory were discussed in
\cite{horava,bgh,os}.
Finally, we conclude our analysis in section 7 with a description of the
modifications of
the previous constructs when global topology of the spacetime and of the
worldvolume
embeddings is taken into account. Here we present the global version
of the bound state construction, as described in \cite{witten}, highlighting
the extra
structures and care that must be taken into account as compared to previous
cases of flat
manifolds. We then move on to describe the appropriate K-theory for dealing
with general
superstring compactifications \cite{bgh}, and introduce a useful connection to
the index
theory of Dirac operators which has been at the forefront of many applications
in
theoretical physics (see \cite{eguchi} for a comprehensive review). We then
apply these
ideas to describe how the celebrated $T$-duality transformations of D-branes
are represented
as natural isomorphisms of K-theory groups \cite{hori,bgh}. We end the review
with
what can be considered as the origin of the material discussed in this paper,
the
derivation of the K-theoretic
charge formula of \cite{minmoore}. This formula gives the explicit relationship
between the
K-theoretic and cohomological descriptions of RR charge, and it thereby allows
one to
explicitly compute D-brane charges in terms of densities integrated over the
spacetime
manifold.

\newsection{Elements of Topological K-Theory}

K-theory was first introduced in the 1950's by Grothendieck in an alternative
formulation of the Riemann-Roch theorem (see \cite{borel}).
It was subsequently developed in the
1960's by Atiyah and Hirzebruch who first introduced the general K-theory
group $\K(X)$ of a topological space $X$ \cite{ah3}. Since then, K-theory has
become an indispensable tool in many areas of topology, differential
geometry and algebra. Generally speaking, topological K-theory can be regarded
as a
cohomology theory for vector bundles that emphasizes features which become
prominent as the ranks of the vector bundles become large. Actually, it is an
example of a {\it generalized} cohomology theory, in that K-theory does not
satisfy all of the Eilenberg-Steenrod axioms \cite{eilensteen} of a
cohomology theory (it satisfies all axioms except the dimension axiom which
defines in advance the cohomology of the topological space consisting of a
single point, e.g. $H^n(\point, \zed)=\delta^{n,0}\,\zed$). This extensive
section will review the core of the mathematical material that we will need
later on in this paper and no mention of physics will be made until section
3. We will not give a complete review of the material, but rather focus only
on those aspects that are useful in superstring applications. For more complete
expositions of the subject, the reader is refered to the books
\cite{atiyah}--\cite{husemoller}, where the proofs of the theorems quoted in
the following may also be found.

\subsection{The Grothendieck Group}

In this subsection we shall start with an abstract formulation that will
naturally lead to the definition of the group $\K(X)$. Although the
formalism is not really required for this definition, it will be of use
later on, and moreover it is this definition which allows one to generalize
the K-theory of topological spaces to more exotic groups, such as the K-theory
of vector spaces, $C^*$-algebras, etc., which could prove important in
future applications of the general formalism of K-theory to string theory and
M-theory. Let ${\cal A}$ be an abelian monoid, i.e. a set with an
addition which satisfies all the axioms of a group except possibly the
existence of an inverse.
One can naturally associate to ${\cal A}$ an abelian group $S({\cal A})$ by
the following construction. Consider the equivalence relation $\sim$ on the
Cartesian product monoid ${\cal A}\times{\cal A}$ with $(E,F)\sim (E', F')$
if there exists an element $G\in {\cal A}$ such that
\beq
E+F'+G=F+E'+G\ .
\eeq
The abelian group $S({\cal A})$, called the {\it symmetrization} of
${\cal A}$, is
then defined to be the set of equivalence classes of such pairs:
\beq
S({\cal A})={\cal A}\times{\cal A}/\sim\ .
\eeq
The equivalence class of the pair $(E,F)$ is denoted by $[(E,F)]$ and the
inverse of such an element in $S({\cal A})$ is $[(F,E)]$.
This follows from the fact that for any $E\in{\cal A}$, $[(E,E)]$ is a
representative of the zero element in $S({\cal A})$. An alternative
definition of $S({\cal A})$ is obtained by using in ${\cal A}\times{\cal A}$
the equivalence relation
$(E,F)\sim (E',F')$ if there exist $G,H\in{\cal A}$ such that
\beq
(E,F)+(G,G)=(E',F')+(H,H)\ .
\eeq
As a simple example, for ${\cal A}=\zed^+$ (the non-negative integers under
addition), we have $S({\cal A})=\zed$. Also, for ${\cal A}=\zed-\{0\}$
(an abelian monoid under multiplication), we have $S({\cal A})=\rat -\{0\}$.

The completion $S({\cal A})$ of the monoid ${\cal A}$ can be
characterized by the following universal property. For any abelian
group $\cal G$, and any homomorphism $f:{\cal A}\to{\cal G}$ of the
underlying monoids, there exists a unique homomorphism $\tilde{f}:S({\cal
A})\to\cal G$ such that $\tilde{f}\circ s=f$, where $s$ is the
natural map ${\cal A}\rightarrow S({\cal A})$ defined by
$s(E)=[(E,0)]$. This means that
$S({\cal A})$ is the ``smallest'' abelian group that can be built from the
abelian monoid ${\cal A}$ and it implies, in particular, that if $\cal A$ is
itself a group, then $S({\cal A})=\cal A$.
In general, this property implies that the map ${\cal A}\rightarrow S({\cal
A})$ is a covariant functor from the category of abelian monoids to the
category of abelian groups, i.e. if
$\gamma: {\cal A}\rightarrow {\cal B}$ is any homomorphism of monoids,
then there is a unique group homomorphism
$s(\gamma):S({\cal A})\rightarrow S({\cal B})$ such that the following
diagram is commutative:
\beq
{\begin{array}{ccc}
{\cal A}&\stackrel{\gamma}{\longrightarrow}&{\cal B}\\&
&\\{\scriptstyle s}\downarrow& &\downarrow{\scriptstyle s}\\&
&\\S({\cal A})&\stackrel{s(\gamma)}{\longrightarrow}&S({\cal B})\end{array}}
\eeq
and such that $s(\gamma\circ\gamma')=s(\gamma)\circ s(\gamma')$, $s({\rm
Id}_{{\cal A}})={\rm Id}_{S({\cal A})}$ (here ${\rm Id}_{\cal A}$ denotes
the identity morphism on ${\cal A}$).

An important example to which this construction applies is the case that
${\cal C}$ is an additive category with $\Phi({\cal C})$ the set of
isomorphism classes of elements $E\in{\cal C}$, which we denote by $[E]$.
$\Phi({\cal C})$ becomes an abelian monoid if we define
$[E]+[F]\equiv [E\oplus F]$ (this is well-defined since
the isomorphism class of $E\oplus F$ depends only on the isomorphism classes
of $E$ and $F$). The {\it Grothendieck group} of
${\cal C}$ is defined as $\K({\cal C})=S(\Phi({\cal C}))$.
Note that every element of $\K({\cal C})$ can be written as a formal
difference $[E]-[F]$ and that $[E]-[F]=[E']-[F']$ in $\K({\cal C})$ if and
only if there exists a $G\in{\cal C}$ such that
$ E\oplus F'\oplus G \cong E'\oplus F\oplus G$. Notice also that $[E]=[F]$ if
and only if there is a $G\in{\cal C}$ such that
$E\oplus G \cong F \oplus G$. As a simple example,
let ${\bb F}$ be an algebraic field (e.g. ${\bb F}=\real$ or $\complex)$,
and let ${\cal C}$ be the category of finite dimensional vector spaces
over ${\bb F}$ whose morphisms are linear transformations. Then, since such
finite dimensional vector spaces are characterized uniquely by their dimension,
$\Phi({\cal C})=\zed^+$ implying that $\K({\cal C})=\zed$.

\subsection{The Group $\K(X)$}

We will use the construction of the previous subsection for a
classification of vector bundles over compact manifolds.
Let $X$ be a compact manifold and let ${\cal C}={\rm Vect}(X)$ be the
additive category of complex vector bundles
over $X$ with respect to bundle morphisms and Whitney sum (later on we shall
also consider real and  quaternionic vector bundles).
Define $I^k$ to be the trivial bundle of rank $k$ over $X$, i.e.
$I^k\cong X\times \complex^k$.
The space of all vector bundles can
be partitioned into equivalence classes as follows. Bundle $E$ over
$X$ is said to be {\it stably equivalent} to bundle $F$, denoted by $E\sim F$,
if there exists positive  integers $j,k$ such that
\beq
E\oplus I^j\cong F\oplus I^k\ .
\eeq
The corresponding equivalence classes in
${\rm Vect}(X)/\sim$ are called stable equivalence classes. It is easily
seen that if $E,F$ and $G$ are vector bundles over $X$ then
\beq
E\oplus G\cong F\oplus G ~~~~\Rightarrow E\sim F\ ,
\label{EsimF}
\eeq
i.e. $E$ and $F$ are stably equivalent.
In the proof one uses the fact that there exists
a bundle $G'$ such that $G\oplus
G'$ is trivial (according to Swan's Theorem this requires $X$ to be a
compact Hausdorff manifold). However, one cannot conclude from the left-hand
side of (\ref{EsimF}) that $E$ and
$F$ are isomorphic as vector bundles. For example, consider
$E=T{\bf S}^n$, the tangent bundle of the $n$-sphere ${\bf S}^n$,
and $G=N({\bf S}^n,\real^{n+1})$, the normal
bundle of ${\bf S}^n$ in $\real^{n+1}$. $G$  has a global section
given by an outward-pointing unit normal
vector, which implies that it is trivial with $N({\bf
S}^n,\real^{n+1})\cong I^1$. Furthermore, we have the usual relations
\beq
I^{n+1}\cong T\real^{n+1}\cong E\oplus G\cong E\oplus I^1\ .
\eeq
However, $E=T{\bf S}^n$ is generally not trivial and therefore not equal to
$I^n$ (in fact $T{\bf S}^n$ is only trivial for parallelizable spheres
corresponding to $n=1,3,7$). So generally, $T{\bf S}^n$ is only
stably trivial.

This example demonstrates that the space of vector bundles over $X$ is not a
group under the Whitney sum of vector bundles, but
rather a monoid, as there is no subtraction defined for vector bundles.
A group can, using the previous setup, be constructed as follows.
The {\it K-group} of a compact manifold $X$ is defined to be the Grothendieck
group of the category ${\rm Vect}(X)$, $\K(X)\equiv \K({\rm Vect}(X))$, or
\beq
\K(X)={\rm Vect}(X)\times {\rm Vect}(X)/\sim\ ,
\eeq
where we have defined an equivalence relation in ${\rm Vect}(X)\times {\rm
Vect}(X)$ according to $(E,F)\sim (E',F')$ if there exists a vector bundle
$G\in {\rm Vect}(X)$ such that
\beq
E\oplus F' \oplus G \cong E' \oplus F \oplus G\ .
\eeq
An equivalent definition of $\K(X)$ is that the pair of bundles $(E,F)$
is taken to be equivalent to $(E\oplus H,F\oplus H)$ for any bundle $H$.
Often the notation $\K^0(X)$ or $\KU(X)$ is also used for this group.
An element of $\K(X)$ is written as $[(E,F)]$. In $\K(X)$ the
unit (zero) element is $[(E,E)]$ so the inverse of the class $[(E,F)]$ is
$[(F,E)]$. Any element $[(E,F)]$
can therefore be identified with $[E]-[F]$ where $[E]=[(E,I^n)]$.
Furthermore, $[E]=[F]$ in $\K(X)$ if and only if $E$ and $F$ are stably
equivalent. The elements of $\K(X)$ are called {\it virtual bundles}.
The map $X\rightarrow \K(X)$ is a contravariant functor from the category of
compact topological spaces to the category of abelian groups, i.e.
if $f:X\rightarrow Y$ is continuous, then it
induces the usual pullback map on vector bundles over $Y$, thus inducing a map
$f^*:{\rm Vect}(Y)\rightarrow {\rm Vect}(X)$ and hence a homomorphism
$\K(Y)\rightarrow\K(X)$.

The K-groups have the following important homotopy invariance property.
Consider two homotopic maps $f,g: X\rightarrow Y$. Then for any vector bundle
$E\rightarrow Y$, there is an isomorphism of vector bundles over $X$:
\beq
f^*E\cong g^*E\ .
\eeq
{}From this it follows that the maps induced by $f$ and $g$ on K-groups are
the same:
\beq
s(f)=s(g):\K(Y)\longrightarrow \K(X)\ .
\eeq
For example, if $X$ is a compact manifold which is contractible to a
point, then we may deduce that $\K(X)=\K({\rm pt})= \zed$. Geometrically, this
expresses the well-known fact
that any vector bundle over a contractible space $X$ is necessarily
trivial, so that the corresponding K-theory of $X$ is also trivial.

\subsection{Reduced K-Theory}

The fact that a vector bundle over a point is just a vector space, so that
$\K({\rm pt})=\zed$, motivates the introduction of a reduced K-theory
in which the topological space consisting of a single point has
trivial cohomology, $\wt{\K}({\rm pt})=0$, and therefore also $\wt{\K}(X)=0$
for any contractible space $X$.
Let us fix a basepoint of $X$ and consider the collapsing and inclusion
maps:
\beq
p:X\longrightarrow {\rm pt}~~~~,~~~~i:{\rm pt}\hookrightarrow X\ .
\eeq
These maps induce, respectively, an epimorphism and a monomorphism
of the corresponding K-groups:
\beq
p^*:\K({\rm pt})= \zed\longrightarrow \K(X)~~~~,~~~~i^*:
\K(X)\longrightarrow \K({\rm pt})= \zed\ .
\eeq
We then have the exact sequences of groups:
\bea
& &0\longrightarrow \zed\stackrel{p^*}{\longrightarrow} \K(X)\longrightarrow
\wt{\K}(X)
\longrightarrow 0\nonumber \\
& &0\longrightarrow \wt{\K}(X)\longrightarrow
\K(X)\stackrel{i^*}{\longrightarrow}
\zed
\longrightarrow 0\ .
\eea
These sequences have a canonical splitting so that the homomorphism
$i^*$ is a left inverse of $p^*$. The kernel of the map $i^*$, or equivalently
the cokernel of the map $p^*$, is called the {\it reduced K-theory group} and
is denoted by $\wt{\K}(X)$,
\beq
\wt{\K}(X)= \ker i^*={\rm coker}\,p^*
\eeq
and therefore we have the fundamental decomposition
\beq
\K(X)= \zed\oplus \wt{\K}(X)\ .
\label{Kdecomp}
\eeq

Given a vector bundle  $E\rightarrow X$, let $E_x$ denote the fiber of
$E$ over $x\in X$. We define the {\it rank function} $\rk:X\rightarrow
\zed^+$ by $\rk(x)=\dim_{\complexs} E_{x}$.  Since $E$ is locally
trivial, the rank function is locally constant, and the space of all
locally constant $\zed^+$-valued functions on $X$ forms an abelian
monoid $H^0(X, \zed^+)$ under pointwise addition.
The map $\rk$ extends naturally to a group homomorphism
\bea
\rk: \K(X)&\longrightarrow& H^0(X,\zed)\nonumber\\
\rk\Bigl([E]-[F]\Bigr)&=&\rk(E)-\rk(F)\ .
\label{rank}\eea
The integer (\ref{rank}) is called the
{\it virtual dimension} of $[(E,F)]\in \K(X)$.
Let $\K'(X)=\ker\,\rk$. Then the short exact sequence
\beq
0\longrightarrow \K'(X)\longrightarrow \K(X)\stackrel{\rk}
{\longrightarrow}H^0(X,\zed)\longrightarrow 0
\label{rankexact}
\eeq
has a canonical split (i.e. $\rk$ has a right inverse), so that if $X$ is
connected, then $H^0(X,\zed)= \zed$ and
\beq
\wt{\K}(X)= \K'(X)=\ker\,\rk\ .
\eeq
In this case $\wt{\K}(X)$ is the subgroup of $\K(X)$ whose elements
have virtual dimension zero (i.e. consisting of equivalence classes of
pairs of vector bundles $[(E,F)]$ of equal rank).
The fundamental examples are $\wt{\K}({\bf
S}^{2n})=\zed$ and $\wt{\K}({\bf S}^{2n+1})=0$ for any positive
integer $n$ (these groups are computed in section 2.7).
Note that the rank function (\ref{rank}) naturally
gives an assignment $\ch_0(E)$ in the zeroth \v{C}ech cohomology group of $X$
which depends only on the stable equivalence class of the vector
bundle $E$ in $\K(X)$. This is the first, basic example of the Chern
character which will be discussed in section 7.1.

For the physical applications of K-theory, which are presented in
the subsequent sections, we shall mostly work
in K-theory with compact support. This means that for each
class $[(E,F)]$, there is a map $T:E\to
F$ which is an isomorphism of vector bundles outside an open set $U\subset X$
whose closure $\overline U$ is compact. This condition
automatically implies that $E$ and $F$ have the same rank, and hence we shall
mostly deal with the reduced K-group $\widetilde{\K}(X)$.
The corresponding virtual bundle may then be represented as
\beq
\Bigl[(E\,,\,F)\Bigr]=\Bigl[(\ker T\,,\,{\rm coker}\,T)\Bigr]
\label{grothrep}\eeq
When $X$ is not compact, we define $\K(X)=\widetilde{\K}(X^+)$, where
$X^+$ is the one-point compactification of $X$.

\subsection{Higher K-Theory and Bott Periodicity}

Starting with $\K(X)$ there is a
natural way to define so-called  ``higher'' K-groups. These groups are labelled
by a positive integer $n=0,1,2,\ldots$ and are defined according to
\beq
\K^{-n}(X)=\K(\Sigma^nX)\ ,
\label{higherK}
\eeq
where $\Sigma^nX\equiv{\bf S}^n\wedge X$ is the $n$-th reduced suspension
of the topological space $X$.  Here $X\wedge Y= X\times Y/ (X\vee Y)$ is the
smash product of $X$ and $Y$, and $X\vee Y$ is their reduced join,
i.e. their disjoint union with a base point of each space identified,
which can be viewed as the subspace $X\times{\rm pt}\amalg{\rm pt}\times Y$
of the Cartesian product $X\times Y$. For $X={\bf S}^n$, the $n$-sphere,
one has $\Sigma{\bf S}^n={\bf S}^1\wedge{\bf S}^n\cong {\bf
  S}^{n+1}$. Alternatively, higher K-groups can be defined through the
{\it suspension isomorphism}:
\beq
\K^{-n}(X)=\K(X\times \real^n)\ ,
\label{suspension}
\eeq
where it is always understood that K-theory with compact support is used.
In contrast to conventional cohomology theories, one does not in this way
generate an infinite number of higher K-groups because of the fundamental
{\it Bott periodicity theorem}:
\beq
\K^{-n}(X)=\K^{-n-2}(X)\ ,
\label{complexbott}
\eeq
which states that the complex K-theory functor $\K^{-n}$ is periodic
with period two. The same is true for the reduced functor $\wt{\K}^{-n}$
since the analogous definition to (\ref{higherK}) holds for reduced K-theory.
However, the higher reduced and unreduced K-groups differ according to
\beq
\K^{-n}(X)=\wt{\K}^{-n}(X)
\oplus \K^{-n}({\rm pt})\ .
\eeq
Since $\K({\rm pt})=\zed$, $\K^{-1}({\rm
pt})=0$, using Bott periodicity we see that for $n$ even these groups
differ by a subgroup $\zed$ (as in (\ref{Kdecomp})), while for $n$ odd they are
identical, so that $\wt{\K}^{-1}(X)=\K^{-1}(X)$.
Here the basic examples are $\K^{-1}({\bf S}^{2n})=0$ and
$\K^{-1}({\bf S}^{2n+1})=\zed$ for any positive integer $n$.

Note that for any decomposition $X=X_1\amalg X_2\amalg\cdots\amalg X_n$ of
$X$ into a disjoint union of open subspaces, the inclusions of the $X_i$
into $X$ induce a decomposition of K-groups as
$\K^{-n}(X)=\K^{-n}(X_1)\oplus\K^{-n}(X_2)\oplus\cdots\oplus\K^{-n}(X_n)$
(this follows from the fact that a bundle over $X$ may be characterized by its
restriction to $X_i$). However, this is not true
for the reduced K-functor, since for example
$\wt{\K}({\bf S}^0)=\zed$ but $\wt{\K}({\rm pt})=0$. More generally,
given two {\it closed} subspaces $X_1$ and $X_2$ of a locally compact space
$X$ with $X=X_1\cup X_2$, there is the long exact sequence
\bea
& &\ldots\longrightarrow\K^{-n-1}(X_1)\oplus\K^{-n-1}(X_2)
\stackrel{v}{\longrightarrow}
\K^{-n-1}(X_1\cap X_2)\longrightarrow\nonumber\\
& &~~~~~~\stackrel{\zeta}{\longrightarrow}\K^{-n}(X_1\cup X_2)
\stackrel{u}{\longrightarrow}\K^{-n}(X_1)\oplus \K^{-n}(X_2)
\stackrel{v}{\longrightarrow}\K^{-n}(X_1\cap X_2)\longrightarrow\ldots\ ,
\nn\\& &
\eea
where $\zeta$ is the zig-zag homomorphism, and $u$
and $v$ are defined by
$u([E])=\left([E|_{X_1}],[E|_{X_2}]\right)$ and $v([E_1],[E_2])
=[E_1|_{X_1\cap X_2}]-[E_2|_{X_1\cap X_2}]$. The corresponding long exact
sequence for two {\it open} subspaces $U_1$ and $U_2$ of $X$ with
$X=U_1\cup U_2$ is
\bea
& &\ldots\longrightarrow\K^{-n-1}(U_1)\oplus\K^{-n-1}(U_2)\longrightarrow
\K^{-n-1}(U_1\cup U_2)\longrightarrow\nonumber\\
& &~~~~~~\longrightarrow\K^{-n}(U_1\cap U_2)
\longrightarrow\K^{-n}(U_1)\oplus \K^{-n}(U_2)
\longrightarrow\K^{-n}(U_1\cup U_2)\longrightarrow\ldots\ .\nn\\& &
\eea
These latter two sequences are the analogs of the usual Mayer-Vietoris
long exact sequences in cohomology.

\subsection{Multiplicative Structures}

As in any cohomology theory, $\K(X)$ and $\wt{\K}(X)$
are actually rings. In this case the multiplication is induced by the
tensor product $E\otimes F$ of vector bundles over $X\times X$:
\beq
\K(X)\otimes_{\zeds}\K(X)\longrightarrow\K(X)\ ,
\eeq
and is defined by
\beq
\Bigl[(E\,,\,F)\Bigl]\otimes\Bigl[(E'\,,\,F')\Bigr]\equiv
\Delta^*\Bigl[(E\otimes E'\oplus F\otimes F'\,,\, E\otimes F'\oplus F\otimes
E')\Bigr]
\label{ringproduct}\eeq
where $\Delta:X\to X\times X$ is the diagonal map.
This multiplication comes from writing $[(E,F)]=[E]-[F]$ and formally using
distributivity of the tensor product acting on virtual bundles.
Note that it acts on $[(E,F)]$ as if $E$'s are bosonic and
$F$'s are fermionic. It is therefore an example of a $\zed_2$-graded
tensor product. There is another product, called the external tensor
product or {\it cup product}, which is a homomorphism
\beq
\K(X)\otimes_{\zeds}\K(Y)\longrightarrow \K(X\times Y)
\label{cupproduct}\eeq
defined as follows. Consider the canonical projections
$\pi_{X}:X\times Y\rightarrow X$ and $\pi_Y:X\times Y\rightarrow Y$.
These projections induce homomorphisms between K-groups according to
\beq
\pi_X^{*}:\K(X)\rightarrow \K(X\times Y)\ \ , \ \
\pi_Y^{*}:\K(Y)\rightarrow \K(X\times Y)\ .
\eeq
Then the cup product of $([E],[F])\in \K(X)\otimes_\zeds\K(Y)$ is the
class $[E]\otimes [F]$ in $\K(X\times Y)$, with
\beq
[E]\otimes [F]\equiv\pi_X^{*}\Bigl([E]\Bigr)\otimes \pi_Y^{*}\Bigl([F]\Bigr)\ .
\eeq

Consider now the canonical injective
inclusion and surjective projection maps:
\beq
X\vee Y \hookrightarrow X\times Y \longrightarrow X\wedge Y\ .
\eeq
The $\wt{\K}^{-n}$ functor is contravariant, and thus, as in any
cohomology theory, this induces a split short exact sequence of K-groups
\beq
0\longrightarrow \wt{\K}^{-n}(X\wedge Y) \longrightarrow
\wt{\K}^{-n}(X\times Y)
\longrightarrow \wt{\K}^{-n}(X\vee Y) \longrightarrow 0\ ,
\eeq
from which it follows that
\bea
\wt{\K}^{-n}(X\times Y)&=& \wt{\K}^{-n}(X\wedge Y)\oplus
\wt{\K}^{-n}(X\vee Y)\nonumber\\
&=& \wt{\K}^{-n}(X\wedge Y)\oplus \wt{\K}^{-n}(X)\oplus
\wt{\K}^{-n}(Y)\ .
\label{Kiso}
\eea
The formula (\ref{Kiso}) is particularly useful for computing the
K-groups of Cartesian products.
As an important example, consider the case that $Y={\bf S}^1$, for
which we find
\bea
\wt{\K}(X\times {\bf S}^1)&=& \wt{\K}(X\wedge {\bf
S}^1)\oplus \wt{\K}(X)\oplus
\wt{\K}({\bf S}^1)\nonumber\\
&=& \K^{-1}(X)\oplus \wt{\K}(X)\ ,
\label{kxs}
\eea
since $\wt{\K}({\bf S}^1)=0$ and $\wt{\K}({\bf S}^1\wedge X)=\K^{-1}(X)$.
Precisely, the canonical inclusion $i:X\hookrightarrow X\times {\bf S}^1$
induces a projection $i^*: \wt{\K}(X\times {\bf S}^1)\rightarrow
\wt{\K}(X)$ such that $\ker i^*= \K^{-1}(X)$. In other words,
$\K^{-1}(X)$ can be identified with the set of K-theory classes in
$\wt{\K}(X\times {\bf S}^1)$ which vanish when restricted to $X\times
{\rm pt}$. Likewise,
\bea
\K^{-1}(X\times {\bf S}^1)&=& \K^{-1}(X\wedge {\bf
S}^1)\oplus \K^{-1}(X)\oplus
\K^{-1}({\bf S}^1)\nonumber\\
&=& \wt{\K}(X)\oplus\K^{-1}(X)\oplus\zed\ ,
\label{kmxs}
\eea
where we have used Bott periodicity.

The action of the cup product (\ref{cupproduct}) on reduced K-theory
can also be deduced using (\ref{Kdecomp}) and (\ref{Kiso}) to
get
\beq
\left(\wt{\K}(X)\otimes_{\zeds}\wt{\K}(Y)\right)\oplus{\cal R}
\longrightarrow \wt{\K}(X\wedge Y)\oplus {\cal R}\ ,
\label{addR}\eeq
where ${\cal R}=\wt{\K}(X)\oplus\wt{\K}(Y)\oplus\zed$. Since the group
${\cal R}$ appears on both sides of (\ref{addR}), we can eliminate it
by an appropriate restriction and thereby arrive at the homomorphism
\beq
\wt{\K}(X)\otimes_{\zeds}\wt{\K}(Y)\longrightarrow \wt{\K}(X\wedge Y)\ .
\label{redcupproduct}\eeq
When either $\K(X)$ or $\K(Y)$ is a free abelian group,
the mappings in (\ref{cupproduct}) and (\ref{redcupproduct})
are isomorphisms.

One can also calculate $\K^{-n}(X\times Y)$ in a manner that
keeps track of the multiplicative structure of the theory. Define
$\K^\#(X)$ to be the $\zed_2$-graded ring $\K^\#(X)=\K(X)\oplus
\K^{-1}(X)$. Then, whenever $\K^\#(X)$ or $\K^\#(Y)$ is freely generated,
we get the K-theory analog of the cohomological K\"unneth theorem:
\beq
\K^\#(X\times Y)=\K^\#(X)\otimes_{\zeds} \K^\#(Y)\ .
\label{kunneth2}\eeq
(In the general case there are correction terms on the right-hand side of
\eqn{kunneth2} which take into account the torsion subgroups of the K-groups
\cite{atiyahkun}). Explicitly, \eqn{kunneth2} leads to
\bea
\K(X\times Y)&=&\Bigl(\K(X)\otimes_\zeds\K(Y)\Bigr)
\oplus\Bigl(\K^{-1}(X)\otimes_\zeds\K^{-1}(Y)\Bigr)\ ,\nonumber\\
\K^{-1}(X\times Y)&=&\Bigl(\K(X)\otimes_\zeds\K^{-1}(Y)\Bigr)
\oplus\Bigl(\K^{-1}(X)\otimes_\zeds\K(Y)\Bigr)\ .
\label{KXYcup}\eea
For example, since $Y={\bf S}^1$ has freely generated K-groups,
using $\K({\bf S}^1)=\K^{-1}({\bf S}^1)=\zed$, we again arrive at
(\ref{kxs}) and (\ref{kmxs}). Similarly, taking $Y={\bf S}^{2n}$ and
$Y={\bf S}^{2n+1}$ in (\ref{kunneth2}) gives
\bea
\K(X\times {\bf S}^{2n})&=& \K(X)\oplus \K(X)\ ,\\
\K(X\times {\bf S}^{2n+1})&=&\K(X)\oplus\K^{-1}(X)\ ,
\eea
as $\K(X)$-modules.

If we choose $Y={\bf S}^2$, then the maps in (\ref{cupproduct}) and
(\ref{redcupproduct}) are actually isomorphisms which can be identified with
the Bott periodicity property of the reduced and unreduced K-groups.
Replacing $X$ by its $n$-th reduced suspension in \eqn{redcupproduct} gives
\beq
\widetilde{\K}(\Sigma^nX)\otimes_\zeds\widetilde{\K}({\bf S}^2)
=\widetilde{\K}(\Sigma^nX\wedge{\bf S}^2)\ ,
\eeq
which yields the isomorphism
\beq
\alpha\,:\,\widetilde{\K}^{-n}(X)\otimes_\zeds\widetilde{\K}({\bf S}^2)~
{\buildrel\approx\over\longrightarrow}~\widetilde{\K}^{-n-2}(X)\ .
\label{alphaiso}\eeq
The generator $[{\cal N}_\complexs]-[I^1]$ of $\widetilde{\K}({\bf S}^2)=\zed$
may be
described by taking ${\cal N}_\complexs$ to be the canonical line
bundle over the complex projective space $\complex P^1$,
which is associated with the Hopf fibration ${\bf S}^3\to
{\bf S}^2$ that classifies the Dirac monopole \cite{husemoller,trautman}. The
isomorphism (\ref{complexbott}) is then given by the mapping
\beq
\Bigl[(E\,,\,F)\Bigr]\longmapsto \alpha\Bigl[(E\otimes {\cal N}_\complexs\,,\,
F\otimes {\cal N}_\complexs)\Bigr]\ ,
\eeq
for $[(E,F)]\in \wt{\K}^{-n}(X)$.

\subsection{Relative K-Theory}

We will now define a relative K-group $\K(X,Y)$ which depends on a pair of
spaces $(X,Y)$, where $Y$ is a closed submanifold of $X$, and whose
classes can be identified with pairs of bundles over $X/Y$. If $Y\neq
\emptyset$, then the topological coset $X/Y$ is defined to be the
space $X$ with $Y$ shrunk to a point. If $Y$ is empty we identify $X/Y$ with
the one-point compactification $X^+$ of $X$.

First we explain how to describe vector bundles over the quotient
space $X/Y$, given a vector bundle $E$ over $X$. Let $\alpha$ be a
trivialization of $E$ over $Y\subset X$, i.e. an isomorphism $\alpha:
E|_{Y}\stackrel{\approx}{\rightarrow}Y\times V$. Define an equivalence
relation on $E|_{Y}$ by taking $e\in E|_Y$ equivalent to $e'\in E|_Y$
if and only if
\beq
\pi\circ\alpha(e)=\pi\circ\alpha(e')\ ,
\eeq
where $\pi:Y\times V\rightarrow V$ is the canonical projection.
This equivalence relation identifies points in the restriction of $E$ to $Y$
which are ``on the same level'' relative to the trivialization $\alpha$. We
then extend this relation trivially to the whole of $E$. The corresponding
set of equivalence classes $E_{\alpha}$ can be shown to be a vector bundle over
$X/Y$, whose isomorphism class depends only on the homotopy class of the
trivialization $\alpha$ of $E$ over $Y\subset X$. In fact, there is a
one-to-one correspondence between vector bundles over the quotient space
$X/Y$ and vector bundles over $X$ whose restriction to $Y$ is a trivial bundle.

The {\it relative K-group} is now defined as
\beq
\K(X,Y)\equiv \wt{\K}(X/Y)\ .
\label{Krelative}\eeq
Then $\K(X,Y)$ is a contravariant functor of the pair $(X,Y)$ and, since
$\K(X)=\wt{\K}(X^+)$ (recall $\K(X)= \wt{\K}(X)\oplus \K({\rm
pt})$), we have $\K(X,\emptyset)=\K(X)$. The {\it Excision Theorem} states
that the projection $\pi:X\rightarrow X/Y$ induces an isomorphism
\beq
\pi^*: \K(X/Y, {\rm pt})\stackrel{\approx}{\longrightarrow}\K(X,Y)\ .
\label{excision}\eeq
Likewise, one can define higher relative K-groups by
\beq
\K^{-n}(X,Y)=\K(X\times {\bf B}^n,X\times {\bf S}^{n-1}\cup Y\times {\bf
B}^n)\ ,
\eeq
where ${\bf B}^n=\{ x\in \real^n: | x|\leq1\}$ is the unit ball in $\real^n$
and ${\bf S}^{n-1}=\partial{\bf B}^n$. Alternatively, there are the
suspension isomorphisms
\beq
\K^{-n}(X,Y)=\K\Bigl((X-Y)\times \real^n\Bigr)\ .
\eeq
The relative K-groups have the usual Bott periodicity:
\beq
\K^{-n}(X,Y)= \K^{-n-2}(X,Y)\ .
\eeq

Let $i:Y\rightarrow X$ and $j:(X,\emptyset)\rightarrow (X,Y)$ be
inclusions. Then there is an exact sequence
\beq
\K(X,Y)\stackrel{j^*}{\longrightarrow} \K(X)\stackrel{i^*}{\longrightarrow}
\K(Y)\ .
\label{Kexact}\eeq
If $Y$ is further equipped with a base-point, then the sequence
\beq
\K(X,Y)\longrightarrow \wt{\K}(X)\longrightarrow \wt{\K}(Y)
\eeq
is exact. More generally, one of the most important properties of the
K-groups is that they possess the excision property, which means that they
satisfy the Barratt-Puppe long exact sequence:
\beq
\ldots\longrightarrow\K^{-n-1}(X)\longrightarrow
\K^{-n-1}(Y)\stackrel{\partial^*}{\longrightarrow}
\K^{-n}(X,Y)\longrightarrow\nonumber\K^{-n}(X)\longrightarrow\K^{-n}(Y)
\stackrel{\partial^*}{\longrightarrow}\ldots \ ,
\label{poppe}\eeq
where $\partial$ is the boundary homomorphism. The long exact sequence
(\ref{poppe}) connects the K-groups of $X$ and $Y\subset X$, and it is in
precisely this sense that K-theory is similar to a cohomology theory.
Using Bott periodicity, this sequence can be
amazingly truncated to a six-term exact sequence. If $Y$ is a
retract of $X$ (i.e. if the inclusion map $i:Y\to X$ admits a
left inverse), then the sequence (\ref{Kexact}) splits giving
\beq
\K^{-n}(X)=\K^{-n}(X,Y)\oplus\K^{-n}(Y)\ .
\label{Kretract}\eeq

The concept of relative K-theory can be reformulated in a way that will prove
useful later on.
Let $\Gamma(X,Y)$ be the set of triples $(E,F;\alpha)$, where $E,F\in {\rm
Vect}(X)$ and $\alpha: E|_{Y}\stackrel{\approx}{\rightarrow}F|_{Y}$ is an
isomorphism of vector bundles when restricted to $Y$. Two such triples
$(E,F;\alpha)$ and $(E',F';\alpha')$ are said to be isomorphic if there
exist isomorphims $f: E\stackrel{\approx}{\rightarrow} E'$ and $g:
F\stackrel{\approx}{\rightarrow}F'$ such
that the diagram
\beq
{\begin{array}{ccc}
E|_{Y}&\stackrel{\alpha}{\longrightarrow}&F|_{Y}\\&
&\\{\scriptstyle f|_{Y}}\downarrow& &\downarrow{\scriptstyle g|_{Y}}\\&
&\\E'|_{Y}&\stackrel{\alpha'}{\longrightarrow}&F'|_{Y}\end{array}}
\eeq
commutes. A triple $(E,F;\alpha)$ is called {\it elementary} if $E\cong F$
and $\alpha$ is homotopic to ${\rm Id}_{E|_{Y}}$ within automorphisms of
$E|_{Y}$. The sum of $(E,F;\alpha)$ and $(E',F';\alpha')$ is defined to be
\beq
(E,F;\alpha)\oplus(E',F';\alpha')
\equiv(E\oplus E', F\oplus F'; \alpha\oplus\alpha')\ ,
\eeq
under which $\Gamma(X,Y)$ becomes an abelian monoid. Now consider the
following equivalence relation in $\Gamma(X,Y)$. We take the triple
$(E,F;\alpha)$ to be equivalent to $(E',F';\alpha')$ whenever there exist two
elementary triples $(G,H;\beta)$ and $(G',H';\beta')$ such that
\beq
(E,F;\alpha)\oplus(G,H;\beta)\cong(E',F';\alpha')\oplus(G',H';\beta')\ .
\eeq
The set of equivalence classes of such triples (which we denote by
$[E,F;\alpha$]) under the operation
$\oplus$ becomes an abelian group which can be identified with the relative
K-group, $\K(X,Y)=\Gamma(X,Y)/\sim$.
Note that $[E,F;\alpha]=0$ in $\K(X,Y)$ if and only if there exist vector
bundles $G,H\in {\rm Vect}(X)$ and bundle isomorphisms $u:E\oplus G
\rightarrow H$, $v: F\oplus G\rightarrow H$ such that $v|_{Y}\circ
(\alpha\oplus {\rm Id}_{G|_{Y}})\circ u^{-1}|_{Y}$ is homotopic to
${\rm Id}_{H|_{Y}}$ within automorphisms of $H|_{Y}$. Moreover,
$[E,F;\alpha]+[F,E;\alpha^{-1}]=0$. Notice also that the group $\K(X)$ can in
this formalism be described as the set of triples $[E,F;\alpha]$, where
$\alpha:E\stackrel{\approx}{\rightarrow} F$ is a
bundle isomorphism defined in a neighbourhood of the infinity of the one-point
compactification of $X$. This is
precisely the statement that was made in (\ref{grothrep}).

The basic properties of $\K(X,Y)$ are as follows. First, if $\alpha,\alpha'$
are isomorphisms $E|_{Y}\stackrel{\approx}{\rightarrow}F|_{Y}$ as described
above and if $\alpha$ and $\alpha'$ are homotopic within isomorphisms from
$E|_{Y}$ to $F|_{Y}$, then $[E,F;\alpha]=[E,F;\alpha']$.
Also, when $[E,F;\alpha]$ and $[F,G;\beta]$ are elements of $\K(X,Y)$ then
their sum is given by the relation
$[E,F;\alpha]+[F,G;\beta]=[E,G;\beta\circ\alpha]$. Thirdly, two elements
of $\K(X,Y)$ determine the same equivalence class,
$[E,F;\alpha]=[E',F';\alpha']$, in $\K(X,Y)$ if and only if there exist triples
$(G,G;{\rm Id}_{G|_{Y}})$ and $(G',G';{\rm Id}_{G'|_{Y}})$, and maps
$f: E\oplus G\rightarrow E'\oplus G'$, $g: F\oplus G\rightarrow F'\oplus G'$
such that the diagram
\beq
{\begin{array}{ccc}
(E\oplus G)|_{Y}&\stackrel{\alpha\oplus {\rm Id}_{G|_Y}}{\longrightarrow}&
(F\oplus G)|_{Y}\\&
&\\{\scriptstyle f|_{Y}}\downarrow& &\downarrow{\scriptstyle g|_{Y}}\\&
&\\(E'\oplus G')|_{Y}&\stackrel{\alpha'\oplus {\rm Id}_{G'|_Y}}
{\longrightarrow}&(F'\oplus
G')|_{Y}\end{array}}
\eeq
commutes. Furthermore, the cup product naturally extends to relative K-theory
to give the unique bilinear homomorphism
\beq
\K(X,Y)\otimes_{\zeds}\K(X',Y')\longrightarrow
\K(X\times X',X\times Y'\cup Y\times X')\ .
\label{cuprel}\eeq
This agrees with the cup product introduced earlier when $Y=Y'=\emptyset$.
Explicitly, the product of two K-theory classes $[E,F;\alpha]$ and
$[E',F';\alpha']$ is obtained using the product (\ref{ringproduct}) on the
pairs of vector bundles and the product isomorphism
\beq
\beta=\left( \begin{array}{cc}\alpha\otimes{\rm Id} & {\rm Id}
\otimes\alpha'^{\dagger}\\
{\rm Id}\otimes\alpha' &-\alpha^{\dagger}\otimes{\rm Id}
\end{array}\right)\ .
\label{productiso}\eeq
acting on (\ref{ringproduct}) (note that this requires the introduction of
fiber metrics on the bundles involved).

As a simple example, consider the case that
$X={\bf B}^2$ and $Y={\bf S}^1\subset \real^2$. Define $[E,F;\alpha]\in
\K({\bf B}^2, {\bf S}^1)$ by $E=F={\bf B}^2\times
\complex$, and $\alpha(x,z)=(x,xz)$ for $x\in {\bf S}^1\subset {\bf B}^2$.
It then turns out that $[E,F;\alpha]$ is a generator of $\K({\bf B}^2,
{\bf S}^1)= \zed$. This is our first instance of the ABS construction
which will be described in section 2.8. As another example, consider the
complex projective spaces $X=\complex P^2$ and $Y=\complex
P^1$. As mentioned before, a non-trivial generator of $\wt{\K}({\bf S}^2)=
\zed$ is given by the canonical line bundle over $\complex P^1$ which is the
restriction of the canonical line bundle over $\complex P^2$. It follows that
the map $\K(X)\rightarrow \K(Y)$ is surjective. Furthermore,
$\K^{-1}({\bf S}^2)=0$ and $\K(X,Y)=\wt{\K}({\bf S}^4)=\zed$. From
(\ref{poppe}) we then obtain the split short exact sequence
\beq
0\longrightarrow \wt{\K}({\bf S}^4)\longrightarrow \K(\complex P^2)
\longrightarrow\K(\complex P^1)\longrightarrow 0\ ,
\eeq
giving $\K(\complex P^2)=\wt{\K}({\bf S}^4)\oplus\K(\complex P^1)
=\zed\oplus\zed\oplus\zed$.

\subsection{Computing the K-Groups}

In this subsection we will show how the K-groups can be computed as homotopy
groups of certain classifying spaces, for which there is often
a finite dimensional approximation. The basic case is the reduced K-groups
$\wt{\K}(X)$, since unreduced K-groups are computed from the decomposition
(\ref{Kdecomp}) and since higher K-groups are given by suspensions as in
(\ref{higherK}). Let ${\rm Vect}_k(X)$ be the set of isomorphism classes of
complex vector bundles $E_k\rightarrow X$ of rank $k$. Then we have the
sequence of inclusions, $\ldots\subset {\rm Vect}_k(X)\subset
{\rm Vect}_{k+1}(X)\subset\ldots$, via the mapping $E_k\mapsto E_k\oplus I^1$.
If $[E_k]\in {\rm Vect}_k(X)$, then $[E_k]-[I^k]\in \ker\,\rk=
\K'(X)$. The map $[E_k]\mapsto [E_k]-[I^k]$ is actually an isomorphism
${\rm Vect}(X)\rightarrow \K'(X)$ of abelian monoids, and hence
${\rm Vect}(X)\equiv\bigcup_{k=0}^{\infty}{\rm Vect}_k(X)$ is an abelian group.

A complex vector bundle $E_k$ of rank $k$ has structure group $GL(k,\complex)$
(the
fiber automorphism group), which upon choosing a metric on $X$,
and thereby inducing a Hermitian inner product on the fibers of $E_k$,
is reducible to the unitary subgroup $U(k)$. The {\it classifying space} for
$E_k$ is the complex Grassmannian manifold:
\beq
\Grass(k,m;\complex)=\frac{U(m)}{U(m-k)\times U(k)}\ ,  \ \ \ m>k+n\ ,
\eeq
where $n\equiv\dim X$. According to a standard theorem of differential
geometry, there exists a so-called universal bundle $Q(k,m;\complex)$ over
$\Grass(k,m;\complex)$ of rank $k$ whose pullbacks generate vector bundles
such as $E_k$. This means that $f^*Q(k,m;\complex)\cong E_k$ for some
continuous map $f: X\rightarrow\Grass(k,m;\complex),\ m>k+n$. Moreover, this
isomorphism depends only on the homotopy class of $f$. Therefore, bundles
$E_k$ are classified according to homotopy classes in
$[X,\Grass(k,m;\complex)]$.

Again, we have natural inclusions $\ldots \subset\Grass(k,m;\complex)\subset
\Grass(k,m+1;\complex)\subset \ldots$, and thus taking the inductive limit we
arrive at the classifying space for $U(k)$ bundles:
\beq
BU(k)\equiv \bigcup_{m=k+n+1}^{\infty}\Grass(k,m;\complex)
\eeq
such that
\beq
{\rm Vect}_k(X)=\Bigl[X\,,\,BU(k)\Bigr]
\eeq
and
\beq
\K'(X)= {\rm Vect}(X)=\Bigl[X\,,\,BU(\infty)\Bigr]\ ,
\eeq
with $BU(\infty)=\bigcup_{k=1}^{\infty}BU(k)$. Note that if $X$ is
compact, then $\K(X)= H^0(X,\zed)\oplus \K'(X)$ (according to
(\ref{rankexact})) with $H^0(X,\zed)= [X,\zed]$. This implies that
\beq
\K(X)=\Bigl[X\,,\,\zed\Bigr]\oplus
\Bigl[X\,,\,BU(\infty)\Bigr]=\Bigl[X\,,\,\zed\times BU(\infty)\Bigr]\ .
\eeq
It turns out, however,
that things simplify somewhat as the rank $k$ is increased. This leads
to the notion of {\it stable range}. Let $k_0=[(n+1)/2]$. Then for all
$k>k_0$ there exists a bundle $F_{k_0}$ of rank $k_0$ such that
$E_k\cong F_{k_0}\oplus I^{k-k_0}$. This means that any vector bundle
$E_k$ in the stable range is stably equivalent to some other bundle
$F_{k_0}$ of lower rank $k_0$. Then $E_k$ and $F_{k_0}$ belong to the
same stable equivalence class and correspond to exactly the same element
of $\K'(X)$, i.e. as far as K-theory is concerned, nothing is gained
by considering bundles of very high rank, because once the stable range
is reached no new K-theory elements are obtained by increasing the rank
$k$. Notice that, in the stable range, two vector bundles of the same rank
are stably equivalent if and only if they are isomorphic. This
implies that for all $k>\frac{1}{2}n$, $\K'(X)= {\rm Vect}_k(X)$,
or $\K'(X)= [X, BU(k)]$. Therefore, whenever $X$ is connected, we have
\beq
\wt{\K}(X)=\Bigl[X\,,\,BU(k)\Bigr]\ .
\label{KBU}\eeq

Let us consider some simple examples. The case of immediate interest is where
$X={\bf S}^{n}$, for which $\wt{\K}(X)= [{\bf S}^n,BU(k)]= \pi_n(BU(k))$ for
all $k> n/2$. We may cover ${\bf S}^n$ with upper and lower
hemispheres ${\bf S}^n_{\pm}$. Since the ${\bf S}^n_{\pm}$ are
contractible, all bundles $E_k|_{{\bf S}^n_{\pm}}$ are trivial and hence
determined by the single $U(k)$-valued transition function $g$ on the overlap
${\bf S}^n_{+}\cap {\bf S}^n_{-}$. But  ${\bf S}^n_{+}\cap {\bf S}^n_{-}\cong
{\bf S}^{n-1}$, so $g$ determines a map from ${\bf S}^{n-1}$ to $U(k)$,
i.e. an element of $\pi_{n-1}(U(k))$. It is this element of
$\pi_{n-1}(U(k))$ which determines the bundle $E_k\in {\rm Vect}({\bf
S}^n)$, and hence an element of $\wt{\K}({\bf S}^n)$, so that
\beq
\wt{\K}({\bf S}^n)= \pi_{n-1}\Bigl(U(k)\Bigr)\ , \ \ \ \ k>n/2\ .
\label{Khomotopy}\eeq
In particular, we have
\beq
\wt{\K}({\bf S}^n)= \pi_{n-1}\Bigl(U(\infty)\Bigr)\ ,
\eeq
where $U(\infty)=\bigcup_{k=1}^{\infty}U(k)$. The homotopy groups of
classical Lie groups such as $U(k)$ have been extensively studied.
Although $\pi_{n-1}(U(k))$ is not known for all $n,k$, it is precisely in
the stable range $k>n/2$ that we have a complete classification.
Note that by (\ref{KBU}), eq. (\ref{Khomotopy}) is actually the assertion that
$[{\bf S}^n,BU(k)]\cong[{\bf S}^{n-1},U(k)]$. This follows from the following
facts. First of all,
\beq
\Bigl[{\bf S}^n\,,\,BU(k)\Bigr]=\Bigl[\Sigma{\bf S}^{n-1}\,,\,BU(k)\Bigr]
=\Bigl[{\bf S}^{n-1}\,,\,\Omega BU(k)\Bigr]\ ,
\label{loopspace}\eeq
where $\Omega^n Y$ denotes the $n$-th iterated
loop space of the topological space $Y$. The isomorphism $[{\bf S}^{n-1},U(k)]
\cong[{\bf S}^{n-1},\Omega BU(k)]$ now follows from the fact that the space
$\Omega BU(k)$ is of the same homotopy type as $U(k)$. This means that the
loop space operand $\Omega$ may be thought of as a type of homotopic inverse
to the classifying space operand $B$. It is precisely this statement which
was the original content of the Bott periodicity theorem for
the classical Lie groups \cite{bott}.

As another example, take $X={\bf B}^n$ and $Y={\bf S}^{n-1}=\partial {\bf
B}^n$ in $\real^n$. The topological coset ${\bf B}^n/{\bf S}^{n-1}$ can be
identified with ${\bf S}^n$, which induces a homeomorphism from
${\bf S}^{n}_{+}/{\bf S}^{n-1}\cong {\bf B}^{n}/{\bf S}^{n-1}$ to
${\bf S}^{n}$. It then follows from the excision theorem that
\beq
\K({\bf B}^{n},{\bf S}^{n-1})=\K({\bf B}^{n}/{\bf S}^{n-1},{\rm
pt})=\wt{\K}({\bf S}^{n})=\pi_{n-1}\Bigl(U(\infty)\Bigr)\ .
\eeq
For example, $\K({\bf B}^{2},{\bf S}^{1})=\pi_1(U(\infty))= \pi_1(U(1))= \zed$.

\subsection{Clifford Algebras and the Atiyah-Bott-Shapiro Construction}

In this subsection we will discuss the relation of Clifford algebras and
spinor representations to K-theory. The analysis of Clifford algebras is very
simple and many K-theoretic results become transparent when translated into
this algebraic language. Let us start by describing the Clifford algebra
$\cliff_{r,s}$ associated with the vector space $V=\real^{r+s}$ and the
quadratic form $q(x)=x_1^{2}+\ldots +x_r^{2}-x_{r+1}^{2}-\ldots -x_{r+s}^{2}$
on $V$, which is invariant under $O(r,s)$-rotations.
There is a natural embedding $V\hookrightarrow \cliff_{r,s}$, and the abstract
unital algebra $\cliff_{r,s}$ is generated by any $q$-orthonormal basis
$\Gamma_1,\ldots ,\Gamma_{r+s}$ of $V$ subject to the relations
\beq
\Gamma_i\Gamma_j+\Gamma_j\Gamma_i~=~\left\{\begin{array}{cll}-2\delta_{ij}~~~~&,
&~~i\leq r\\+2\delta_{ij}~~~~&,&~~i > r\end{array}\right.
\label{cliffrs}\eeq
The minimal representation of the algebra (\ref{cliffrs}) consists of
Dirac matrices of dimension $2^{\left[\frac{r+s}{2}\right]}$.
The reflection map $x\mapsto -x$ for $x\in V$ extends to an
automorphism $\eta: \cliff_{r,s}\rightarrow\cliff_{r,s}$. Since
$\eta^2={\rm Id}$, this leads to the decomposition
\beq
\cliff_{r,s}=\cliff_{r,s}^+\oplus \cliff_{r,s}^-\ ,
\eeq
where $\cliff_{r,s}^{\pm}=\left\{\phi\in\cliff_{r,s}\,:\,
\eta(\phi)=\pm\phi\right\}$ are the eigenspaces of $\eta$. It follows
that
\beq
\cliff_{r,s}^{\alpha}\cdot\cliff_{r,s}^{\beta}
\subset\cliff_{r,s}^{\alpha\beta}\ ,
\eeq
where $\alpha,\beta=\pm$. The associated graded algebra of
$\cliff_{r,s}$ is then naturally isomorphic to the exterior algebra
$\Lambda^*V$, i.e. Clifford multiplication defined by (\ref{cliffrs})
is a natural enhancement of exterior multiplication which is
determined by the quadratic form $q$. In fact, there is a canonical
vector space isomorphism
$\Lambda^*V\stackrel{\approx}{\rightarrow}\cliff_{r,s}$, and hence the
natural embeddings $\Lambda^nV\subset\cliff_{r,s}$ for all $n\geq 0$.

The spin group $Spin(r,s)\subset \cliff_{r,s}$, of dimension
$2^{r+s}$, is obtained from the
group of multiplicative units of the Clifford algebra through the
embedding ${\bf S}^{r+s-1}\subset V\subset \cliff_{r,s}$.
It is a double cover of the group $SO(r,s)$,
as is expressed by the exact sequence:
\beq
1\longrightarrow\zed_2\longrightarrow Spin(r,s)\longrightarrow SO(r,s)
\longrightarrow 1
\eeq
The spin group associated to
$\cliff_{n,0}$ is $Spin(n)$ which is a double cover of the isometry group
$SO(n)$ of the sphere ${\bf S}^{n-1}$. We will use the short-hand notation
$\cliff_n\equiv\cliff_{n,0}$ and $\cliff_n^{*}\equiv\cliff_{0,n}$.
As a simple example, $\cliff_1$ is generated by the unit element and
an element $\Gamma$ obeying $\Gamma^2=-1$, so that $\cliff_1\cong\complex$.
Similarly it is easily seen that $\cliff^*_{1}\cong\real\oplus\real$.

Under the canonical isomorphism $\cliff_n\cong \Lambda^*\real^n$,
Clifford multiplication has a particularly nice form. For
$x\in\real^n$, we define the {\it interior product}
$x\,\neg:\Lambda^p\real^n\rightarrow\Lambda^{p-1}\real^n$ by
\beq
x\,\neg\,(x_1\wedge\cdots\wedge x_p)=\sum_{m=1}^{p}(-1)^{m+1}\,
\sum_{i=1}^{n}x^i(x_m)^i\,
x_1\wedge\cdots\wedge x_{m-1}\wedge x_{m+1}\wedge\cdots\wedge x_p\ .
\eeq
This defines a skew-derivation of
the algebra since $x\,\neg\,(\omega\wedge y)=(x\,\neg\,\omega)\wedge y+(-1)^p
\omega\wedge(x\,\neg\,y)$ for all $\omega\in\Lambda^p\real^n$ and all
$y\in\Lambda^q\real^n$. Furthermore, $(x\,\neg)^2=0$ for all
$x\in\real^n$, so that the interior product extends universally to a
bilinear map $\Lambda^*\real^n\otimes\Lambda^*\real^n\rightarrow
\Lambda^*\real^n$. It is now elementary to show that the Clifford
multiplication between $x\in\real^n$ and $\phi\in\cliff_n$ can be
written as
\beq
x\cdot\phi=x\wedge\phi-x\,\neg\,\phi\ ,
\label{cliffmul}\eeq
with respect to the canonical isomorphism $\cliff_n\cong\Lambda^*\real^n$.

For any pair of positive integers $(r,s)$ there is an explicit presentation of
the algebra $\cliff_{r,s}$ as a matrix algebra over one of the fields
$\real$, $\complex$ or $\quater$. The first few examples are easy to
construct by hand, for example
\bea
\cliff_{1,0}~=~\complex\ \ \ &,& \ \ \ \cliff_{0,1}~=~\real\oplus\real
\nonumber\\
\cliff_{2,0}~=~\quater\ \ \ &,& \ \ \ \cliff_{0,2}~=~\real(2)\nonumber\\
\cliff_{1,1}&=&\real(2)\ ,
\label{cliffhand}\eea
where ${\bb F}(m)$ denotes the
$\real$-algebra of $m\times m$ matrices with entries in the algebraic field
$\bb F$. The complete classification of Clifford algebras is then obtained by
using the periodicity relations (valid for any $n,r,s\geq 0$):
\bea
\cliff_{n,0}\otimes\cliff_{0,2}&=&\cliff_{0,n+2}\ ,\\
\cliff_{0,n}\otimes\cliff_{2,0}&=&\cliff_{n+2,0}\ ,\\
\cliff_{r,s}\otimes\cliff_{1,1}&=&\cliff_{r+1,s+1}\label{cliff11}\ ,
\eea
and
\bea
\cliff_{n,0}\otimes\cliff_{8,0}&=&\cliff_{n+8,0}\label{cliffper}\ ,\\
\cliff_{0,n}\otimes\cliff_{0,8}&=&\cliff_{0,n+8}\ ,
\eea
where
\beq
\cliff_{8,0}=\cliff_{0,8}=\real(16)\ .
\eeq
Using these relations and (\ref{cliffhand}) it possible to write
down the complete set of Clifford algebras $\cliff_{r,s}$ which are
summarized in table \ref{tablerealcliff2}.
\begin{table}
\begin{center}
\begin{tabular}{|c||c|c|c|c|c|} \hline
$s$ & $r=0$ & $r=1$ & $r=2$ & $r=3$ & $r=4$\\ \hline\hline
0 & $\real$ & $\complex$ & $\quater$ & $\quater\oplus\quater$ &
$\quater(2)$\\ \hline
1 & $\real\oplus\real$ & $\real(2)$ & $\complex(2)$ & $\quater(2)$
& $\quater(2)\oplus\quater(2)$\\ \hline
2 & $\real(2)$ & $\real(2)\oplus\real(2)$ & $\real(4)$ & $\complex(4)$
& $\quater(4)$\\ \hline
3 & $\complex(2)$ & $\real(4)$ & $\real(4)\oplus\real(4)$ & $\real(8)$
& $\complex(8)$\\ \hline
4 & $\quater(2)$ & $\complex(4)$ & $\real(8)$ &
$\real(8)\oplus\real(8)$
& $\real(16)$\\ \hline
5 & $\quater(2)\oplus\quater(2)$ & $\quater(4)$ & $\complex(8)$
& $\real(16)$
& $\real(16)\oplus\real(16)$\\ \hline
6 & $\quater(4)$ & $\quater(4)\oplus\quater(4)$ & $\quater(8)$
& $\complex(16)$ &
$\real(32)$\\ \hline
7 & $\complex(8)$ & $\quater(8)$ & $\quater(8)\oplus\quater(8)$ & $\quater(16)$
&
$\complex(32)$\\ \hline
8 & $\real(16)$ & $\complex(16)$ & $\quater(16)$ &
$\quater(16)\oplus\quater(16)$
& $\quater(32)$\\ \hline
\end{tabular}
\vskip 1cm
\begin{tabular}{|c||c|c|c|c|} \hline
$s$ & $r=5$ & $r=6$ & $r=7$ & $r=8$\\ \hline\hline
0 & $\complex(4)$ & $\real(8)$ & $\real(8)\oplus\real(8)$ & $\real(16)$ \\
\hline
1 & $\quater(4)$ & $\complex(8)$ & $\real(16)$ & $\real(16)\oplus\real(16)$\\
\hline
2 & $\quater(4)\oplus\quater(4)$ & $\quater(8)$ & $\complex(16)$ &
$\real(32)$\\ \hline
3 & $\quater(8)$ & $\quater(8)\oplus\quater(8)$ & $\quater(16)$ &
$\complex(32)$\\ \hline
4 & $\complex(16)$ & $\quater(16)$ & $\quater(16)\oplus\quater(16)$ &
$\quater(32)$\\ \hline
5 & $\real(32)$ & $\complex(32)$ & $\quater(32)$
& $\quater(32)\oplus\quater(32)$\\ \hline
6 & $\real(32)\oplus\real(32)$ & $\real(64)$ & $\complex(64)$
& $\quater(64)$\\ \hline
7 & $\real(64)$ & $\real(64)\oplus\real(64)$ & $\real(128)$ & $\complex(128)$\\
\hline
8 & $\complex(64)$ & $\real(128)$ & $\real(128)\oplus\real(128)$ &
$\real(256)$
\\ \hline
\end{tabular}
\end{center}
\caption{\it The real Clifford algebras $\cliff_{r,s}$ for $0\leq r,s\leq 8$.}
\label{tablerealcliff2}\end{table}
{}From this table one observes some extra intrinsic symmetries of the Clifford
algebras, for example
\bea
\cliff_{r,s}&=&\cliff_{r-4,s+4}\ ,\\
\cliff_{r,s+1}&=&\cliff_{s,r+1}\ ,
\eea
which can also be proven directly from the definition of $\cliff_{r,s}$.

We will now describe the complexified Clifford algebras which are related to
the K-theory of complex vector bundles over spheres.
The complexification of the real Clifford algebra $\cliff_{r,s}$ is the
$\complex$-algebra $\ccliff_{r,s}=\cliff_{r,s}\otimes_{\reals}\complex$,
which can also be viewed as the Clifford algebra associated with the vector
space $\complex^{r+s}$ and the complexification $q\otimes\complex$ of the
quadratic form $q$. Since all non-degenerate quadratic forms over $\complex$
are equivalent, we have the sequence of isomorphisms
\beq
\ccliff_n\cong \cliff_{n,0}\otimes_{\reals} \complex \cong
\cliff_{n-1,1}\otimes_{\reals}\complex
\cong \cdots \cong \cliff_{0,n}\otimes_{\reals}\complex\ ,
\eeq
which makes the classification of the complexified Clifford algebras much
simpler, since
it means that $\ccliff_{r,s}$ only depends on the sum of $r$ and $s$:
$\cliff_{r,s}\otimes_{\reals}\complex\cong \ccliff_{r+s}$. From this it also
follows
that the periodicity of $\ccliff_n$ is
\beq
\ccliff_{n+2}\cong \ccliff_n\otimes_{\complexs}\ccliff_2\ ,
\label{ccliffper}\eeq
where $\ccliff_2=\complex(2)$. (We shall see that this periodicity is
related to Bott periodicity of the complex K-theory of spheres). Using
these identities one can easily deduce the list of complexified
Clifford algebras in table \ref{tablecomplexcliff}.
\begin{table}
\begin{center}
\begin{tabular}{|c||c|c|c|c|c|c|c|c|} \hline
$n$ & 1 & 2 & 3 & 4 & 5 & 6 & 7 & 8\\ \hline\hline
$\ccliff_n$ & $\complex\oplus\complex$ & $\complex(2)$ &
$\complex(2)\oplus\complex(2)$ &
$\complex(4)$ & $\complex(4)\oplus\complex(4)$ & $\complex(8)$ &
$\complex(8)\oplus\complex(8)$
& $\complex(16)$\\ \hline
\end{tabular}
\end{center}
\caption{\it The complexified Clifford algebras $\ccliff_n$ for $1\leq n
\leq 8$.}
\label{tablecomplexcliff}\end{table}

Most of the important applications of Clifford algebras come through a detailed
understanding of their representations and, by restriction, of the
representation theory of their corresponding spin groups. Such properties
follow rather easily from the classification just presented.
For any algebraic field ${\bb F}$ we define an ${\bb F}$-representation of
the Clifford algebra to be a homomorphism $\rho: \cliff\longrightarrow
{\rm End}_{\bbs F}(W)$ into the endomorphism algebra of linear
transformations of a finite dimensional vector space $W$ over ${\bb F}$ (here
$\cliff$ could be either $\cliff_{r,s}$ or $\ccliff_n$). In particular $\rho$
satisfies the property $\rho(\phi\psi)=\rho(\phi)\circ\rho(\psi)$ for all
$\phi,\psi\in \cliff$. In this way $W$ becomes a Clifford-module over
${\bb F}$. For $\phi\in\cliff$ the action of $\rho(\phi)$ on $w\in W$
is denoted by
\beq
\rho(\phi)(w)\equiv\phi\cdot w\
\label{cliffmul2}\eeq
and is customarily refered to as Clifford multiplication.

As shown above, the tensor products of irreducible representations of certain
Clifford algebras gives another irreducible Clifford module (see e.g.
(\ref{cliffper}) and (\ref{ccliffper})). In general, however,
$\cliff_n\otimes\cliff_m$ is not a Clifford algebra, and so to find a
multiplicative structure in the representations of Clifford algebras it is
natural to consider a special class of Clifford modules. For this, we define
a $\zed_2$-graded module $W$ for $\cliff_n$
as one with a decomposition $W=W^+\oplus W^-$ such that
\beq
\cliff_n^{\alpha}\cdot W^{\beta}\subset W^{\alpha\beta}\ ,
\eeq
where $\alpha,\beta=\pm$. An important grading comes from the chirality
grading of the corresponding spin groups.
Given a positively oriented, $q$-orthonormal basis $\Gamma_i$ of the
oriented vector space $V$, we define an oriented volume element
$\Gamma_c$ of $\cliff_{r,s}$ by the chirality element
\beq
\Gamma_c=\Gamma_1\cdots\Gamma_{r+s}\ .
\eeq
Setting $n=r+s$, this volume element satisfies
\bea
(\Gamma_c)^2&=&(-1)^{\frac{n(n+1)}{2}+s}\ ,\nonumber\\
x\,\Gamma_c&=&(-1)^{n-1}\,\Gamma_c\,x \ , \ \ \forall x\in\real^n\ ,
\eea
showing that for $n$ odd, $\Gamma_c$ lies in the center of
$\cliff_{r,s}$, whereas for $n$ even, $\Gamma_i\Gamma_c=\Gamma_c
\,\eta(\Gamma_i)$. Therefore, when $n$ is even there is a chirality grading
induced by the $\pm 1$ eigenspaces of~$\Gamma_c$.

Let us start by classifying the representations of the real Clifford algebra
$\cliff_{r,s}$. A real representation of this algebra is constructed in the
obvious way. A $\complex$-representation, on the other hand,
is constructed as follows. Recall that a complex vector space is just a real
vector space $W$ together with a real linear map $J:W\rightarrow W$ such that
$J^2=-{\rm Id}$. Then, a complex representation of
$\cliff_{r,s}$ is a real representation $\rho: \cliff_{r,s}\rightarrow {\rm
End}_{\reals}(W)$ that commutes with the complex structure:
\beq
\rho(\phi)\circ J=J\circ\rho(\phi)\ .
\eeq
Similarly one defines quaternionic representations of $\cliff_{r,s}$.
By restriction the representations of the algebras $\cliff_n$ give rise to
important representations of the spin group.
The real spinor representation of $Spin(n)$ is defined as a homomorphism
\beq
\Delta_n: Spin(n)\longrightarrow GL_\reals(W)\ ,
\eeq
given by restricting an irreducible real representation
$\cliff_n\rightarrow{\rm End}_{\reals}(W)$ to $Spin(n)\subset\cliff_n$.
It can be shown that when $n\neq 0$ (mod 4) the representation $\Delta_n$ is
either irreducible or a direct sum of two equivalent irreducible
representations, and that the second possibility occurs exactly when
$n\equiv 1$ or 2 (mod 8). In the other cases there is a decomposition
\beq
\Delta_{4m}=\Delta_{4m}^+\oplus\Delta_{4m}^-\ ,
\eeq
where $\Delta_{4m}^{\pm}=\frac12(1\pm\Gamma_c)\Delta_{4m}$
are inequivalent irreducible representations of $Spin(4m)$.
The reality properties of these spinor modules are then easily deduced.
The only real spinor modules (or, more precisely, the only ones which are
complexifications of real representations)
are $\Delta_{8k\pm1},\Delta_{8k}^\pm$, while the representations
$\Delta_{8k+3},\Delta_{8k+4}^\pm,\Delta_{8k+5}$ are the restrictions
of quaternionic Clifford modules. The remaining modules
$\Delta_{8k+2}^\pm,\Delta_{8k+6}^\pm$ are complex.

We can similarly classify the representations of the complexified Clifford
algebra $\ccliff_n$. We define the complex representation of $Spin(n)$ to be
the homomorphism
\beq
\Delta^{\complexs}_n: Spin(n)\longrightarrow GL_{\complexs}(W)\ ,
\eeq
given by restricting an irreducible complex representation $\ccliff_n
\rightarrow {\rm End}_{\complexs}(W)$ to $Spin(n)\subset\ccliff_n$.
Similarly to the real case, it is possible to show that when $n$ is odd the
representation $\Delta_n^{\complexs}$ is irreducible, whereas when $n$ is
even there is a decomposition
\beq
\Delta_{2m}^{\complexs}=\Delta_{2m}^{\complexs +}\oplus
\Delta_{2m}^{\complexs -}\ ,
\eeq
with $\Delta_{2m}^{\complexs\pm}=\frac12(1\pm i^m\,\Gamma_c)
\Delta_{2m}^{\complexs}$, into a direct sum of two inequivalent irreducible
complex representations of $Spin(n)$.

We finally come to the connection with K-theory, via the classic
Atiyah-Bott-Shapiro (ABS) construction \cite{abs} which relates the
Grothendieck groups of Clifford modules to the K-theory of spheres.
For this we will use the definition introduced in section 2.6
of the relative K-group $\K(X,Y)$ as
the group of equivalence classes $[E,F;\alpha]$, where $\alpha$ is an
isomorphism of the vector bundles $E$ and $F$ when restricted to $Y$.
Let ${\rm R}[Spin(n)]$ be the complex representation ring of
$Spin(n)$, i.e. the Grothendieck group constructed from the abelian monoid
generated by the irreducible complex representations, with respect to
the direct sum and tensor product of $Spin(n)$-modules.
(We will describe representation rings in more generality in section 6.1).
Let $W=W^+\oplus W^-$ be a $\zed_2$-graded module over the Clifford algebra
$\cliff_n$. We then associate to the graded module $W$ the element
\beq
\varphi(W)=[E^+,E^-;\mu]\in \K({\bf B}^n,{\bf S}^{n-1})\ ,
\label{phiW}
\eeq
where $E^{\pm}\equiv {\bf B}^n\times W^{\pm}$ is the trivial product bundle,
and $\mu:E^+\rightarrow E^-$ is the isomorphism
over ${\bf S}^{n-1}$ given by Clifford multiplication:
\beq
\mu(x,w)\equiv (x,x\cdot w)\ \ , \ \ x\in {\bf S}^{n-1}\ .
\eeq
Note that, since the ball ${\bf B}^n$ is contractible, all bundles over it
are trivial and the topology all lies in the winding of the homotopically
non-trivial map $\mu:E^+\rightarrow E^-$ over ${\bf S}^{n-1}$.
It is now straightforward to show that the element $\varphi(W)$ depends only
on the isomorphism class of the graded module $W$, and furthermore that the
map $W\mapsto\varphi(W)$ is an additive homomorphism. Thus the map
(\ref{phiW}) gives a homomorphism
\beq
\varphi:{\rm R}\Bigl[Spin(n)\Bigr]
\longrightarrow \K({\bf B}^n,{\bf S}^{n-1})\ .
\label{phiR}\eeq
By restriction, the natural inclusion $i:\real^n\hookrightarrow \real^{n+1}$
induces an epimorphism $i^*:{\rm R}[Spin(n+1)]\rightarrow {\rm R}[Spin(n)]$.
It then follows that the homomorphism (\ref{phiR}) descends to a homomorphism
$\varphi_n:{\rm R}[Spin(n)]/i^*\,{\rm R}[Spin(n+1)]\rightarrow
\K({\bf B}^n,{\bf S}^{n-1})$, which turns out to be
a graded ring isomorphism \cite{abs}:
\beq
{\rm R}\Bigl[Spin(n)\Bigr]\,/\, i^*\,{\rm R}\Bigl[Spin(n+1)\Bigr]
\stackrel{\approx}{\longrightarrow}\wt{\K}({\bf S}^{n})
\label{Kabs}\eeq
The groups in (\ref{Kabs}) are isomorphic to $\zed$ for $n$ even,
while they vanish when $n$ is odd according to the above classification of
Clifford modules. The isomorphism (\ref{Kabs}) is generated
by the principal $Spin(n)$ bundle over ${\bf S}^n$:
\beq
Spin(n)\hookrightarrow Spin(n+1)\longrightarrow {\bf S}^n \ .
\label{Snprincipal}\eeq

This theorem also gives us explicit generators for $\wt{\K}({\bf S}^{2n})$
defined via representations of Clifford algebras. For example, let
${\cal S}={\cal S}^+\oplus{\cal S}^-$ be the fundamental $\zed_2$-graded
representation space for $\ccliff_{2n}$. There is an isomorphism
${\rm R}[Spin(2n)]\cong\zed\oplus\zed$ with generators given by ${\cal S}$ and
its ``flip'' $\wt{\cal S}$, the same graded module with the factors
interchanged (this correponds to a reversal of the orientation in
$\real^{2n}$). The generator of $i^*\,{\rm R}[Spin(2n+1)]\cong\zed_{\rm diag}$
is then $[{\cal S}]+[\wt{\cal S}]$. Thus the group $\wt{\K}({\bf
S}^{2n})=\zed$ is generated by the element
\beq
\varphi_{\complexs}^{(n)}=[{\cal S}^+,{\cal S}^-;\mu]\ ,
\eeq
where $\mu_x:{\cal S}^+\rightarrow {\cal S}^-$ denotes Clifford
multiplication by $x\in\real^{2n}$. Denoting the generators of ${\cal S}$ by
$\Gamma_i$, the inclusion $\real^{2n}\hookrightarrow \ccliff_{2n}$ along with
the definition (\ref{cliffmul2}) of Clifford multiplication shows that
$\mu_x$ can be represented via
ordinary matrix multiplication by $x=\sum_ix^i\,\Gamma_i\in\real^{2n}$:
\beq
\mu_x(w)=\left(
\sum_{i=1}^{2n}x^i\,\Gamma_i\right)w\ \ , \ \ x^i\in\real\ .
\eeq
Moreover, from (\ref{cliffmul}) it follows that the square of the isomorphism
$\mu_x$ is just
multiplication by the norm of the vector $x\in\real^{2n}$:
\beq
\mu_x\circ\mu_x(w)=-|x|^2\,w\ .
\eeq
Note that the Bott periodicity of spheres, $\wt{\K}({\bf S}^n)=\wt{\K}({\bf
S}^{n+2})$, can now be derived from the periodicity property
(\ref{ccliffper}) of complexified Clifford algebras. Furthermore, using the
structure of Clifford modules it is straightforward
to show using the cup product that
\beq
\varphi_{\complexs}^{(n)}=\left(\varphi_{\complexs}^{(1)}
\right)^n\ .
\eeq

\newsection{Type IIB D-Branes and $\K(X)$}

We will now begin describing the systematic applications of K-theory to the
classification of D-brane charges in superstring theory. We start in this
section by considering the Type IIB theory, for which the simplest analysis
can be carried out. Type II superstrings are oriented and therefore have
Chan-Paton bundles with unitary structure groups.
Except for the new ways of thinking about and constructing
D-branes, the K-theory formalism merely reproduces the known spectrum of
stable brane charges. However, the analysis we present in the following
easily generalizes to more complicated situations where we will see that
K-theory makes genuinely new predictions, and it moreover provides a nice
consistency check that the mathematical formalism is indeed the correct one.

We will show in this section that the group $\K(X)$ classifies D-branes in
Type IIB superstring theory on the spacetime manifold $X$ \cite{witten}.
More precisely, the RR-charge of a Type IIB D-brane is measured by the
K-theory class of its transverse space, so that $\wt{\K}({\bf S}^n)$
classifies $(9-n)$-branes in Type IIB string theory on flat $\real^{10}$,
for example. The corresponding K-groups are determined by homotopy theory as
described in section 2.7:
\beq
\wt{\K}({\bf S}^n)=\pi_{n-1}\Bigl(U(k)\Bigr)\ ,\ \ \ \ k>n/2\ .
\eeq
Taking the inductive limit one has
\beq
\wt{\K}({\bf S}^n)=\pi_{n-1}\Bigl(U(\infty)\Bigr)\ ,
\eeq
where $U(\infty)=\bigcup_kU(k)$ is the infinite unitary group. Bott
periodicity states that the corresponding homotopy groups
$\pi_n(U(\infty))$ are periodic with period {\sl two}:
\beq
\pi_n\Bigl(U(\infty)\Bigr)=\pi_{n+2}\Bigl(U(\infty)\Bigr)\ ,
\eeq
or
\beq
\wt{\K}({\bf S}^n)=\wt{\K}({\bf S}^{n+2})\ .
\eeq
{}From this and the fact that $\wt{\K}({\bf S}^0)=\zed, \wt{\K}({\bf S}^1)=0$
follows the complete classification of D-branes in Type IIB superstring
theory, which is summarized in table \ref{tableKgroups}.
\begin{table}
\begin{center}
\begin{tabular}{|c|c|c|c|c|c|c|c|c|c|c|c|} \hline
\ D-brane\ & \ D9\ & \ D8\ & \ D7\ & \ D6\ & \ D5\ & \ D4\ & \ D3\
& \ D2\ & \ D1\ & \ D0\ & \ D(--1)\ \\ \hline
Transverse & {} & {} & {} & {} & {} & {} & {} & {} & {} & {} & {}\\
space & \raisebox{1.5ex}[0pt]{${\bf S}^0$} & \raisebox{1.5ex}[0pt]{${\bf S}^1$}
&
\raisebox{1.5ex}[0pt]{${\bf S}^2$} & \raisebox{1.5ex}[0pt]{${\bf S}^3$} &
\raisebox{1.5ex}[0pt]{${\bf S}^4$} & \raisebox{1.5ex}[0pt]{${\bf S}^5$} &
\raisebox{1.5ex}[0pt]{${\bf S}^6$} & \raisebox{1.5ex}[0pt]{${\bf S}^7$} &
\raisebox{1.5ex}[0pt]{${\bf S}^8$} & \raisebox{1.5ex}[0pt]{${\bf S}^9$} &
\raisebox{1.5ex}[0pt]{${\bf S}^{10}$}\\\hline
$\wt{\K}({\bf S}^n)$ & $\zed$ & 0 &$\zed$ & 0
& $\zed$ & 0 & $\zed$ & 0 & $\zed$ & 0 & $\zed$ \\ \hline
\end{tabular}
\end{center}
\caption{\it D-brane spectrum in Type IIB superstring theory from
$\wt{\K}({\bf S}^n)$.}
\label{tableKgroups}\end{table}
This table just reflects the fact that the Type IIB theory has stable
D$p$-branes only for $p$ odd. In this way one recovers the usual spectrum of
IIB BPS brane charges.

\subsection{The Brane-Antibrane System}

The physics behind the K-theory description of D-brane charges hinges
on a new interpretation of branes in terms of higher-dimensional
branes and antibranes. We shall therefore start by briefly reviewing the
properties of brane-antibrane pairs in superstring theory. This system
is unstable due to the presence of a tachyonic mode in the open string
excitations that start on the brane (respectively antibrane) and end
on the antibrane (respectively brane) \cite{gbs}.
The simplest way to see this property is by appealing to the boundary
state formalism (see \cite{divecchia} and references therein).
A stable supersymmetric D$p$-brane can be represented
and described by a boundary state
\beq
|Dp\rangle=|Dp\rangle_{\NS}\pm |Dp\rangle_\RR\ ,
\label{Dp}\eeq
which is a particular coherent state in the Hilbert space of the {\it
  closed} string theory. It represents a source for the closed string modes
  emitted by a D$p$-brane. The boundary state (\ref{Dp}) consists of a part
$|Dp\rangle_{\NS}$ which is a source for the closed string states of the
NS-NS sector of the fundamental string worldsheet, and a piece
$|Dp\rangle_\RR$ for the RR sector. The relative sign in (\ref{Dp})
distinguishes a brane from its antibrane which have opposite RR charges.
Taking into account the closed string GSO projection gives the decompositions
\bea
|Dp\rangle_{\NS}&=&\mbox{$\frac{1}{2}$}\,\Bigl(|Dp,+\rangle_{\NS}-
|Dp,-\rangle_{\NS}\Bigr)\nonumber\ ,\\
|Dp\rangle_{\RR}&=&\mbox{$\frac{1}{2}$}\,\Bigl(|Dp,+\rangle_{\RR}+
|Dp,-\rangle_{\RR}\Bigr)\ ,
\label{Dp+-}\eea
where the $\pm$ label the two possible implementations of the boundary
conditions appropriate for a D$p$-brane. The decompositions
(\ref{Dp+-}) take into account the sum over the four spin structures
on the string worldsheet.

The boundary state formalism allows one to easily compute the spectrum
of open strings which begin and end on a D$p$-brane. This can be found
by computing a tree-level two-point function of the boundary state
with itself (the cylinder amplitude) and via a modular transformation
re-expressing the result as a one-loop trace over open string states
(the annulus amplitude) according to
\beq
\int\limits_0^{\infty}d\tau~\left\langle Dp,\alpha\left|
\,\e^{-\pi\tau(L_0+\tilde{L}_0)}\right|Dp,\beta\right\rangle
=V_p\int\limits_0^{\infty}\frac{dt}{t}~{\rm Tr}_{\rm open}\left(\e^{-2\pi
    tL_0}\right) \ ,
\label{ampl}\eeq
where $V_p$ is the (infinite) worldvolume and $\alpha,\beta=\pm$. The
open string sectors which appear in (\ref{ampl}) depend both on the
closed string sectors and on the spin structures $\alpha,\beta$. Of
particular importance are the open string NS and NS$(-1)^F$ even spin
structures, which correspond respectively to the closed string NS-NS
and RR sectors with $\alpha=\beta$. The NS sector is thereby GSO
projected in the usual way as NS-NS+RR=NS+NS$(-1)^F$. This leads to
the well-known fact that the sum over the contributions from all even
spin structures vanishes, and thus the spectrum of open strings which
start and end on a D$p$-brane is supersymmetric and free from
tachyons.

However, if one considers instead a system composed of one D$p$-brane
and one anti-D$p$-brane (which we will also call a
D$\overline{p}$-brane), then the contribution to (\ref{ampl}) from the
RR sector changes sign, and the NS open string sector has the
``wrong'' GSO projection, NS-NS -- RR=NS--NS$(-1)^F$. The open string
spectrum therefore exhibits a tachyon, and this fact is responsible
for the instability of the brane-antibrane pair. This feature is a
consequence of the fact that the system sits at the top of a potential
well, and it is precisely the presence of this tachyon field in a
$p-\overline{p}$ system that makes the connection to K-theory.

We may choose a suitable basis of the open string Chan-Paton gauge group
$U(2)$ of the brane-antibrane pair in which diagonal matrices represent the
open string excitations which start and end on the same brane or antibrane,
while off-diagonal matrices represent the string states which stretch between
the brane and its antibrane of a given orientation. The tachyon vertex
operators which create the appropriate $p-\overline{p}$
tachyonic open string states are therefore given by
\bea
V_T(z)&=&\e^{ik_aX^a(z)}\otimes\left(\begin{array}{cc}0 & 1\\ 0 &
 0\end{array}\right)\nonumber\ ,\\
V_{T^\dagger}(z)&=&\e^{ik_aX^a(z)}\otimes\left(\begin{array}{cc}0 &
0\\ 1 & 0\end{array}\right)\ ,
\label{tachyonops}\eea
where $X^a(z)$ are worldsheet boson fields and $k_a$ is the momentum
along the D$p$-brane worldvolume. From the structure of the Chan-Paton
matrices in (\ref{tachyonops}), it is straightforward to see that the
only non-vanishing correlation functions are those involving an equal
number of $T$ and $T^\dagger$ vertex operators. If $T(x)$ and
$T^\dagger(x)$ denote the complex tachyon fields living on the worldvolume of
the D$p$-brane anti-D$p$-brane system, then there is a tachyon
potential of the form
\beq
V(TT^\dagger)=\sum_{n=2}^{\infty}c_n\left(TT^\dagger\right)^n \ .
\eeq
This implies that the tachyon potential depends only on the modulus of $T$,
\beq
V(T)=V(|T|^2)\ .
\label{potmodulus}\eeq
The presence of a non-trivial tachyon potential $V(T)$ implies that a
stable configuration cannot be reached by simply superimposing a
D$p$-brane and a D$\overline{p}$-brane, since the system is sitting on top of a
tachyon well. The lowest energy configuration (i.e. the stable
configuration) of the system is obtained by allowing the tachyon to
roll down to the minimum $T_0$ of its potential. From (\ref{potmodulus}) it
follows that these points live on a circle described by the equation
\beq
|T|=T_0 \ .
\eeq
Note that in terms of the {\it real} tachyon field $t=T+T^\dagger$, the
tachyon potential is an even function of $t$,
\beq
V(t)=V(-t)\ ,
\eeq
and the corresponding minima always come in pairs $\pm\,t_0$.

Furthermore, one may argue that, when the tachyon condenses into
one of its vacuum expectation values, the negative
potential energy density of the condensate cancels exactly with the positive
energy density associated with the tension of the $p-\overline{p}$ pair:
\beq
2{\cal T}_p+V(T_0)=0\ ,
\label{finiteenergy}\eeq
where ${\cal T}_p$ is the $p$-brane tension. This shows that the
tachyon ground state is indistinguishable from the supersymmetric
vacuum configuration, since it carries neither any charge nor any
energy. Thus, under these circumstances, the stable configuration of a
brane-antibrane pair which is reached by tachyon condensation is
nothing but the vacuum state. However, instead of considering the
tachyon ground state, one can also construct tachyonic soliton
solutions on the brane-antibrane worldvolume. This will be done in the
next subsection, where we will see that one of the astonishing
features of the K-theory formalism is that is provides a very explicit
form for the classical tachyonic soliton field~$T(x)$.

The reversal of the GSO projection described above may be formalised as
follows. The endpoints of the open string excitations of the $p-\overline{p}$
pair carry a charge which takes values in a two-dimensional quantum Hilbert
space. The first component of such a wavefunction may be regarded as bosonic
and representing, say, the open strings which end on the $p$-brane, while the
second component is fermionic and represents the open strings which end on the
$\overline{p}$-brane. The $p-\overline{p}$ open strings have off-diagonal
Chan-Paton wavefunctions which are odd under the adjoint action of the operator
\beq
(-1)^F=\sigma_3=\left(\begin{array}{cc}1
& 0\\ 0 &- 1\end{array}\right)\ ,
\label{GSOop}\eeq
and are thereby removed by the GSO projection operator
\beq
P_{\rm GSO}=\mbox{$\frac12$}\,\Bigl(1+(-1)^F\Bigr) \ .
\label{PGSO}\eeq
On the other hand, the $p-p$ and $\overline{p}-\overline{p}$ open strings
have diagonal Chan-Paton wavefunctions. They are even under $(-1)^F$ and are
therefore selected by the GSO projection (\ref{PGSO}). Having
one bosonic and one fermionic Chan-Paton state leads to a $p-\overline{p}$
worldvolume gauge symmetry with gauge supergroup $U(1|1)$. However, because of
the GSO projection, the off-diagonal fermionic gauge fields of $U(1|1)$ are
absent, leading to the usual elimination of the massless vector multiplet.
The remaining bosonic fields on the $p-\overline{p}$ worldvolume form instead
a structure whose lowest modes correspond to the superconnection~\cite{quillen}
\beq
{\cal A}=\left( \begin{array}{cc}A^+ & T\\ T^{\dagger} &A^-
\end{array}\right)\ ,
\label{superconn}\eeq
on $X$, where $A^\pm$ are the gauge fields on the bundles $E$ and $F$ of the
bosonic
and fermionic Chan-Paton states of the $p$-brane and $\overline{p}$-brane,
respectively. The $p-\overline{p}$ tachyon field $T$ is regarded as a map
$T:E\rightarrow F$, while its adjoint $T^{\dagger}$ is a map
$T^{\dagger}: F\rightarrow E$. Alternatively, $T$ may be regarded as a section
of $E\otimes F^*$ and $T^{\dagger}$ of $E^*\otimes F$, where
$E^*={\rm Hom}_{\complexs}(E,\complex)$ is the dual vector bundle to $E$. The
superconnection (\ref{superconn}) has been used recently in \cite{kenwil} for
a generalization, to the brane-antibrane system, of the usual Wess-Zumino
couplings of RR fields to worldvolume gauge fields (see also \cite{bcr}). It
will play a crucial role in section 7 when we discuss index theory.

\subsection{The Bound State Construction}

We will now discuss how to construct tachyonic soliton solutions and show
that this construction is equivalent to the ABS homomorphism which maps classes
in $\K(Y)$ to classes in $\K(X)$, where the D-branes wrap around a submanifold
$Y$ of the spacetime $X$. Until section 7 we shall deal only with flat
spacetimes and topologically trivial worldvolume embeddings
$Y\hookrightarrow X$. We will start by constructing a stable $p$-brane in
Type IIB superstring theory as the bound state of a $(p+2)$-brane and a
coincident $\overline{(p+2)}$-brane. For this, we shall consider an infinite
$(p+2)$ brane-antibrane pair stretching over a submanifold
$\real^{p+3}\subset X$. Due to the tachyon, this system tends to annihilate
itself unless there is some topological obstruction. This obstruction is
measured by the K-theory group $\K(X)$.

On the $(p+2)-\overline{(p+2)}$ pair, there is a $U(1)\times U(1)$ gauge
field $(A^+,A^-)$ and a tachyon field $T$ of corresponding charges $(1,-1)$.
This means that the kinetic energy term for the tachyon field in the
worldvolume field theory is of the form $|(\partial_i-iA_i^++iA_i^-)T|^2$.
We consider a vortex in which $T$ vanishes on a codimension two submanifold
$\real^{p+1}\subset \real^{p+3}$, which we interpret as the $p$-brane
worldvolume. We suppose that $|T(x)|$ approaches its vacuum expectation
value $T_0$ at $|x|\rightarrow \infty$ (up to a gauge transformation). $T$ is
a complex scalar field, so it can have a winding number around the
codimension 2 locus where it vanishes, or equivalently at $|x|=\infty$. The
basic case is where the winding number is 1, and $T$ breaks the
$U(1)\times U(1)$ gauge symmetry of the brane-antibrane pair down to the
diagonal $U(1)$ subgroup. To keep the energy per unit $p$-brane worldvolume
finite (i.e. to have finite tension), there is a unit of magnetic flux in the
broken $U(1)$ group, which is achieved by giving the gauge field $A^+-A^-$ on
the worldvolume of the $(p+2)-\overline{(p+2)}$ pair a unit of topological
charge at infinity. The non-vanishing asymptotic field configuration
therefore takes the form
\beq
T\simeq T_0~\e^{i\theta}~~~,~~~A_\theta^+-A_\theta^-\simeq1~~~~~~{\rm for}
{}~~r\to\infty \ ,
\label{codim2asympt}\eeq
where $(r,\theta)$ are polar coordinates on the two-dimensional transverse
space $\real^{p+3}-\real^{p+1}$. Then both the kinetic and potential energy
terms in the worldvolume field theory vanish sufficiently fast as
$r\to\infty$, leading to a static finite energy vortex configuration for the
tachyon field. This system has one unit of $p$-brane charge, but its
$(p+2)$-brane charge is zero between the brane and antibrane. With $T$
approaching its vacuum expectation value everywhere except close to the core
$\real^{p+1}$ of the vortex, the system looks like the vacuum everywhere
except very close to the locus where $T$ vanishes. This soliton thereby
describes a stable, finite energy $p$-brane in Type IIB string theory. By
studying the boundary conformal field theory describing this solution, one
can prove that this soliton is indistinguishable from the D$p$-brane of Type
IIB superstring theory and is simply a different representation of the same
topological defect in the spacetime $X$.

One can easily generalize this construction to a $(p+2k)$ brane-antibrane
pair for $k>1$. First we construct a $p$-brane from a $(p+2)$ brane-antibrane
pair, then we construct the $(p+2)$ brane and antibrane each as a bound state
of a $(p+4)$ brane-antibrane pair, and so on. After $k-2$ more steps, we get a
$p$-brane built from $2^{k-1}$ pairs of $(p+2k)$-branes and antibranes.
However, such a ``stepwise'' bound state construction breaks the manifest
spacetime symmetries and limits the possible applications of this formalism.
A more direct construction exhibiting the full symmetries of the system is
desired. This is precisely where the formalism of K-theory plays a central
role.

To relate these constructions to K-theory, we recall from section 1
that the D-brane charges of tadpole anomaly cancelling Type IIB
$9-\overline{9}$-brane configurations are classified by the reduced K-theory
group $\wt{\K}(X)$ of the spacetime $X$. Each class in $\wt{\K}(X)$ is
represented by an equal number $N$ of 9-branes and $\overline{9}$-branes
wrapping $X$, with the class in $\K(X)$ given by the difference $[E]-[F]$ of
the Chan-Paton gauge bundles on the 9-branes and $\overline{9}$-branes.
Open strings ending on all possible pairs of these branes give rise to a
$U(N)\times U(N)$ gauge field, and a tachyon field $T$ in the bifundamental
${\bf N}\otimes\overline{\bf N}$ representation of the gauge group.
Although we don't know the precise form of the tachyon potential, we
may argue that at the minima $|T|=T_0$ all eigenvalues of $T_0$ are equal.
This follows from the possibility of separating the brane-antibrane
pairs. It then follows that the tachyon condensate $T_0$ breaks the worldvolume
gauge symmetry from $U(N)\times U(N)$ down to the diagonal $U(N)$ subgroup.

We will now construct a stable D-brane of the Type IIB theory as a bound
state of a system of $N$ 9-branes and $N$ $\overline{9}$-branes which locally
near $Y$ resembles a topologically stable vortex of the tachyon field. The
number $N$ will be fixed below by the mathematics of the ABS construction.
The stable values of $T_0$ (i.e. the gauge orbits of the tachyon field with
minimum energy) live in the vacuum manifold
\beq
{\cal V}_{\rm IIB}(N)=\frac{U(N)\times U(N)}{U(N)_{\rm diag}}\cong U(N)\ .
\label{IIBvac}\eeq
Therefore, when viewed as a Higgs field in this description, $T$ supports
stable topological defects in codimension $2k$ which are
classified by the non-trivial homotopy groups of the vacuum manifold:
\beq
\pi_{2k-1}\Bigl({\cal V}_{\rm IIB}(N)\Bigr)= \pi_{2k-1}\Bigl(U(N)\Bigr)=
\zed\ , \ \ \ \ N>k\ .
\eeq
So for a $p$-brane wrapping a submanifold $\real^{p+1}\subset X$, we take
$T(x)$ to vanish in codimension $2k=9-p$, and let it approach its vacuum orbit
at $|x|\rightarrow \infty$, with a non-zero topological twist around the
locus $\real^{p+1}$ on which it vanishes, and of a given winding number at
infinity. These configurations are classified topologically by the homotopy
classes of maps ${\bf S}^{2k-1}\rightarrow U(N)$, or by the K-theory classes
\beq
\wt{\K}({\bf S}^{2k})= \pi_{2k-1}\Bigl(U(N)\Bigr)= \zed\ , \ \ \ \
\forall N>k\ .
\eeq
Note that D-brane charges are labelled by reduced K-groups of the transverse
spaces to the worldvolumes (compactified by adding a point at infinity).
This result makes manifest the relation between homotopy theory (i.e. the
classification of topological defects) and K-theory (i.e. the classification
of configurations of spacetime-filling branes up to pair creation and
annihilation). As discussed in the previous subsection, the negative energy
density corresponding to the vacuum condensate of $T$ is equal in magnitude
to the positive energy density due to the non-zero tension of the
$9-\overline{9}$ brane system wrapping $X$. This implies that the total
energy density away from the core of the bound state approaches zero rapidly,
and the configuration is very close to the supersymmetric vacuum. Therefore,
tachyon condensation leaves behind an object wrapped on $Y=\real^{p+1}$ that
carries the charge of a supersymmetric D$p$-brane wrapping $Y$.

The embedding $\K(Y)\hookrightarrow \K(X)$ is realized
mathematically by the K-theoretic ABS construction that was described in
section 2.8. (This is an example of a push-forward map that we will return to
in section 7). It corresponds to the mapping of a non-trivial class
describing a D-brane wrapping $Y$ into a class where it corresponds to the
bound state of a 9-brane $\overline{9}$-brane configuration wrapping the
spacetime $X$. For $Y$ of codimension $2k$ in $X$, this construction selects
the prefered value $N=2^{k-1}$ of the number of $9-\overline{9}$-brane pairs
(recall that this was precisely the prediction of the previous ``stepwise''
construction), and it moreover gives a particularly simple, natural and
useful representation of the tachyon vortex configuration (i.e. of the
generator of $\pi_{2k-1}(U(N))$) via Clifford multiplication.
Consider the rotation group $SO(2k)$ of the transverse space, which is
the group of orientation-preserving automorphisms of the normal bundle of
$Y\subset X$. It has two inequivalent positive and negative
chirality complex spinor representations $\Delta_{2k}^{\complexs\pm}$ of
dimension $2^{k-1}$. They give rise to two spin bundles
${\cal S}^{\pm}\rightarrow Y$, which can be extended to a neighbourhood of
$Y$ in $X$ (modulo some global obstructions, as we will describe in section
7). They therefore define a K-theory class
$[{\cal S}^+]-[{\cal S}^-]\in \wt{\K}(X)$, where $E={\cal S}^+$ is the
Chan-Paton bundle carried by the 9-branes and $F={\cal S}^-$ by the
$\overline{9}$-branes in the above bound state construction.

The gauge symmetry of the 9-brane worldvolume $X$ is $U(2^{k-1})\times
U(2^{k-1})$, and the tachyon field is a map $T: {\cal S}^+\rightarrow
{\cal S}^-$. Let $\Gamma_1, \ldots , \Gamma_{2k}$ be the generators of
$\Delta_{2k}^{\complexs+}\oplus\Delta_{2k}^{\complexs-}$, which
can be regarded as maps ${\cal S}^+\oplus {\cal S}^-
\rightarrow {\cal S}^+\oplus {\cal S}^-$. Let $(x^1,\ldots ,
x^{2k})\in {\bf S}^{2k-1}\subset\real^{2k}$. Then, using the construction of
section 2.8, we define the tachyon field via Clifford multiplication
\beq
T(x)=f(x)\,\mu_x=f(x)\,\sum_{i=1}^{2k}\Gamma_i\,x^i\ ,
\eeq
where $f(x)$ is a convergence factor with the asymptotic behaviours
\beq
\lim_{x\in Y}f(x)={\rm const.}~~~~~~,~~~~~~
\lim_{|x|\rightarrow\infty}f(x)=\frac{T_0}{|x|}\ ,
\eeq
which ensures that far away from the core of the vortex, $T(x)$ takes
values in the IIB vacuum manifold (\ref{IIBvac}), whereas the tachyonic
soliton is located on the submanifold $x^i=0, i=1,\ldots, 2k$. In the
sequel we shall usually not write such convergence factors explicitly.
The field $T(x)$ has winding number 1 \cite{bfn}, and according to the ABS
construction, it generates $\pi_{2k-1}(U(2^{k-1}))= \zed$, or
equivalently $\K({\bf B}^{2k},{\bf S}^{2k-1})=\zed$.
The precise mapping $\K(Y)\hookrightarrow \K(X)$ of K-theory classes is
given by the cup product (\ref{cuprel}):
\bea
\lambda\,:\,\widetilde{\K}(Y)\otimes_\zeds
\K({\bf B}^{2k},{\bf S}^{2k-1})&{\buildrel\approx\over\longrightarrow}&
\K(Y\times{\bf B}^{2k},Y\times{\bf S}^{2k-1})\nn\\
\Bigl[(E\,,\,F)\Bigr]&\longmapsto&\lambda\left[(E\otimes{\cal S}^+
\oplus F\otimes{\cal S}^-\,,\,E\otimes{\cal S}^-\oplus
F\otimes{\cal S}^+)\right]\nn\\& &
\label{absmapK}\eea
where $[(E,F)]\in\widetilde{\K}(Y)$ and we have used the
fact that $\K({\bf S}^m)$ for any $m$ is a free abelian group.

One can verify that this ``all at once'' construction is equivalent to the
previous ``stepwise'' construction. This fact follows from the periodicity
property (\ref{ccliffper}) of the complexified Clifford algebras, or
equivalently from Bott periodicity of complex K-theory. Namely, the process
of tachyon condensation of the bound state of a $p$-brane $\overline{p}$-brane
pair into a $p-2$-brane may be regarded as the Bott periodicity isomorphism
on the spacetime K-theory group $\wt{\K}(X)\rightarrow \wt{\K}(X)$, which can
in turn be described by the ABS map
\bea
\left[({\cal S}_{2k}^+\,,\,{\cal
S}_{2k}^-)\right]&\longmapsto&\left[\left({\cal S}_{2k}^+\otimes({\cal
S}_2^+\oplus{\cal S}_2^-)\,,\,{\cal S}_{2k}^-\otimes({\cal
S}_2^+\oplus{\cal
S}_2^-)\right)\right]\nn\\T_{2k}&\longmapsto&\pmatrix{T_{2k}\otimes
\id_2&\id_{2k}\otimes T_2^\dagger\cr \id_{2k}\otimes
T_2&-T_{2k}^\dagger\otimes \id_2\cr}\ ,
\label{Bottspinmap}\eea
where $\id_N$ denotes the $N\times N$ identity matrix, and
$[{\cal S}_{2k}^+,{\cal S}_{2k}^-;T_{2k}]$ is the generator of
$\wt{\K}(X)$ above. Here
\beq
T_2(x)=\sigma_1\,x^1+\sigma_2\,x^2=
\left( \begin{array}{cc}0 & x^1+ix^2\\
x^1-ix^2&0
\label{2tachyon}\end{array}\right)
\eeq
is the codimension 2 tachyon field which generates the stable homotopy groups
of the vacuum manifold
\beq
{\cal V}_{\rm IIB}(1)=\frac{U(1)\times U(1)}{U(1)_{\rm diag}}\cong
U(1)\ ,
\eeq
and $\sigma_i$ will always denote the standard $SU(2)$ Pauli
spin matrices. Alternatively, as shown at the end of section 2.5, the
$p-2$-brane may be identified with a Dirac magnetic monopole vortex
\cite{trautman} in the $p-\overline{p}$-brane worldvolume. This
identification is consistent with the topological stability $\pi_1(U(1))=\zed$
of the worldvolume soliton. Moreover, it identifies the explicit form of the
gauge field configuration in the worldvolume field theory as \cite{wuyang}
\beq
A_\phi^\pm=0~~~~~~,~~~~~~A_\theta^\pm=\pm\,\frac{1\mp\cos\phi}{\sin\phi}
\label{diracgauge}\eeq
where $(\theta,\phi)$ are angular coordinates on the transverse space
${\bf S}^2$, and the brane-antibrane indices $\pm$ now label the corresponding
upper and lower hemispheres ${\bf S}_\pm^2$.

Thus, the $p$-brane charge of the above configuration equals one, while all
higher and lower dimensional charges vanish (this can also be verified by
using formulas for brane charges induced by gauge fields). Notice that
$T: E\rightarrow F$ is trivial at infinity (where the system resembles the
vacuum), and is an isomorphism $E\cong F$ in a neighbourhood of
infinity. This means that the K-theory class $[(E,F)]$ is assumed to be
equivalent to the vacuum at infinity, i.e. that one can relax to the
vacuum by tachyon condensation at infinity. Thus the RR charge of an
excitation of a given supersymmetric vacuum configuration is best measured by
subtracting from its K-theory class the K-theory class of the vacuum. The RR
charge of an excitation of the vacuum therefore takes values in
K-theory with compact support.

\newsection{Type IIA D-Branes and $\K^{-1}(X)$}

In this section we will show that D-brane charges in Type IIA superstring
theory are classified by the higher K-group $\K^{-1}(X)$ \cite{witten,horava}.
Again we shall simply reproduce the well-known spectrum of the Type IIA
theory, but we shall gain many new insights into the constructions of D-branes
as well as the interrelationships between branes in the Type II theories.
Furthermore, we shall uncover some remarkable applications of the bound state
construction.

\subsection{The Group $\K^{-1}(X)$}

There are a number of equivalent definitions of $\K^{-1}(X)$, each of which
are useful in different situations. The 11-dimensional ``M-theory'' definition
was given in sections 2.4 and 2.5. In this definition, the higher K-group
$\K^{-1}(X)$ is the subgroup of $\wt{\K}(X\times {\bf S}^1)$ which classifies
RR charges in Type IIA string theory on the ten-dimensional spacetime $X$.
It is therefore tempting to interpret the ${\bf S}^1$ here as the
compactification circle used in relating 11-dimensional M-theory and
ten-dimensional Type IIA superstring theory. However, there are no
spacetime-filling M-branes,  i.e. no known 10-branes, and also no hierarchy
of branes in M-theory. So at present it is unclear how to interpret the
eleven-dimensional extension of $X$ required to classify D-brane charges
of Type IIA string theory. It is for these physical reasons that alternative
formulations of the group $\K^{-1}(X)$ are desired.

The ``string theory'' definition of $\K^{-1}(X)$, i.e. with no reference to an
11-dimensional extension of $X$, is similar to the definition of relative
K-theory introduced in section 2.6 and can be given as follows.
Let $E\in{\rm Vect}(X)$ and let $\alpha: E\stackrel{\approx}
{\longrightarrow} E$ be an automorphism of the vector bundle
$E$. Two pairs $(E,\alpha)$ and $(F,\beta)$ are called isomorphic if
there exists an isomorphism of vector bundles $h:E\stackrel{\approx}
{\longrightarrow} F$ such that the following diagram commutes:
\beq
{\begin{array}{ccc}
E&\stackrel{h}{\longrightarrow}&F\\&
&\\{\scriptstyle\alpha}\downarrow& &\downarrow{\scriptstyle\beta}\\&
&\\E&\stackrel{h}{\longrightarrow}&F\end{array}}
\eeq
i.e. $\beta\circ h=h\circ\alpha$. Define the sum of two pairs $(E,\alpha)$
and $(F,\beta)$ by $(E\oplus F, \alpha\oplus\beta)$. A pair $(E,\alpha)$ is
called elementary if $\alpha$ is homotopic to ${\rm Id}_{E}$ within the
automorphisms of $E$. Two pairs $(E,\alpha)$ and $(F,\beta)$ are
equivalent, $(E,\alpha) \sim (F,\beta)$, if there exists two elementary
pairs $(G,\gamma)$ and $(H,\delta)$ such that
\beq
(E\oplus G,\alpha\oplus\gamma)\cong
(F\oplus H, \beta\oplus\delta)\ .
\eeq
The set of equivalence classes of pairs $[(E,\alpha)]$ defines an abelian
group, which is precisely $\K^{-1}(X)$. The inverse of a
class $[(E,\alpha)]$ is $-[(E,\alpha)]=[(E,\alpha^{-1})]$. To prove
this, we need to show that $(E\oplus E,\alpha\oplus\alpha^{-1})$ is an
elementary pair, where we may write
\beq
\alpha\oplus\alpha^{-1}=\left( \begin{array}{cc}\alpha & 0\\
0&\alpha^{-1}\end{array}\right)=
\left(\begin{array}{cc}0 & -\alpha\\ \alpha^{-1}&0
\end{array}\right)
\left(\begin{array}{cc}0 & 1\\ -1&0\end{array}\right)\ .
\eeq
Using the decomposition
\beq
\left(\begin{array}{cc}0 & -\alpha\\ \alpha^{-1}&0\end{array}\right)
=\left(\begin{array}{cc}1 & -\alpha\\ 0&1\end{array}\right)
\left(\begin{array}{cc}1 & 0\\ \alpha^{-1}&1\end{array}\right)
\left(\begin{array}{cc}1 & -\alpha\\ 0&1\end{array}\right)\ ,
\label{alphamat}\eeq
we may define a continuous map $\sigma:[0,1]\rightarrow {\rm Aut}(E\oplus
E)$ by
\beq
\sigma(t)=\left(\begin{array}{cc}1 & -t\alpha\\ 0&1\end{array}\right)
\left(\begin{array}{cc}1 & 0\\ t\alpha^{-1}&1\end{array}\right)
\left(\begin{array}{cc}1 & -t\alpha\\ 0&1\end{array}\right)
\eeq
with $\sigma(0)={\rm Id}_{E\oplus E}$ and $\sigma(1)$ coinciding with
(\ref{alphamat}). It follows that (\ref{alphamat}) is homotopic to
${\rm Id}_{E\oplus E}$ within automorphisms of $E\oplus E$, and hence so is
$\alpha\oplus\alpha^{-1}$. More generally, it can be shown that
$[(E,\alpha)]+[(E,\beta)]=[(E,\alpha\circ\beta)]=[(E,\beta\circ\alpha)]$.
Note that the analogous statement for relative
K-theory in section 2.6 can be proven in a similar way.

To show that this abelian group is indeed $\K^{-1}(X)$, we need only prove
that two automorphisms $\alpha,\beta$ determine the same class in
$\K^{-1}(X)$, i.e. $[(E,\alpha)]=[(F,\beta)]$, if and only if there exists
a vector bundle $G\in {\rm Vect}(X)$ such that
$\alpha\oplus {\rm Id}_{F}\oplus {\rm Id}_{G}$ and
${\rm Id}_{E}\oplus \beta \oplus {\rm Id}_{G}$ are homotopic within the
automorphisms of $E\oplus F\oplus G$. For this, we first demonstrate that
$[(E,\alpha)]=0$ if and only if there exists a vector bundle
$G\in{\rm Vect}(X)$ such that $\alpha\oplus {\rm Id}_{G}$ is homotopic to
${\rm Id}_{E\oplus G}$ within the automorphisms of $E\oplus G$. Indeed, if
$[(E,\alpha)]=0$ then there exists elementary pairs $(G,\gamma)$ and
$(G',\gamma')$ and an isomorphism
$h:E\oplus G\stackrel{\approx}{\longrightarrow}G'$ such that the diagram
\beq
{\begin{array}{ccc}
E\oplus G&\stackrel{h}{\longrightarrow}&G'\\&
&\\{\scriptstyle\alpha\oplus\gamma}\downarrow& &
\downarrow{\scriptstyle\gamma'}\\&
&\\E\oplus G&\stackrel{h}{\longrightarrow}&G'\end{array}}
\eeq
is commutative. Thus $\alpha\oplus {\rm Id}_G$ is homotopic to
$\alpha\oplus\gamma=h^{-1}\circ\gamma'\circ h$. This, in turn, is homotopic
to $h^{-1}\circ{\rm Id}_{G'}\circ h={\rm Id}_{E\oplus G}$, which proves the
assertion. The converse statement is obvious. Going back to the original
assertion, we consider two classes with $[(E,\alpha)]=[(F,\beta)]$.
Then, $[(E,\alpha)]-[(F,\beta)]=[(E\oplus F,\alpha\oplus\beta^{-1})]=0$ and,
as just argued, there exists a vector bundle $G$ such that
$\alpha\oplus\beta^{-1}\oplus {\rm Id}_G$ is homotopic to
${\rm Id}_{E\oplus F\oplus G}$. By composing this homotopy equivalence with
${\rm Id}_E\oplus\beta\oplus{\rm Id}_G$ one sees that
$\alpha\oplus{\rm Id}_F\oplus{\rm Id}_G$
and ${\rm Id}_E\oplus\beta\oplus{\rm Id}_G$ are homotopic. The converse
statement is again obvious.

This ``string theory'' definition of $\K^{-1}(X)$, as well as its properties
described above, generalize to give the higher Grothendieck group
$\K^{-1}({\cal C})$ associated to any category ${\cal C}$ which is an abelian
monoid. Our third and final definition of $\K^{-1}(X)$ is one that relates
the ``M-theory'' and ``string theory'' definitions, thereby showing their
precise equivalence. Going back again to the definition of $\K^{-1}(X)$ as a
subgroup of $\wt{\K}(X\times{\bf S}^1)$, we identify $(E,\alpha)$ with
$(E_{\alpha},E_{{\rm Id}_{E}})$, where $E_{\alpha}$ is the vector bundle over
${\bf S}^1\times X$ with total space $[0,1]\times E$ modulo the
identification $(0,v)\equiv(1,\alpha(v))$ for all $v\in E$.

To relate $\K^{-1}(X)=\K(\Sigma X)$ to homotopy theory, we use the
observation stated after eq. (\ref{loopspace})
in section 2.7. It follows that there is a natural isomorphism
\beq
\K^{-1}(X)=\Bigl[X\,,\,U(\infty)\Bigr]\ ,
\eeq
where $U(\infty)=\bigcup_{k=1}^{\infty}U(k)$ is the infinite unitary group.
In particular, it is possible to show that
\beq
\K^{-1}({\bf S}^n)= \pi_{n-1}\Bigl(\Grass(k,2k;\complex)
\Bigr)\ , \ \ \ \ k>n\ ,
\eeq
where
\beq
\Grass(k,2k;\complex)=\frac{U(2k)}{U(k)\times U(k)}\ ,
\eeq
and where $k>n$ defines the stable range for $\K^{-1}(X)$.

\subsection{Unstable 9-Branes in Type IIA String Theory}

To describe supersymmetric $p$-branes of the Type IIA theory as elements of
a K-theory group of the spacetime $X$, we have to resort to looking at
bound states of unstable 9-branes. If we relax the usual requirements
that D-branes preserve half of the original supersymmetries and that
they carry one unit of the corresponding RR
charge, then Type IIA $p$-branes with $p$ odd are allowed and in
particular we have spacetime-filling 9-branes.
These states are non-supersymmetric unstable excitations in the
superstring theory, as there is always a tachyon in the
spectrum of open strings connecting a 9-brane to itself. Thus the
9-branes of Type IIA are highly unstable, and we expect
that they should rapidly decay to the supersymmetric vacuum by
tachyon condensation on the spacetime-filling worldvolume (there are
no RR fields in the corresponding Type IIA supergravity that would
couple to any such conserved charges).
But as before, the unstable D-brane configurations can carry lower-dimensional
D-brane charges, so that when the tachyon rolls down to the minimum of
its potential and the state decays, it leaves behind a supersymmetric
state that differs from the vacuum by a lower-dimensional D-brane
charge,i.e. the state decays into a supersymmetric D-brane configuration and
one can represent a supersymmetric D-brane state as the bound state of the
original system of unstable branes.
Note that a representation in terms of bound states of 8-branes and
$\overline{8}$-branes is possible using the constructions of the previous
section. However, such a construction is undesirable, as it breaks some of
the manifest spacetime symmetries (in the choice of an 8-brane worldvolume
submanifold of $X$), and it limits the kinematics of branes that can be
studied in this way. We shall therefore present a string theoretical
construction that keeps all spacetime symmetries manifest.

The D9-brane boundary state $|D9\rangle$, as a coherent state in the Type
IIA closed string Hilbert space, is of the form
\beq
|D9\rangle=|D9,+ \rangle_{\NS}-|D9,-\rangle_{\NS}\ ,
\label{D9IIA}\eeq
where $|D9,\pm\rangle_{\NS}$ are the two possible implementations of
Neumann boundary conditions on all spacetime coordinates of $X$. Since
\beq
(-1)^{F_{L,R}}|D9,\pm\rangle_{\RR}=|D9,\mp\rangle_{\RR}\ ,
\eeq
no combination of the states $|D9,\pm\rangle_{\RR}$ is invariant under
the Type IIA GSO projection operator $\frac14(1-(-1)^{F_L})(1+(-1)^{F_R})$
and hence there is no RR component in the D9-brane boundary state.
But this just means that there is no RR tadpole, and thus no
spacetime anomalies related to RR tadpoles can arise for the IIA
9-branes, i.e. unlike the Type IIB case, where the number of 9-branes must
equal the number of $\overline{9}$-branes, there is no restriction on the
number of Type IIA 9-branes.  Moreover, the 9-branes carry no conserved
charge, and there is no distinction between 9-branes and
$\overline{9}$-branes in IIA.

There is no GSO projection in the open string channel of the torus
amplitude $\langle D9|D9\rangle$ and therefore, in the NS sector, the
open strings connecting a 9-brane to itself will contain both the
$U(1)$ gauge field $A_{\mu}$ that a supersymmetric D-brane would
contain, and the tachyon field $T$ which would be otherwise
eliminated by the GSO projection for supersymmetric branes.
Furthermore, in the Ramond sector of the open string both spacetime
chiralities $\chi, \chi'$ of the ground state spinors are retained. We can
generalize this construction to the case of $N$ coincident 9-branes. Then the
free open string spectrum of
massless and tachyonic states gives rise to the following low-energy field
content on the spacetime-filling worldvolume: a $U(N)$ gauge field $A_{\mu}$,
a tachyon field $T$ in the adjoint representation of $U(N)$, and two
chiral fermion fields $\chi,\chi'$ of opposite spacetime chirality in the
adjoint representation of $U(N)$. This can be compared with the Type IIB case,
where $N$ pairs of 9-branes and $\overline{9}$-branes give rise to a
spectrum consisting of a $U(N)\times U(N)$ gauge field and a tachyon
field in the bifundamental representation of $U(N)\times U(N)$.

Note that the case $N=1$ is ``degenerate'' and will be dealt with
separately later on in section 4.4. Notice also that the field content on
$N$ 9-branes of Type IIA superstring theory coincides with the
ten-dimensional decomposition of an 11-dimensional system, with
$A_M=(A_{\mu},T)$ an 11-dimensional $U(N)$ gauge field, and
$\Psi=(\chi,\chi')$ a 32-component spinor field in the adjoint
representation of $U(N)$. This indicates a hidden 11-dimensional
symmetry of the lowest lying open string states. It hints at a
possible connection to M-theory which is in agreement with the properties of
the K-group $\K^{-1}(X)$ described in the previous subsection.

Consider the configurations of $N$ 9-branes in Type IIA superstring
theory up to possible creation and annihilation of 9-branes to and
from the vacuum. An {\it elementary} 9-brane configuration is one which
rapidly decays to the supersymmetric vacuum, and therefore does not contain any
lower-dimensional D-brane charges. Any elementary configuration of
$N'$ 9-branes wrapping the spacetime $X$ gives rise to a $U(N')$
bundle $F$, together with a $U(N')$ gauge field $A$ on $F$ and a
tachyon field $T$ in the adjoint representation of $U(N')$. The
presence or absence of lower D-brane charges is thereby measured by the
tachyon condensate $T_0$. Thus, as before, we assume that a bundle $E$ with
tachyon field $T$ can be deformed by processes involving only creation
and annihilation of 9-branes into a bundle isomorphic to $E\oplus F$
where $F$ is the Chan-Paton bundle of an elementary 9-brane configuration.

We therefore consider the set of equivalence classes of 9-branes with tachyon
condensate, up to creation and annihilation of elementary 9-brane
configurations to and from the vacuum. A 9-brane configuration thereby
defines an element $[(E,\alpha)]\in \K^{-1}(X)$ where $E$ is the
rank-$N$ Chan-Paton bundle carried by the system of $N$ unstable Type
IIA 9-branes. We will see in the following that the automorphism $\alpha$ is
given by
\beq
\alpha=-\exp(\pi i\,T) \ ,
\label{automorph}\eeq
and it acts by the natural adjoint action (conjugation) on $E$. Here $T$ is the
adjoint $U(N)$ tachyon field on the 9-brane worldvolume. The possible 9-brane
configurations up to creation and annihilation of elementary 9-branes are
therefore classified by $\K^{-1}(X)$. It is instructive to compare this to
the situation in Type IIB, where $\K(X)= \zed\oplus \wt{\K}(X)$ and D-brane
charges are classified by the reduced K-theory group $\wt{\K}(X)$ with tadpole
anomaly cancellation requiring that the number of 9-branes equals the number of
$\overline{9}$-branes (here the integer in $\zed$ is in general the difference
between the number of 9-branes and the number of $\overline{9}$-branes).
In the Type IIA theory, we have $\wt{\K}^{-1}(X)=
\K^{-1}(X)$, with no tadpole restriction on the number of IIA 9-branes. Also,
as previously computed, we have
\bea
\wt{\K}({\bf S}^{2n})&=& \zed\ \ , \ \
\wt{\K}({\bf S}^{2n+1})~=~0\nonumber\\
\K^{-1}({\bf S}^{2n+1})&=& \zed\ \ , \ \
\K^{-1}({\bf S}^{2n})~=~0\ .
\eea
This represents the fact that Type IIB contains supersymmetric $p$-branes for
$p$ odd, while Type IIA has supersymmetric $p$-branes for $p$ even. Note that
in this language, Bott periodicity is the statement that there are only two
Type II superstring theories.

\subsection{The Bound State Construction}

We shall now present an explicit bound state construction of $p$-branes with
worldvolume $Y$ of odd codimension in the spacetime $X$ as bound states of
unstable Type IIA 9-branes. This will show that $\K^{-1}(X)$ indeed does
classify D-brane charges in Type IIA superstring theory. This bound state
construction is simply the analog of the ABS construction, now mapping
classes in $\wt{\K}(Y)$ to classes in $\K^{-1}(X)$ in K-theory. This
shows that whatever can be done with stable lower-dimensional branes
can be done with unstable 9-branes of the Type IIA theory.

Consider a system of $N$ unstable 9-branes. The gauge group is $U(N)$
and the tachyon field lives in the adjoint representation ${\bf N}\otimes
\overline{\bf N}$ of $U(N)$ with
tachyon potential $V(T)=V(-T)$. If we assume that $T$ condenses into
one of its vacuum expectation values $T=T_0$, and that
the negative energy density associated with the condensate cancels the
positive energy density associated with the 9-brane tension, as in
(\ref{finiteenergy}), then the system of 9-branes completely annihilates into
the supersymmetric vacuum and is therefore an elementary configuration. In
general, $T$ has the tendency to roll down to the minimum of its potential
$V(T)$ and break part of the $U(N)$ gauge symmetry. The precise
symmetry breaking pattern depends on the structure
of the eigenvalues of $T_0$, i.e. on the precise form of the tachyon potential.
For example, consider the symmetric tachyon potential
\beq
V(T)=-m^2\,{\rm tr}\,T^2+\lambda^2\,{\rm tr}\,T^4+\ldots\ ,
\label{potential}\eeq
which is anticipated from the structure of the disc amplitudes at tree-level
in open string perturbation theory. In this case,
$T_0=T_v\cdot{\rm diag}(\pm 1,\pm1,\ldots, \pm 1)$ after diagonalization.
Corrections to (\ref{potential}) from worldsheets
with more than one boundary give terms of the form
\beq
\delta V(T)=\tilde{\lambda}^2\left[{\rm tr}\,T^2\right]^2+\ldots
\eeq
It can be shown that if $\lambda^2\geq 0$ and $\tilde{\lambda}^2>0$, then the
minimum $T_0$ of the tachyon potential still has only two distinct
eigenvalues $\pm\,T_v$.

We will henceforth assume that the 9-brane system under consideration has
tachyon condensate $T_0$ with the same number of positive and negative
eigenvalues. The number of 9-branes is therefore $2N$, and the gauge group
$U(2N)$ is broken down to $U(N)\times U(N)$. The Type IIA vacuum manifold is
thus
\beq
{\cal V}_{\rm IIA}(2N)=\frac{U(2N)}{U(N)\times U(N)}\ ,
\label{IIAvac}\eeq
and it parametrizes the stable vortex-like configurations of the tachyon
field. Away from the core of such a stable vortex (at
$|x|\rightarrow\infty$), the tachyon field approaches its vacuum
expectation values. This defines a map
${\bf S}^m\rightarrow {\cal V}_{\rm IIA}(2N)$, where the sphere ${\bf S}^m$
asymptotically surrounds the core of a stable vortex of codimension $m+1$ in
the spacetime $X$. Therefore, the stable tachyon vortices are parametrized by
classes in
\beq
\K^{-1}({\bf S}^{m+1})=
\pi_{m}\Bigl({\cal V}_{\rm IIA}(2N)\Bigr)=\left\{
\begin{array}{cll}\zed~~~~&,&~~m=2k\\
0~~~~&,&~~m=2k+1\end{array}\right.
\eeq
{}From this we see that the Type IIA system exhibits stable bound states in odd
codimension $2k+1$. Note that the Type IIA vacuum manifold (\ref{IIAvac})
of the tachyon field on the 9-branes is a finite-dimensional approximation to
the classifying space $BU(\infty)$ for complex vector bundles over $X$.

We shall now explicitly construct the bound state tachyon vortices, which we
will interpret as supersymmetric D$(2p)$-branes of the Type IIA theory.
As before, K-theory selects a prefered natural value for the
number $2N$ of 9-branes used to build the bound state (and is the same
number that would arise in a ``stepwise'' construction). Namely,
bound states in codimension $2k+1$ are most efficiently described by
$2N=2^k$ 9-branes. Then the stable tachyon vortices are classified
topologically by the homotopy groups
\beq
\pi_{2k}\Bigl({\cal V}_{\rm IIA}(2^k)\Bigr)= \zed\ .
\eeq
Again we can explicitly construct the classical tachyon soliton field
corresponding to the generator of this homotopy group. The spacetime-filling
worldvolume of $2^k$ 9-branes supports a $U(2^k)$ Chan-Paton bundle,
which we identify as the spinor bundle ${\cal S}$ of the group
$SO(2k+1)$ of rotations in the transverse space whose spinor
representation of dimension $2^k$ is irreducible. The tachyon field is then
given by
\beq
T(x)=\sum_{i=1}^{2k+1}\Gamma_i\, x^i\ ,
\eeq
where $\Gamma_i$ are the Dirac matrices of $SO(2k+1)$ and $x^i$ are
local coordinates in the transverse space. The tachyon field is a map
$T: {\cal S}\rightarrow{\cal S}$ and it asymptotically takes values in
the Type IIA vacuum manifold (\ref{IIAvac}). This should
be contrasted with the Type IIB case, where the tachyon field asymptotically
took values in the IIB vacuum manifold (\ref{IIBvac}),
because of the different structure of the Clifford algebra representations
corresponding to the rotation groups $SO(2k)$ and $SO(2k+1)$.

In the present case we can go even further, and construct
explicitly the non-trivial $U(2^k)$ gauge field configuration that
lives on the 9-branes and must accompany the tachyon vortex above
due to the finite energy conditions imposed on the system as a whole.
There is a natural map
\beq
\pi_{2k-1}\left(U(2^{k-1})\right)\longrightarrow
\pi_{2k}\left({\cal V}_{1}(2^k)\right)\ ,
\eeq
defined by the transformation of tachyon generators
\beq
T_{\rm IIB}(x)=\sum_{i=1}^{2k}\Gamma_i\,x^i~~\longmapsto~~
T_{\rm IIA}(x)=T_{\rm IIB}(x)+x^{2k+1}\,\sigma_3\otimes\id_{2^{k-1}}\ .
\label{IIAIIBmap}\eeq
Here $T_{\rm IIB}(x)$ is the IIB tachyon field, and it is constructed as the
unbroken part of the non-trivial $U(2^k)$ gauge field. Decomposing the sphere
as before as ${\bf S}^{2k}={\bf S}_+^{2k}\cup {\bf S}_-^{2k}$, with
${\bf S}^{2k-1}={\bf S}_+^{2k}\cap {\bf S}_-^{2k}$, gauge fields on
${\bf S}_{\pm}^{2k}$ are topologically trivial and can be patched together to
give a global gauge field, with the appropriate magnetic charge on
${\bf S}^{2k}$, using the transition function on the equator ${\bf S}^{2k-1}$.
This large gauge transformation is just $T_{\rm IIB}(x)$, and the unbroken
long-ranged gauge field of $U(2^{k-1})\times U(2^{k-1})$ corresponds to
that of a generalized magnetic monopole.

Using $T_{\rm IIA}^2=|x|^2$, it is possible to show \cite{bfn,horava} that
the bundle automorphism (\ref{automorph}) is actually the generator of
the homotopy group $\pi_{2k+1}(U(2^k))$, and that far away from the
core of the vortex, $T_{\rm IIA}(x)\in{\cal V}_{\rm IIA}(2^k)$. This induces
the natural map
\beq
\pi_{2k}\left({\cal V}_{\rm IIA}(2^k)\right)\longrightarrow
\pi_{2k+1}\left(U(2^k)\right)\ ,
\eeq
defined by
\beq
T_{\rm IIA}~~\longmapsto~~\alpha=-\exp(\pi i\,T_{\rm IIA})\ .
\eeq
Thus the tachyon condensate represents the generator of the relative K-theory
group $\K^{-1}({\bf B}^{2k+1},{\bf S}^{2k})= \zed$, and the above bound state
construction is precisely the analog of the ABS construction, now mapping
classes in $\wt{\K}(Y)\hookrightarrow \K^{-1}(X)$ for $Y$
of odd codimension in the spacetime manifold $X$ wrapped by the unstable
9-branes of the Type IIA theory. Again the precise embedding is given
by the cup product (\ref{cuprel}) as
\bea
\widehat{\lambda}\,:\,\wt{\K}(Y)\otimes_\zeds
\K^{-1}({\bf B}^{2k+1},{\bf S}^{2k})&{\buildrel\approx\over\longrightarrow}&
\K^{-1}(Y\times{\bf B}^{2k+1},Y\times {\bf S}^{2k})
\nn\\\Bigl[(E\,,\,F)\Bigr]&\longmapsto&\widehat{\lambda}
\left[\Bigl((E\otimes{\cal S})_{{\rm Id}_E\otimes\alpha}
\,,\,(F\otimes{\cal S})_{{\rm Id}_F\otimes\alpha}\Bigr)\right]
\label{absmapK1}\eea
for $[(E,F)]\in \wt{\K}(Y)$.

In this way we get a hierarchy of bound state constructions in IIA
and IIB, represented by brane systems of increasing dimensions which
support worldvolume gauge groups that form a natural hierarchy
\beq
U(1)\subset U(1)\times U(1)\subset U(2)\subset U(2)\times U(2)\subset U(4)
\subset U(4)\times U(4)\subset \cdots \ .
\eeq
This property leads to the usual descent relations among D-branes
\cite{sendescent,os}. In this hierarchy, the bound state construction
in terms of pairs of stable branes alternates with the bound state
construction in terms of unstable branes. It shows that a
supersymmetric D$p$-brane of Type II superstring theory can be
constructed as the tachyonic kink in the worldvolume of an unstable
D$(p+1)$-brane (see the next subsection), or alternatively as a bound
state vortex in a $(p+2)$-brane $\overline{(p+2)}$-brane pair, or yet as
a bound state of two unstable $(p+3)$-branes, and so on. This
procedure continues until one reaches the spacetime filling dimension,
thereby ending up with a construction in terms of 9-branes in which all
spacetime symmetries are manifest.

As a simple example, consider the case of codimension $2k+1=3$. The K-theory
gauge group is then $U(2)$, acting on two unstable 9-branes whose Chan-Paton
bundle in the ${\bf2}$ representation of $U(2)$ is identified with the spinor
bundle ${\cal S}$ of the rotation group $SO(3)$ of the transverse space.
Using standard Pauli spin matrices $\sigma_i$ for the Dirac matrices of
$SO(3)$ gives
\beq
T(x)=\sum_{i=1}^3\sigma_i\, x^i =
\left( \begin{array}{cc}x^3 & x^1+ix^2\\ x^1-ix^2&-x^3
\end{array}\right)\ .
\label{codim3t}\eeq
The tachyon field (\ref{codim3t}) represents a vortex of vorticity 1.
The finite energy condition ties it to the non-trivial $U(2)$ gauge field
\bea
A_i(x)&=&\frac1{|x|^2}\left(1-\frac{|x|}{\sinh|x|}\right)
\sum_{j=1}^3\, \Gamma_{ij}\,x^j\nn\\
\Gamma_{ij}&\equiv&\frac{1}{2}\Bigl[\sigma_i\,,\,\sigma_j\Bigr]
=\sum_{k=1}^3\epsilon_{ijk}\,\sigma^k\ .
\label{tpmonopole}\eea
Up to the trivial lift from $SU(2)$ to $U(2)$ gauge theory this is
nothing but the 't~Hooft-Polyakov magnetic monopole in $3+1$ dimensions
\cite{tpmon}, with the convergence factor in (\ref{tpmonopole}) the
usual BPS solution of the 't~Hooft-Polyakov ansatz. This monopole represents
a supersymmetric stable D$(2p)$-brane of Type IIA
superstring theory as a bound state of two unstable D$(2p+3)$-branes.
Alternatively, the diagonal block in (\ref{codim3t}) represents the
construction of one D$(2p+2)$-brane and one D$\overline{(2p+2)}$-brane from a
pair of D$(2p+3)$-branes (see the next subsection), while the off-diagonal
block represents a  D$(2p)$-brane as the bound state of a
D$(2p+2)$-brane-antibrane pair (c.f. eq. (\ref{2tachyon})). Again, this is an
example of the
descent relations in Type II superstring theory \cite{sendescent,os}.

\subsection{Domain Walls in Type IIA String Theory}

The case of codimension 1 (i.e. $k=0$) is ``degenerate'', as
we shall now discuss. According to the general prescription, this
represents a stable 8-brane (or $\overline{8}$-brane) of the Type IIA theory
constructed as the tachyonic kink of $2^k=1$ 9-brane. The gauge group is
$U(1)$ and the tachyon field is a real scalar field of charge 0 which is given
by
\beq
T(x)=\frac{\pm\,T_0\,x^9}{\sqrt{1+(x^9)^2}}\ ,
\label{kink}\eeq
since there is now only one $\Gamma$-matrix which can be taken to be the
$1\times1$ identity matrix. Here $x^9$ is the coordinate transverse to the
core of the kink which represents a domain wall in spacetime. The sign in
(\ref{kink}) distinguishes an 8-brane from an $\overline{8}$-brane and it
corresponds to the sign of the difference $T(-\infty)-T(+\infty)$ between the
asymptotic values of the tachyon field on the two sides of the domain wall.
Note that only one 8-brane or $\overline{8}$-brane may be constructed from
one 9-brane. In this case there is no symmetry breaking of the $U(1)$ gauge
group, and we are left with a $U(1)$ gauge theory and a tachyon field that
can condense into either one of the two vacuum expectation values $\pm\,T_0$.
The relevant homotopy group of the vacuum manifold for one 9-brane is
\beq
\pi_0\Bigl(O(1)\Bigr)=\pi_0\Bigl(\{ \pm\,T_0\}\Bigr)= \zed_2\ ,
\label{O1homotopy}\eeq
and so there is not enough room for the anticipated conserved 8-brane RR
charge that should be classified by $\zed$. Therefore, each
individual 8-brane and $\overline{8}$-brane requires its own 9-brane, and so
the smallest 9-brane system that would accommodate the full
K-theory group $\K^{-1}({\bf S}^1)= \zed$ of 8-brane charges has an infinite
number of spacetime filling branes.

More generally, if the tachyon potential is arranged so that the tachyon
field on the worldvolume of $N$ 9-branes condenses into its vacuum
expectation value $T_0$ with $N-n$ positive eigenvalues and $n$ negative
eigenvalues, then
\beq
T_0=T_v
\left( \begin{array}{cc}\id_{N-n} & 0\\ 0 & -\id_{n}
\end{array}\right)
\eeq
and the $U(N)$ gauge symmetry is broken to $U(N-n)\times U(n)$. As in the
$N=1$ case, for $N>1$ the tachyon field forms kinks of codimension 1.
Suppose that all eigenvalues of $T$ correspond to a kink localized at
a common domain wall $Y$ of codimension 1 in $X$. Then locally
near $Y$ we can write the tachyon field as
\beq
T(x)=
\left( \begin{array}{cc}\frac{T_v\,x^9}{\sqrt{1+(x^9)^2}}\,\id_{N-n}& 0\\
0&-\frac{T_v\,x^9}{\sqrt{1+(x^9)^2}}\,\id_{n}\end{array}\right)\ ,
\eeq
which describes $N-n$ 8-branes and $n$ $\overline{8}$-branes with coinciding
worldvolumes wrapping $Y$. More general configurations of separated
8-branes and $\overline{8}$-branes may be constructed by letting each
eigenvalue vanish along separate submanifolds of codimension 1 in the
spacetime $X$. Again one cannot represent more than $N$ 8-branes and
$\overline{8}$-branes as a bound state of $N$ 9-branes, as one would have to
do a K-theoretic ``stabilization'' by adding other 9-branes in order to keep
the relevant homotopy groups in the stable range.

\subsection{Application to Matrix Theory}

Consider $N$ D0-branes in Type IIA superstring theory on
$\real^{10}$ (or some compactification thereof). Each D0-brane can
be represented as a bound state of 16 unstable spacetime-filling
9-branes, whose worldvolume field theory contains a $U(16)$ gauge
field, a tachyon field $T$ in the adjoint representation of $U(16)$,
and two chiral fermion fields $\chi, \chi'$ of opposite spacetime
chirality in the adjoint representation of $U(16)$. The tachyon field near
the core of each stable point-like soliton can be represented as
\beq
T(x)=\sum_{i=1}^9\Gamma_i\, x^i\ ,
\label{Tparticle}\eeq
where $\Gamma_i$ are the Dirac matrices of the $SO(9)$ group of rotations
of the transverse space to the core of the vortex. This field generates the
homotopy group
\beq
\pi_8\Bigl({\cal V}_{\rm IIA}(16)\Bigr)= \zed\ \ \ \ ,\ \ \ \
{\cal V}_{\rm IIA}(16)=\frac{U(16)}{U(8)\times U(8)}\ .
\eeq
Moreover, the non-trivial long-ranged gauge field gives rise to a ``magnetic
charge'' of each D0-brane, in addition to the unit vorticity from $T(x)$.

The 16 9-branes live in the ${\bf16}$ representation of the
$U(16)$ gauge group, which, in the background of the generalized
magnetic monopole-vortex configuration representing the D0-brane, is
identified with the spinor ${\bf16}$ representation of $SO(9)$. (This
generalizes the three-dimensional 't Hooft-Polyakov monopole, where
the ${\bf3}$ representation of the $SU(2)$ gauge group is identified with the
spinor representation of the space rotation group $SO(3)$). K-theory
implies that the stable string-theoretical soliton carries one unit of
D0-brane charge, and therefore represents a D0-brane as the bound
state of 16 unstable 9-branes. K-theory also implies that
the trivial topology of the D0-brane worldlines in $\real^{10}$ does not
require ``stabilization'' of the configuration by adding extra 9-branes
(this is true even in compactifications of $\real^{10}$). Thus in this
bound state construction, we never need to assume that the worldline $Y$ is
connected, and the spinor bundle ${\cal S}$ in this case is actually trivial
along $Y$, and is thus extendable over $X$ as the trivial
bundle. In other words, we do not need to introduce a new set of 16
9-branes for each additional D0-brane and therefore {\it any} system of $N$
D0-branes can be represented as bound states in a fixed system of 16
unstable spacetime-filling 9-branes.

Thus given a multi-D0-brane state described in terms of just 16 Type
IIA 9-branes, we can follow this 9-brane configuration as we take the
usual Sen-Seiberg scaling limit that defines Matrix theory \cite{seiberg}.
Matrix theory can then be formulated as a theory of stable solitons on the
spacetime-filling worldvolume of 16 unstable 9-branes.
This interpretation of Matrix theory, in terms of vortices in a gauge
theory with fixed gauge group, allows one to change the number $N$ of
D0-branes in the system without changing the rank of the gauge group.
In the conventional formulation of Matrix theory, whereby the
large-$N$ limit requires relating theories with gauge groups of
different ranks, it is very difficult to understand how systems with
different values of $N$ are related (for instance by some
renormalization group approach). But the K-theoretic construction
of D0-branes as magnetic vortices keeps the gauge group fixed for
arbitrary values of $N$. In summary, the dynamics of Matrix theory
appears to be contained in the physics of magnetic vortices on the
worldvolume of 16 unstable 9-branes, described at low energies by a
$U(16)$ gauge theory.

\newsection{Type I D-Branes and $\KO(X)$}

Having used the Type II theories to become well-acquainted with the bound state
constructions and the use of them to describe D-branes in terms of new
solitonic objects, we shall now start considering more complicated superstring
theories in which the K-theory formalism will make some unexpected predictions.
In this section we shall deal with the Type I theory, and thereby make contact
with the original constructions of non-BPS states in string theory. Type I
superstrings are unoriented and their supersymmetric vacuum configuration has
gauge group $SO(32)$ which requires there to be always 32 spacetime filling
branes in the vacuum state. The K-theory of the corresponding Chan-Paton gauge
bundles therefore requires a refinement of what was described in section 2.
It is precisely this difference that will lead to a much richer spectrum of
D-brane charges in the Type I theory.

\subsection{The Group $\KO(X)$}

Consider a system of $N$ 9-branes and $M$ $\overline{9}$-branes in Type I
superstring theory. Tadpole anomaly cancellation now requires that $N-M=32$,
and the branes support an $SO(N)$ bundle $E$ and the antibranes an $SO(M)$
bundle $F$. Because of brane-antibrane creation and annihilation, we identify
pairs of bundles $(E,F)$ with $(E\oplus H,F\oplus H)$ for any $SO(K)$ bundle
$H$. Pairs $(E,F)$ with this equivalence relation define the real K-group of
the spacetime $X$, $\KO(X)$. By replacing $F$ with $F\oplus I^{32}$, it follows
that the configurations of tadpole anomaly cancelling 9-branes and
$\overline{9}$-branes are classified by the reduced real K-theory group
$\wt{\KO}(X)$. Furthermore, it follows from the bound state construction that
D-brane configurations of Type I superstring theory are classified by
$\KO(X)$ with compact support \cite{witten}.

Almost everything we said about $\K(X)$ carries through for real
K-theory. The important change, however, is the relation to homotopy theory.
Namely, for $k>n=\dim X$ (the stable range for $\KO(X)$), we have
\beq
\wt{\KO}(X)=\Bigl[X\,,\,BO(k)\Bigr]\ ,
\eeq
where $BO(k)=\bigcup_{m>k+n}\Grass(k,m;\real)$
is the classifying space for real vector bundles with structure group $O(k)$,
and the real Grassmannian manifold is
\beq
\Grass(k,m;\real)=\frac{O(m)}{O(m-k)\times O(k)}\ , \ \ \ \ m>k+n\ .
\eeq
For $X={\bf S}^n$ we now have that
\beq
\wt{\KO}({\bf S}^n)=\pi_n\Bigl(BO(k)\Bigr)=\pi_{n-1}\Bigl(O(k)\Bigr)\ ,\
\ \ \  k>n\ ,
\eeq
and this group classifies $(9-n)$-brane charges in Type I string theory on flat
$\real^{10}$. The stable homotopy of the orthogonal groups is much different
than that of the unitary groups. For example,
\bea
\pi_0\Bigl(O(k)\Bigr)~=&\zed_2& \ ,\ \ \ \ k\geq 1\nonumber\\
\pi_1\Bigl(O(k)\Bigr)~=&\zed_2& \ ,\ \ \ \ k\geq 3\nonumber\\
\pi_2\Bigl(O(k)\Bigr)~=&0& \ ,\ \ \ \ k\geq 4\nonumber\\
\pi_3\Bigl(O(k)\Bigr)~=&\zed& \ , \ \ \ \ k\geq 5\ .
\label{Ohomotopy}\eea
Note that the identification of 9-brane configurations
with $\KO(X)$ does not really require brane-antibrane
annihilation. This follows from the fact that as $\dim X=10$, $SO(32)$
bundles on $X$ are classified by $\pi_n(SO(32))$ for $n\leq 9$. These
homotopy groups always lie within the stable range, so that all $SO(32)$
bundles on $X$ are automatically classified by $\wt{\KO}(X)$.

The Bott periodicity theorem now states that the homotopy groups of
$O(\infty)$ are periodic with period {\sl eight}:
\beq
\pi_n\Bigl(O(\infty)\Bigr)=\pi_{n+8}\Bigl(O(\infty)\Bigr)\ ,
\eeq
and there are accordingly eight higher-degree $\KO$-groups, defined by
using suspensions as described in section 2.4, with
\beq
\wt{\KO}^{-n}(X)=\wt{\KO}^{-n-8}(X)\ ,
\label{KOperiod}\eeq
and as usual $\KO^{-n}(X)=\wt{\KO}^{-n}(X)\oplus\KO^{-n}({\rm pt})$.
In particular, for $X={\bf S}^n$, we have that
\beq
\wt{\KO}({\bf S}^n)=\wt{\KO}({\bf S}^{n+8})\ .
\label{realperiod8}\eeq
The periodicity (\ref{realperiod8}) can be derived from the ABS
construction for KO-theory. Let ${\rm RO}[Spin(n)]$ be the real
representation ring of the spin group $Spin(n)$, which is generated by the
irreducible real representations. Then following section 2.8, there
is a natural homomorphism
\beq
\varphi: \, {\rm RO}\Bigl[Spin(n)\Bigr]\longrightarrow
\KO({\bf B}^n, {\bf S}^{n-1})
\eeq
which descends to a graded ring isomorphism
\beq
{\rm RO}\Bigl[Spin(n)\Bigr]\,/\, i^*\,{\rm RO}\Bigl[Spin(n+1)\Bigr]
\stackrel{\approx}{\longrightarrow}\wt{\KO}({\bf S}^{n}) \ .
\label{ABSKO}\eeq
Using the periodicity property (\ref{cliffper}) of real Clifford algebras,
eq. (\ref{realperiod8}) is immediate.

The isomorphism (\ref{ABSKO}) can also be used to show how extra
$\zed_2$-valued charges such as those in (\ref{Ohomotopy}) appear in the
D-brane spectrum of the Type I theory. For example, consider the case $n=1$
in (\ref{ABSKO}). Since $\cliff_1=\complex$, the $\cliff_1$-modules are just
complex vector spaces, and the isomorphism
${\rm RO}[Spin(1)]\stackrel{\approx}{\rightarrow}\zed$ is
generated by taking the complex dimension. Similarly, since
$\cliff_2=\quater$, the $\cliff_2$-modules are quaternionic vector
spaces and ${\rm RO}[Spin(2)]\stackrel{\approx}{\rightarrow}\zed$ comes from
taking the quaternionic dimension. The map
$i^*:\, {\rm RO}[Spin(2)]\rightarrow{\rm RO}[Spin(1)]$ is realized by
regarding a quaternionic vector space as a complex vector space
under restriction of scalars. This is just the map
$\zed\rightarrow\zed$ given by multiplication by 2, since the complex
dimension is twice the quaternionic one. This leads to $\wt{\KO}({\bf
  S}^1)=\zed/2\zed=\zed_2$. Moreover, the generators of the right-hand side of
(\ref{ABSKO}) can be conveniently represented in terms of spinor
modules and Clifford multiplication maps. For example, if ${\cal
  S}={\cal S}^+\oplus{\cal S}^-$ is the fundamental graded module for
$\cliff_{4m}$, then
\beq
\varphi^{(4m)}_{\reals}=[{\cal S}^+,{\cal S}^-;\mu]
\eeq
is a generator of the group $\wt{\KO}({\bf S}^{4m})=\zed$, where again
$\mu_x:{\cal S}^+\rightarrow {\cal S}^-$ denotes Clifford
multiplication by $x\in\real^{4m}$. Using the Clifford
module structure and the cup product we can again easily compute that
\beq
\varphi_{\reals}^{(8n)}=\left(\varphi_{\reals}^{(8)}\right)^n \ \ \ ,
\ \ \ 4\varphi_{\reals}^{(8)}=\left(\varphi_{\reals}^{(4)}\right)^2\ .
\eeq
The new torsion KO-groups also modify various product relations that
were described in section 2. For instance, the K\"unneth formula
(\ref{kunneth2}) need not hold in general for spheres, because $\KO({\bf S}^n)$
is not necessarily freely generated. Nevertheless, the analog of
(\ref{kxs}), for example, follows using (\ref{Kiso}) to get
\beq
\widetilde{\KO}(X\times {\bf S}^1)=\widetilde{\KO}^{-1}(X)\oplus
\widetilde{\KO}(X)\oplus\zed_2\ .
\label{KOXS1}\eeq

\subsection{The Bound State Construction}

We need only know the first eight KO-groups to determine the
complete spectrum of (BPS and non-BPS) D-branes in Type I superstring
theory. This spectrum may be found in table \ref{tableKOgroups}.
\begin{table}
\begin{center}
\begin{tabular}{|c|c|c|c|c|c|c|c|c|c|c|c|} \hline
\ \rm D-brane\ & \ D9\ & \ D8\ & \ D7\ & \ D6\ & \ D5\ & \ D4\ & \ D3\ & \ D2\
& \ D1\ & \ D0\ & \ D(--1)\ \\ \hline
Transverse & {} & {} & {} & {} & {} & {} & {} & {} & {} & {} & {}\\
\rm space & \raisebox{1.5ex}[0pt]{${\bf S}^0$} & \raisebox{1.5ex}[0pt]{${\bf
S}^1$} &
\raisebox{1.5ex}[0pt]{${\bf S}^2$} & \raisebox{1.5ex}[0pt]{${\bf S}^3$} &
\raisebox{1.5ex}[0pt]{${\bf S}^4$} & \raisebox{1.5ex}[0pt]{${\bf S}^5$} &
\raisebox{1.5ex}[0pt]{${\bf S}^6$} & \raisebox{1.5ex}[0pt]{${\bf S}^7$} &
\raisebox{1.5ex}[0pt]{${\bf S}^8$} & \raisebox{1.5ex}[0pt]{${\bf S}^9$} &
\raisebox{1.5ex}[0pt]{${\bf S}^{10}$}\\\hline
$\wt{\KO}({\bf S}^n)$ & $\zed$ & $\zed_2$ & $\zed_2$ & 0
& $\zed$ & 0 & 0 & 0 & $\zed$ & $\zed_2$ & $\zed_2$ \\ \hline
\end{tabular}
\end{center}
\caption{\it Type I spectrum of D-branes from $\wt{\KO}({\bf S}^n)$.}
\label{tableKOgroups}\end{table}
The list contains the well-known stable BPS D9-branes, D5-branes and
D-strings of the Type I theory (of integer-valued charges).
The D0-brane is the $\zed_2$-charged D-particle, which is stable but
non-BPS, originally discovered in \cite{senbps}. The D8, D7 and D(--1)-branes
are new predictions of K-theory \cite{witten} which imply that the spectrum of
the Type
I theory should contain new $\zed_2$-charged stable, but non-BPS,
8-branes, 7-branes and instantons. This new spectrum of the Type I theory has
been computed in \cite{frau} using the boundary state formalism,
thus explicitly confirming the K-theory predictions.
Since the D9, D5 and D1 branes all carry RR charge, they are described by
boundary states of the form
\beq
|Dp\rangle =|Dp\rangle_{\NS}
\pm |Dp\rangle_{\RR}\ ,\ \ \ \ p=1,5,9 \,
\eeq
The other non-BPS D-branes do not carry any RR charge, and so have boundary
states
\beq
|Dp\rangle = |Dp\rangle_{\NS}\ ,\ \ \ \ p=-1,0,7,8\ ,
\eeq
and they are their own antibranes. Note that the explicit boundary state
descriptions of the non-BPS D-branes proves that all charges in table
\ref{tableKOgroups} are carried by spacetime defects onto which open strings
can attach.

The Type I theory can be considered as the orientifold projection of Type IIB
superstring theory by the worldsheet parity operator $\Omega$ which reverses
the orientation of the fundamental string worldsheet. The action of the
orientifold group $\Omega$ on Chan-Paton bundles is anti-linear, i.e.
$E\stackrel{\Omega}{\rightarrow} \overline{E}$, where $\overline{E}\cong E^*$
is the conjugate bundle defined by complex-conjugating the transition
functions of $E$. Thus only real bundles survive the orientifold projection,
leading to the $\KO$-theory of real virtual bundles for Type I systems.
For $p=1,5,9$, the corresponding Type IIB RR charge is invariant
under the $\Omega$-projection, i.e. $\Omega|Dp\rangle_\RR=|Dp\rangle_\RR$.
The associated Type I bound state constructions are then just the
orientifold projections of the Type IIB ones. One can describe the non-BPS
branes in terms of bound states of a single BPS brane-antibrane pair of
lowest possible dimension. For $p=0,8$, there is no IIB RR charge, and the
boundary state is automatically even under $\Omega$. The D0 (respectively D8)
brane are topologically stable kinks in the tachyon field on the worldvolumes
of Type I D1--D$\overline{1}$  (respectively D9--D$\overline{9}$) systems,
with $\zed_2$-valued Wilson lines (c.f. section 4.4). For $p=-1,7$, $\Omega$
exchanges IIB $p$-branes with $\overline{p}$-branes, i.e.
$\Omega|Dp\rangle_\RR=-|Dp\rangle_\RR$, so that the $p$-brane-antibrane
configuration is $\Omega$-invariant. Thus the Type I D$(-1)$ (respectively
D7) brane is the orientifold projection of the D(--1)--D$\overline{(-1)}$
(respectively D7--D$\overline{7}$) system in IIB. In these latter two cases,
we may write the corresponding Type I boundary states in terms of those of
the IIB theory as
\beq
|Dp\rangle_{\rm I}= |Dp\rangle_{\rm IIB} + |D\overline{p}\rangle_{\rm IIB}
=|Dp\rangle_{\NS}\ ,\ \ \ \ p=-1,7\ .
\eeq
One may then show that the Type IIB tachyon present in the unstable
$p-\overline{p}$ state is eliminated by the $\Omega$ orientifold projection
\cite{frau}, leading to a stable solitonic state. From these bound state
constructions one may also immediately
deduce the worldvolume field theories of the non-BPS D-branes, and in
particular the worldvolume gauge groups listed in table
\ref{tablegaugegroups}, as we will demonstrate explicitly in the following.
\begin{table}
\begin{center}
\begin{tabular}{|c|c|c|c|c|c|c|} \hline
\ D-brane\ & \ D0\ & \ D1\ & \ D5\ & \ D7\ & \ D8\ & \ D9 \\ \hline
Gauge group & $\zed_2$ & $\zed_2$ & $USp(2)$ & $U(1)$
& $\zed_2$ & $\zed_2$\\ \hline
\end{tabular}
\end{center}
\caption{\it Worldvolume gauge groups of Type I D-branes.}
\label{tablegaugegroups}\end{table}
In the remainder of this section we shall describe some aspects of the
D-brane spectrum of Type I superstring theory using its KO-theory
structure. We will consider each type of soliton separately and discuss the
features unique to each dimensionality.

\subsection{Type I D-Instantons}

The perturbative symmetry group of the Type I superstring should really be
considered as $O(32)$, rather than $SO(32)$, because orthogonal
transformations ${\cal O}$ with $\det{\cal
O}=-1$ are symmetries of Type I perturbation theory, i.e. the
central element $-1$ of $O(32)$ acts trivially on the perturbative spectrum,
so that the corresponding symmetry group is $O(32)/\zed_2$.
This fact makes a connection with how the perturbative gauge group of the Type
I superstring appears, which is locally isomorphic to $SO(32)$.
However, $S$-duality with the $SO(32)$ heterotic string implies that
transformations ${\cal O}$ of determinant $-1$ are actually not
symmetries. This then implies that there must exist some non-perturbative
effect that breaks the group $O(32)$ to its connected subgroup $SO(32)$, and
this is precisely the $\zed_2$-charged gauge instanton associated with
$\pi_9(SO(32))=\zed_2$. This is proven in \cite{witten} using
index-theoretical arguments, namely the fact that a non-trivial bundle on the
sphere ${\bf S}^{10}$ is characterized by having an odd number of fermionic
zero modes of the corresponding chiral Dirac operator (the relationship
between index theory and K-theory will be discussed in section~7.4). The Type
I D-instanton comes from a bound state of Type IIB 9-brane-antibrane pairs
with Chan-Paton bundles $({\cal S}^+,{\cal S}^-)$, while the anti-D-instanton
has gauge bundles $({\cal S}^-,{\cal S}^+)$. Here ${\cal S}^{\pm}$ are the
usual 16-dimensional complex chiral spinor representations of $SO(10)$. The
orientifold projection acts by complex conjugation, so it reverses the
chiralities ${\cal S}^+\leftrightarrow {\cal S}^-$ (equivalently the 9-branes
and $\overline{9}$-branes) and thereby identifies the instanton
and anti-instanton in the Type I theory. For Type I superstrings and KO-theory,
the gauge bundles must be {\it real}, and so we take the Type I 9-brane
Chan-Paton bundles to transform as the spinor module
${\cal S}={\cal S}^+\oplus {\cal S}^-$ which, by regarding complex
representation vector spaces as real ones under restriction of the scalars,
becomes the unique irreducible, real 32-dimensional spinor representation of
$SO(32)$. The $(-1)$-brane is therefore
described by 32 $9-\overline{9}$ brane pairs with Chan-Paton bundles
$({\cal S},\overline{\cal S})$ and a tachyon field $T(x)=\sum_i\Gamma_i\,x^i$.

\subsection{Type I D-Particles}

An element of $\KO(\real^9)$ (or $\wt{\KO}({\bf S}^9)$) is described by a
pair of trivial $SO(N)$ bundles $(E,F)$ over $\real^9$ with a bundle map
$T:E\rightarrow F$ that is an isomorphism near infinity and such that
the rotation group $SO(9)$ acts on the fibers of $E$ and $F$ in the spinor
representation. For KO-theory, we must use real spinor representations, and
for $SO(9)$ there is a unique such irreducible representation ${\cal S}$ of
dimension 16. Thus $E$ and $F$ have rank 16 and transform under $SO(9)$
rotations like ${\cal S}$. The tachyon field is then given
by~(\ref{Tparticle}).

We can compare this K-theoretical construction to the original
construction of the Type I D-particle in \cite{seninst}. For this, we make an
$8+1$ dimensional split of the coordinates and $\Gamma$-matrices. Pick an
$SO(8)$ subgroup of $SO(9)$, and let $x=(x^a,x^9)$ under this split,
with $a=1,\ldots, 8$. The spinor representation ${\cal S}$ of $SO(9)$
breaks up under this split into $SO(8)$ as ${\cal S}={\cal S}^+\oplus
{\cal S}^-$, with ${\cal S}^{\pm}$ the real eight-dimensional chiral spinor
representations of $SO(8)$. Write the $SO(8)$ Dirac matrices as
$\Gamma_a: {\cal S}^-\rightarrow {\cal S}^+$ and
$(\Gamma_a)^{\top}:{\cal S}^+\rightarrow{\cal S}^-$. It then follows from
(\ref{Tparticle}) that the tachyon field decomposes as (see (\ref{IIAIIBmap}))
\beq
T(x)=\left( \begin{array}{cc}x^9\,\id_{16}& \sum_a\Gamma_a\,x^a\\
\sum_a(\Gamma_a)^{\top}\,x^a&-x^9\,\id_{16}
\end{array}\right)\ .
\label{SO8tachyon}\eeq
Changing the basis of Chan-Paton factors on the $\overline{9}$-branes by the
matrix
\beq
\sigma_1\otimes\id_{16}=\left(
\begin{array}{cc}0&\id_{16}\\\id_{16}&0
\end{array}\right)
\eeq
leads to
\beq
T(x)~\longmapsto~(\sigma_1\otimes\id_{16})\,T(x)\,(\sigma_1\otimes\id_{16})=
\left( \begin{array}{cc} \sum_a\Gamma_a\,x^a & x^9\,\id_{16}\\ -x^9\,\id_{16}
&\sum_a(\Gamma_a)^{\top}\,x^a
\end{array}\right)\ .
\eeq
The diagonal blocks here represent two decoupled systems each containing eight
$9-\overline{9}$ pairs. The first set of eight $9-\overline{9}$ pairs has
tachyon field
\beq
T_s(x^a)=\sum_{a=1}^8\Gamma_a\,x^a\ ,
\label{Dstringtachyon}\eeq
and the second one has $(T_s)^\top$. But $T_s$ describes a
D-string located at $x^1=\ldots=x^8=0$ (see section 5.6 below),
and, since $(T_s)^\top$ is made from $T_s$ by exchanging
9-branes with $\overline{9}$-branes, the tachyon field $(T_s)^\top$ describes
an anti-D-string located at $x^1=\ldots=x^8=0$. This is just the construction
in \cite{senbps} of the Type I D-particle from a coincident D-string and
anti-D-string. The off-diagonal blocks correspond to a codimension one
tachyon field which connects the D-string and anti-D-string and is odd under
the reflection $x^9\rightarrow -x^9$. This is precisely the solitonic
configuration of the D1--D$\overline{1}$ tachyon field constructed in
\cite{seninst}. Thus, the K-theory formalism can also be used to
produce string theoretical constructions of non-BPS states.

The spinor quantum numbers carried by the D-particle also appear naturally in
this framework. As shown in \cite{seninst}, the Type I 0-brane transforms in
the spinor representation of $SO(32)$, which agrees with the fact that
the non-perturbative gauge group of the Type I superstring is really the spin
cover $Spin(32)/\zed_2$ of $SO(32)$. In the above construction this can
be seen from the fact that the $\overline{9}$-brane Chan-Paton factors produce
an $SO(32)$ vector of fermionic zero modes, whose quantization gives a spinor
representation of $SO(32)$ (again one uses the index theoretical fact that a
non-trivial $SO(32)$ bundle on ${\bf S}^9$ is characterized by having an odd
number of fermionic zero modes of the corresponding Dirac operator)
\cite{witten}. Furthermore, given $N$ coincident Type I
0-branes, the tachyon vertex operators have the form (in the zero-picture)
\beq
V(\Lambda)=\psi~\e^{ik_0X^0(z)}\otimes\Lambda\ ,
\eeq
where $\Lambda$ is an $N\times N$ matrix which acts on the Chan-Paton factors
and $\psi$ is a worldsheet fermion field. The $\Omega$ projection maps
$\Lambda\rightarrow \Lambda^\top$. Thus, if $\Lambda^\top=-\Lambda$,
then $V(\Lambda)$ is odd under $\Omega$ and the tachyon state survives the
$\Omega$-projection. An antisymmetric matrix $\Lambda$ always has an even
number of non-zero eigenvalues, such that each pair describes the flow toward
annihilation of a pair of 0-branes. This means that the D-particle number is
conserved only modulo 2, in agreement with the fact that
$\wt{\KO}({\bf S}^9)=\zed_2$.
A similar argument applies to the Type I D-instantons.

\subsection{Domain Walls in Type I String Theory}

The Type I D8-brane is described by $\wt{\KO}({\bf S}^1)=\pi_0(O(32))=\zed_2$,
which is represented via trivial bundles $E, F\rightarrow \real^1$ and a
tachyon field $T:E\to F$ that is invertible at infinity. As in section 4.4,
the 8-brane is a domain wall, located at $x^9=0$, and constructed from a single
$9-\overline{9}$ pair with a tachyon field (\ref{kink}) that is positive on
one side and negative on the other side of the wall. In contrast to the
situation of section 4.4, however, the $\zed_2$-valued charges that arise
here from the bound state construction are very natural. One way to see this is
by appealing to the Bott periodicity map (\ref{realperiod8}) which may be
described as follows. Take $[(E_0,F_0)]\in \wt{\KO}({\bf S}^n)$ with tachyon
map $T_0:E_0\rightarrow F_0$, and construct
$[(E,F)]\in \wt{\KO}({\bf S}^{n+8})$ by setting
\beq
E=E_0\otimes ({\cal S}^+\oplus {\cal S}^-)\ \ \ \ , \ \ \ \
F=F_0\otimes ({\cal S}^+\oplus {\cal S}^-)\ ,
\eeq
where ${\cal S}^{\pm}$ are the chiral spinor representations of $SO(8)$ with
Dirac matrices $\Gamma_a: {\cal S}^-\rightarrow {\cal S}^+$.
Let $x^a$ denote the last eight coordinates of $\real^{n+8}$. The tachyon
field is then given as before by the cup product:
\beq
T(x)=\left( \begin{array}{cc}T_0\otimes\id_{16}&{\rm Id}\otimes
\sum_a\Gamma_a\,x^a\\
{\rm Id}\otimes\sum_a(\Gamma_a)^\top\,x^a&-T_0\otimes\id_{16}
\end{array}\right)\ .
\eeq
For example, setting $n=1$ we obtain the tachyon field (\ref{SO8tachyon})
with the diagonal matrix representing 16 8-branes and $\overline{8}$-branes
and the off-diagonal ones corresponding to the bound state construction of a
D-particle in terms of the $8-\overline{8}$ brane pairs. So the relation
(\ref{SO8tachyon}) between the 8-brane and the 0-brane is a typical example
of the Bott periodicity map in Type I superstring theory.

Alternatively, the Bott periodicity isomorphism (\ref{KOperiod}) of KO-groups
comes from taking the cup product of an element of $\wt{\KO}^{-n}(X)$ with
the generator $[{\cal N}_{\reals}]-[I^8]$ of $\wt{\KO}({\bf
  S}^8)=\zed$, where ${\cal N}_\reals$ is the rank 7 Hopf bundle over the
real projective space $\real P^8$ associated with the real Hopf
fibration ${\bf S}^{15}\rightarrow {\bf S}^8$. This shows that the
construction of a $p$-brane in terms of $p+8$-$\overline{p+8}$ brane
pairs in Type I superstring theory is determined by a D-string
solitonic configuration which gives an explicit physical realization
of the $Spin(8)$ instanton. The corresponding eight-dimensional
non-trivial gauge connections, and the associated spinor structures,
may be found in \cite{hopf8}. This identifies the explicit form of the
worldvolume gauge fields living on the $p+8$-$\overline{p+8}$ brane
pair, required to ensure that the tachyon field is covariantly constant near
infinity and hence to produce the finite energy solitonic $p$-brane
configuration, as \cite{hopf8}
\beq
A_i^-(x)=0~~~~~~,~~~~~~
A_i^+(x)=-2i\,\sum_{j=1}^8\Gamma_{ij}\,\frac{x^j}{(1+|x|^2)^2}\ ,
\label{8gaugefield}\eeq
where $\Gamma_{ij}$ are the generators of $Spin(8)$. These gauge field
configurations are $Spin(9)$ symmetric (thereby preserving the
manifest spacetime symmetries) and carry unit topological charge. Similar
arguments apply to the non-BPS D7-brane.

\subsection{Type I D-Strings}

We will now exhibit the Type I D-string as a bound state of 9-branes and
$\overline{9}$-branes, located at $x^1=\ldots =x^8=0$ in $\real^{10}$ with
worldvolume coordinates $(x^0,x^9)$. The group of rotations keeping the
D-string worldsheet fixed is $SO(8)$, which rotates the vector
$x=(x^1,\ldots, x^8)$. The two spinor representations ${\cal S}^{\pm}$ of
$SO(8)$ are both eight-dimensional, with Dirac matrices $\Gamma_a:
{\cal S}^+\rightarrow {\cal S}^-$. Thus we consider a configuration of eight
9-branes and eight $\overline{9}$-branes with trivial gauge bundles, but with
the rotation group $SO(8)$ acting on the Chan-Paton bundles with the rank
eight bundle of the 9-branes transforming as ${\cal S}^+$ and of the
$\overline{9}$-branes as ${\cal S}^-$. The tachyon field is then given by
(\ref{Dstringtachyon}) and the map $x\mapsto\sum_a\Gamma_a\,x^a/|x|$ is
the generator of $\pi_7(SO(8))=\zed$. As in section 4.5, there are no global
obstructions that occur in this bound state construction, since the Type I
spacetime $X$ is a spin-manifold and so is the orientable two-dimensional
D-string worldsheet (a global version of the bound state construction will be
presented in section 7.3). There is also no need to assume in the above
construction that the D-string worldvolume is connected. This implies that any
collection of (disjoint) D-strings can be represented by a configuration of
eight $9-\overline{9}$ pairs and there is no need to introduce eight more pairs
for every D-string. This is in contrast to the Type IIB case, where the
spacetime $X$ need not admit a spin structure and in general one would have to
carry out a K-theoretic stabilization by adding extra $9-\overline{9}$ pairs.

It is interesting to examine some of the gauge solitons we have described
above in light of the $S$-duality between the Type I theory and
$Spin(32)/\zed_2$ heterotic string theory. The Type I D-instanton, D-particle
and D-string are all manifest in heterotic string perturbation theory. The
D-string is equivalent to the perturbative heterotic string \cite{polwitten},
so that the
second quantized Fock space of perturbative heterotic strings can be
described completely by configurations of eight $9-\overline{9}$ brane pairs.
(Note that similar conclusions as those of section 4.5 can also be reached for
heterotic Matrix string theory.) The D-particle is a gauge soliton in the
spinor representation of $SO(32)$, just like some of the particles in the
elementary heterotic string spectrum. Finally, the D-instanton gives a
mechanism that breaks the disconnected component of $O(32)$, and this
symmetry breaking is manifest in heterotic string perturbation theory. Thus,
from the point of view of the heterotic string, these three non-perturbative
objects can be continuously connected to ordinary perturbative objects. We note
also that all of the above bound state constructions, like those of the
previous sections, preserve the manifest symmetries of the transverse spaces
to the D-branes. Moreover, the extra 32 9-branes which must be added for
anomaly cancellation yields an $SO(32)$ gauge symmetry that plays no role in
the above constructions. The bound state construction uses extra
brane-antibrane pairs to enlarge the gauge group, so that $SO(32)$ invariance
is manifest.

\subsection{Type I D5-Branes and the Group $\KSp(X)$}

The group $\KSp(X)$ classifies Type I D-branes which are quantized using
symplectic gauge bundles. The appearance of symplectic gauge symmetry can be
understood from the analysis of \cite{gimonpol} (see also \cite{quivers})
where the requirement of closure of the worldsheet operator product expansion
was shown to put stringent restrictions on the actions of discrete gauge
symmetries on Chan-Paton bundles. In particular, the square of the worldsheet
parity operator $\Omega$ acts on Chan-Paton indices as
\beq
\Omega^2\,:\,~~~~|Dp;ab\rangle~\longmapsto~\sum_{a',b'}\left
(\gamma_\Omega^2\right)^{a'}_a\,\left|
Dp;a'b'\right\rangle\,\left(\gamma_\Omega^{-2}\right
)^{b'}_b=(\pm i)^{(9-p)/2}\,|Dp;ab\rangle
\label{Om2action}\eeq
where $a,b$ are the open string endpoint Chan-Paton labels of a D$p$-brane
state of the IIB theory, and $\gamma_{\Omega}$ denotes the adjoint
representation of the orientifold group in the Chan-Paton gauge group. While
the 9-branes have the standard orthogonal subgroup projection (as required by
tadpole anomaly cancellation), eq. \eqn{Om2action} shows that $\Omega^2=-1$
when acting on, for example, 5-branes (and also on the corresponding tachyon
vertex operators \cite{witten}). The 5-branes must therefore be quantized
using pseudo-real gauge bundles, i.e. Chan-Paton bundles with structure group
$Sp(2N)$ on the 9-branes and $\overline{9}$-branes. An alternative
explanation \cite{witten} uses the fact that a Type I 5-brane is
equivalent to an instanton on the spacetime filling 9-branes that
occupy the vacuum \cite{witteninst}.
The tachyon field breaks the $SO(4N)\times SO(4N)$ gauge symmetry of the
9-$\overline{9}$ brane configuration to the diagonal subgroup
$SO(4N)_{\rm diag}$, which is then further broken down to $Sp(2N)$ by the
instanton field. (Note that for a configuration of unit 5-brane number one
needs at least $4N=4$ spacetime filling brane-antibrane pairs).
Notice that eq.~(\ref{Om2action}) also explains the standard spectrum of
stable BPS D-branes in the Type I theory, as well as the worldvolume gauge
groups listed in table~\ref{tablegaugegroups}.

For $\KSp(X)$ the connection with homotopy theory is given by
\beq
\wt{\KSp}({\bf S}^n)=\pi_{n-1}\Bigl(Sp(k)\Bigr) \ , \ \ \ \ k>n/4\ ,
\eeq
where $k>n/4$ defines the stable range for $\KSp(X)$. As previously,
\beq
\wt{\KSp}({\bf S}^n)=\pi_{n-1}\Bigl(Sp(\infty)\Bigr)\ ,
\eeq
where $Sp(\infty)=\bigcup_kSp(k)$ is the infinite symplectic group. In
this case Bott periodicity takes the special form
\beq
\pi_n\Bigl(Sp(\infty)\Bigr)=\pi_{n+4}\Bigl(O(\infty)\Bigr)\ ,
\eeq
so that
\beq
\wt{\KSp}({\bf S}^n)=\wt{\KO}({\bf S}^{n+4})\ .
\label{KSpBott}\eeq
Thus, any calculation in symplectic K-theory can be reduced to one in real
K-theory. The complete spectrum of corresponding brane charges
can be found in table \ref{tableKSpgroups}, which shows that while the
spectrum of supersymmetric D-branes remains unchanged, that of the stable
non-BPS states differs from before.
\begin{table}
\begin{center}
\begin{tabular}{|c|c|c|c|c|c|c|c|c|c|c|c|} \hline
\ D-brane\ & \ D9\ & \ D8\ & \ D7\ & \ D6\ & \ D5\ & \ D4\ & \ D3\
& \ D2\ & \ D1\ & \ D0\ & \ D(--1)\ \\ \hline
Transverse & {} & {} & {} & {} & {} & {} & {} & {} & {} & {} & {}\\
space & \raisebox{1.5ex}[0pt]{${\bf S}^0$} & \raisebox{1.5ex}[0pt]
{${\bf S}^1$} &
\raisebox{1.5ex}[0pt]{${\bf S}^2$} & \raisebox{1.5ex}[0pt]{${\bf S}^3$} &
\raisebox{1.5ex}[0pt]{${\bf S}^4$} & \raisebox{1.5ex}[0pt]{${\bf S}^5$} &
\raisebox{1.5ex}[0pt]{${\bf S}^6$} & \raisebox{1.5ex}[0pt]{${\bf S}^7$} &
\raisebox{1.5ex}[0pt]{${\bf S}^8$} & \raisebox{1.5ex}[0pt]{${\bf S}^9$} &
\raisebox{1.5ex}[0pt]{${\bf S}^{10}$}\\\hline
$\wt{\KSp}({\bf S}^n)$ & $\zed$ & 0 & 0 & 0
& $\zed$ & $\zed_2$ & $\zed_2$ & 0 & $\zed$ & 0 & 0 \\ \hline
\end{tabular}
\end{center}
\caption{\it D-brane spectrum in Type I string theory with symplectic gauge
bundles.}
\label{tableKSpgroups}\end{table}
The isomorphism (\ref{KSpBott}) comes from taking the cup product with
the class of the rank 2 instanton bundle ${\cal N}_{\quaters}$
associated with the pseudo-real Hopf fibration ${\bf S}^7\rightarrow{\bf
  S}^4$, i.e. the holomorphic vector bundle of rank 2 over $\complex
P^3$ \cite{trautman}. Thus the relationships between a BPS $p$-brane and a BPS
$p+4$-brane is a 5-brane soliton which may be identified with an
$SU(2)$ Yang-Mills instanton field. For example, consider a Type I
D-string in the worldvolume of a $5-\overline{5}$ brane pair \cite{senbps}. The
worldvolume gauge symmetry is $SO(4)=SU(2)\times SU(2)$ and the tachyon
field transforms in its ${\bf2}\otimes\overline{\bf2}$
representation. The $\Omega$-projection identifies the vacuum manifold
of the 5-brane configuration as $SU(2)=Sp(1)$. The topological
stability of the D-string is guaranteed by the homotopy group
$\pi_3(SU(2))=\zed$. A finite energy, static string-like solution in the
corresponding 5+1 dimensional worldvolume field theory is possible when one
imposes the following asymptotic forms on the fields (analogously
to~\eqn{8gaugefield}):
\beq
T\simeq T_v\,U~~~~,~~~~A^-\simeq0~~~~,~~~~A^+\simeq i\,dU\,U^{-1} \ .
\eeq
Here $U$ is an $SU(2)$ matrix-valued function corresponding to the identity
map (of unit winding number) from the asymptotic boundary ${\bf S}^3$, of the
string-like configuration in five dimensions, to the $SU(2)$ group manifold
${\bf S}^3$. Then the string soliton carries 1 unit of instanton number
(living on the 5-brane) which is known to be a source of D-string charge in
Type I string theory \cite{douglasinst,witteninst}. Further applications of the
K-theory
of symplectic gauge bundles will be discussed in section 6.

\subsection{Relationships between Type I and Type II Superstring Theory}

The K-theory formalism has given us many different relations between D-branes
in a given superstring theory. It turns out that it also provides new
relationships between the Type I and Type II theories, which we shall now
proceed to briefly describe. Let us first note that the codimension 1 cases
described in sections 4.4 and 5.5 are actually realizations of the
elementary Hopf fibration ${\bf S}^1\rightarrow {\bf S}^1$ with
discrete fiber $\zed_2$ \cite{trautman}. An example of the construction of a
Type I non-BPS brane as a kink of brane-antibrane pairs is of course the
original construction \cite{seninst} of the Type I D-particle from a D-string
anti-D-string pair. The double cover of ${\bf S}^1$ corresponds to the
pair of branes, and the winding number of the tachyon field is
labelled by the homotopy group (\ref{O1homotopy}) of the fiber corresponding
to the discrete gauge transformation $T\rightarrow -T$ (so that the D-string
carries a $\zed_2$-valued Wilson line). The cup product with
the generator $\omega$ of $\KO({\bf B}^1,{\bf S}^0)=\zed_2$ then achieves the
desired ABS mapping of $\zed_2$-valued KO-theory classes on
$\KO(Y)\to\KO(Y\times{\bf B}^1,Y\times{\bf S}^0)$.

Generally, the Hopf fibration
\beq
{\bf S}^{n-1}~\hookrightarrow~{\bf S}^{2n-1}~\longrightarrow~{\bf S}^n
\label{hopffib}\eeq
is non-trivial only for $n=1,2,4,8$ when its fiber ${\bf S}^{n-1}$ is
a parallelizable sphere \cite{karoubi,husemoller}. This topological fact is
related to the algebraic property that there are only four normed division
algebras over the field of real numbers, corresponding respectively to the
reals, the complex numbers, the quaternions and the octonions (or Cayley
numbers). In that case, the classifying map of the fibration, which
determines the corresponding topological soliton field, is determined by the
principal $Spin(n)$ bundle \eqn{Snprincipal} and is a generator of
\beq
\pi_{n-1}\Bigl(Spin(n)\Bigr)\,/\,\pi_{n-1}\Bigl(Spin(n-1)\Bigr)~
=~\left\{\new{\begin{array}{cll}\zed_2~~~~&,&~~n=1\\\zed~~~~&,
&~~n=2,4,8\end{array}}\right.
\label{hopfclassmap}\eeq
As we have seen, the four Hopf fibrations determine all the fundamental bound
state constructions of D-branes in Type I and Type II superstring theory, and
hence the complete spectrum of D-brane charges in these theories rests on the
fact that there are only four such fibrations. For $n\neq 1$, the topological
charge of the corresponding soliton is given by the Pontryagin number density
which is proportional to $\tr(F^{n/2})$, where $F$ is the curvature of the
associated topologically non-trivial gauge field configuration. For $n=1$ the
charge is determined by a $\zed_2$-valued Wilson line, as in \cite{senbps}.
This feature determines string solitons in terms of magnetic
monopoles in the Type II theories, while in the Type I theories we obtain
non-BPS branes as kinks, BPS branes as $SU(2)$ instantons, and both BPS and
non-BPS branes as $Spin(8)$ instantons. This topological property
realizes all D-branes in terms of more conventional solitons, and it
moreover determines the explicit forms of the non-trivial gauge fields living
on the brane worldvolumes. Therefore, all fundamental D-brane constructions,
and hence the complete spectrum of D-brane charges in Type I and Type II
superstring theory, are quite naturally determined by the four non-trivial
Hopf fibrations \cite{os} which thereby provide a non-trivial link between
the two types of string theories.

Some further connections can be deduced from the relationships that exists
between the different types of K-theories. Given a complex
vector bundle $E$, the correspondence $E\mapsto\overline{E}$ induces
an involution on the group $\K(X)$. Furthermore, the realification and
complexification functors $r$ and $c$ on the categories of real and
complex vector bundles induce homomorphisms of the corresponding
K-groups. The first one associates to each complex vector bundle its
underlying real vector bundle, while the second one associates to each
real vector bundle $E$ the complex vector bundle
$E\otimes_{\reals}\complex\cong E\oplus E$. Then there are the natural
homomorphisms between the K-groups of the Type I and Type IIB theories
\beq
\K(X)\stackrel{r^*}{\longrightarrow}\KO(X)\stackrel{c^*}{\longrightarrow}
\K(X)\ .
\label{rc}\eeq
Note that the composition $r^*\circ c^*$ is multiplication by 2, while
$(c^*\circ r^*)([E])=[E\oplus E]$. For example,  consider the
generator of $\wt{\K}({\bf S}^4)=\zed$, which is the pseudo-real $SU(2)$
instanton bundle described above. To realize it as a generator of
$\wt{\KO}({\bf S}^4)=\zed$, which labels Type I 5-brane charge, it
must be embedded in the orthogonal structure group as
$SO(4)=SU(2)\times SU(2)$ to make it real. The embedding in $SO(4)$
doubles the charge, since the natural map in (\ref{rc}) from $\KO({\bf
  S}^4)$ to $\K({\bf S}^4)$ is multiplication by 2, and so the RR
charge of a Type I 5-brane is twice that of a Type IIB 5-brane. These
facts may be viewed as special instances of the natural periodicity
isomorphisms \cite{karoubi}
\bea
\KO^{-n}(X,Y)\otimes_{\zeds}\zed\left[\mbox{$\frac{1}{2}$}\right]&=&
\KO^{-n-4}(X,Y)\otimes_\zeds\zed\left[\mbox{$\frac{1}{2}$}\right]\nonumber\ ,\\
\K^{-n}(X,Y)\otimes_{\zeds}\zed\left[\mbox{$\frac{1}{2}$}\right]&=&
\left(\KO^{-n}(X,Y)\oplus\KO^{-n-2}(X,Y)\right)\otimes_{\zeds}
\zed\left[\mbox{$\frac{1}{2}$}\right]\ ,
\eea
which can be proven using the cup product with the class of the $SU(2)$
instanton bundle. More generally, the Type I and Type II theories are related
by the exact sequence
\beq
\KO^{-n-1}(X,Y)\stackrel{c^*}{\longrightarrow}\K^{-n-1}(X,Y)
\stackrel{r^*\circ\beta}{\longrightarrow}\KO^{-n+1}(X,Y)
\stackrel{\otimes\omega}{\longrightarrow}\KO^{-n}(X,Y)
\stackrel{r^*}{\longrightarrow}\K^{-n}(X,Y)
\label{IIIrels}\eeq
where $\beta:\K^{-n-1}(X,Y)\to\K^{-n+1}(X,Y)$ is the Bott periodicity
isomorphism.

\newsection{D-Branes on Orbifolds and Orientifolds}

In this section we will analyze the properties of D-branes in orbifolds and
orientifolds of the Type II and Type I theories. As we shall see, the natural
K-theoretic arena for this classification is equivariant K-theory which
takes into account of a group action on the spacetime. Equivariant K-theory
is of enormous interest in mathematics because it merges cohomology with
group representation theory. It is therefore of central importance to both
topology and group theory. In the following we will see that it also leads
to some non-trivial aspects of the D-brane spectrum in these theories.

\subsection{Equivariant K-Theory}

Consider Type IIB superstring theory on an orbifold $X/G$, where $G$ is a
finite group of symmetries of $X$. In this subsection we will show that
D-branes on $X/G$ are classified by the so-called $G$-equivariant K-theory
group of $X$ \cite{segal}. This group is defined as
follows. Let $X$ be a smooth manifold and $G$ a group acting on $X$ (in
general $G$ is either a finite group or a compact Lie group). In this
situation we say that $X$ is a {\it $G$-manifold} and write the $G$-action
$G\times X\rightarrow X$ as $(g,x)\mapsto g\cdot x$. A {\it $G$-map}
$f:X\rightarrow Y$ between two $G$-manifolds is a smooth map which commutes
with the action of $G$ on $X$ and $Y$:
\beq
f(g\cdot x)=g\cdot f(x)\ .
\eeq
In other words, $f$ is $G$-equivariant. A {\it $G$-bundle} $E_G\rightarrow X$
is a principal fiber bundle $E\rightarrow X$ with $E$ a $G$-manifold and
canonical fiber projection $\pi$ which is a $G$-map, i.e. $\pi(g\cdot
v)=g\cdot \pi(v)$, for all $v\in E,\ g\in G$. A {\it $G$-isomorphism}
$E_G\rightarrow F_G$ between $G$-bundles over $X$ is a map which is
both a bundle isomorphism and a $G$-map. These conditions define the category
of $G$-equivariant bundles over the $G$-space $X$. The corresponding
Grothendieck group is called the $G$-equivariant K-theory $\K_{G}(X)$, i.e.
$\K_{G}(X)$ consists of pairs of bundles $(E,F)$ with $G$-action, modulo the
equivalence relation $(E,F)\sim (E\oplus H, F\oplus H)$ for any
$G$-bundle $H$ over $X$.

D-brane configurations on $X/G$ are understood as $G$-invariant
configurations of D-branes on $X$ \cite{quivers}, i.e. the orbifold spacetime
is regarded as a $G$-space. We assume that $X$ is endowed with an orientation
and a spin structure, both of which are preserved by $G$. Given a D-brane
configuration, i.e. a virtual bundle $[(E,F)]$, we can assume that $G$ acts on
$(E,F)$, since the gauge bundles can be constructed in a completely
$G$-invariant way. In tachyon condensation, we assume that a pair of bundles
$(H,H)$ can be created and annihilated only if $G$ acts on both copies
of $H$ in the same way (otherwise the requisite tachyon field would
not be $G$-invariant). Thus, we conclude that for Type IIB
superstrings on an orbifold $X/G$, D-brane charge takes values in
$\K_G(X)$.  For Type IIA one similarly has $\K^{-1}_{G}(X)$ and
for Type I we get $\KO_{G}(X)$ (here $\K_G^{-1}(X)\equiv \K_{G}(\Sigma
X)=\K_{G}({\bf S}^1\wedge X)$ with $G$ acting trivially on the ${\bf S}^1$).

Let $V_I$ denote the irreducible, finite-dimensional complex representation
vector spaces of the group $G$. As in section 2.8, the
isomorphism classes $[V_I]$ of the additive category of $G$-modules with
respect to the direct sum of vector spaces, i.e. with
$[V_I]+[V_J]=[V_I\oplus V_J]$, generate an abelian monoid. The corresponding
Grothendieck group ${\rm R}(G)$ is called the {\it representation ring} of the
group $G$. According to the description of section 2.1, each element of
${\rm R}(G)$ can be expressed as a formal difference
$[V_I]-[V_J]$, where $[V_I]$ and $[V_J]$ are equivalence classes of
finite-dimensional representations of $G$. Thus we have
$[V_I]-[V_J]=[V_I']-[V_J']$ if and only if $V_I\oplus V_J'$ is unitarily
equivalent to $V_I'\oplus V_J$. As always, the tensor product
of vector spaces $V_I\otimes V_J$ induces a
commutative ring structure on ${\rm R}(G)$. For example, let $G={\bf
  S}^1$ and let $y^m$ denote the one-dimensional representation defined by
\beq
\varphi_m(\sigma)\,z=\e^{im\sigma}\,z\ , \ \ \ \ z\in {\bf S}^1\ ,
\eeq
with $\sigma\in\real$. Then it is easy to see that the representation ring of
the compact group ${\bf S}^1=U(1)$ is the ring of formal Laurent polynomials
in the variable $y$:
\beq
{\rm R}({\bf S}^1)=\zed\left[y,y^{-1}\right]\ .
\eeq
The representation ring of the cyclic subgroup
$\zed_n\subset {\bf S}^1$ is the direct sum of $n$ integer groups:
\beq
{\rm R}(\zed_n)=\zed^{\oplus n}\ ,
\eeq
while the representation ring of the torus group ${\bf T}^n=U(1)^n$ is the
ring of formal Laurent polynomials in $n$ variables $y_1,\ldots ,y_n$:
\beq
{\rm R}({\bf T}^n)=\zed\left[y_1,y_2,\ldots,y_n,
(y_1y_2\cdots y_n)^{-1}\right]\ .
\eeq
Generally, for any simply connected Lie group $G$, ${\rm R}(G)$ is a
polynomial ring over $\zed$ with ${\rm rank}(G)$ generators \cite{segalR}.

A more familiar description of ${\rm R}(G)$ is in terms of the space of
characters of the group $G$. The isomorphism class of the $G$-module $V_I$ is
completely determined by its character map $\chi_{V_I}:G\to\complex$
defined by $\chi_{V_I}(g)=\tr_{V_I}(g)$. Since the characters enjoy the
properties $\chi_{V_I\oplus V_J}=\chi_{V_I}+\chi_{V_J}$,
$\chi_{V_I\otimes V_J}=\chi_{V_I}\chi_{V_J}$, and
$\chi_{V_I}(hgh^{-1})=\chi_{V_I}(g)$, it follows that the map
$V_I\mapsto\chi_{V_I}$ identifies ${\rm R}(G)$ as a subring of the ring of
$G$-invariant complex-valued functions on $G$. We
shall see that the representation ring correctly incorporates the
structure of the mirror brane charges induced by the action of $G$ on $X$.

If $G$ acts trivially on the spacetime $X$ then
\beq
\K_G(X)=\K(X)\otimes {\rm R}(G)\ ,
\label{trivialaction}\eeq
where $\K(X)$ is the ordinary K-group of $X$. This follows from the fact
that, for trivial $G$-actions, a $G$-bundle $E$ may be decomposed as
\beq
E\cong \bigoplus_{I}{\rm Hom}_G(E_I,E)\otimes E_I\ ,
\eeq
where $E_I=X\times V_I$ is the trivial bundle over $X$ with fiber $V_I$.
More generally, for any compact $G$-space $X$, the collapsing map
$X\rightarrow$ pt gives rise to an ${\rm R}(G)$-module structure on
$\K^\#_{G}(X)$, such that ${\rm R}(G)$ is the coefficient ring in equivariant
K-theory (rather than simply $\zed$ as in the ordinary case). The
$\K_G$-functor enjoys most of the properties of the ordinary K-functor that
we described in the previous sections. In addition, $\K_{G}$ is functorial
with respect to group homomorphisms. In this sense, $\K_{G}(X)$ is a
generalization of the two important classification groups $\K(X)$ and
${\rm R}(G)$, so that equivariant K-theory unifies K-theory and group
representation theory. In fact, the trivial space $X={\rm pt}$ gives
$\K_G(X)={\rm R}(G)$ while the trivial group $G={\rm Id}$ gives
$\K_G(X)=\K(X)$. A useful ``excision type'' computational feature is that if
$H$ is a closed subgroup of $G$, then for any $H$-space $X$, the inclusion
$i:H\hookrightarrow G$ induces an isomorphism
$i^*:\K_{G}(G\times_HX)\stackrel{\approx}{\longrightarrow}\K_H(X)$.

If the group $G$ acts freely on $X$ (i.e. without fixed points), then $X/G$ is
also a topological space and its $G$-equivariant K-theory is just
\beq
\K_G(X)=\K(X/G)\ .
\label{freeaction}\eeq
However, in general $X/G$ is not a topological space (let alone a smooth
manifold) and the $G$-equivariant cohomology is far more intricate.
Then, a useful theorem for computing equivariant K-theory is the
six-term exact sequence that was introduced in section 2.6:
\beq
{\begin{array}{ccccc}
\K_{G}^{-1}(X,Y)&\longrightarrow&\K_{G}^{-1}(X)&\longrightarrow&
\K_{G}^{-1}(Y)\\&
& & &\\{\scriptstyle\partial^*}\uparrow& & &
&\downarrow{\scriptstyle\partial^*}\\& & &
&\\ \K_{G}(Y)&\longleftarrow&\K_{G}(X)&\longleftarrow&\K_{G} (X,Y)\end{array}}
\label{sixterm}\eeq
where $Y$ is a closed $G$-subspace of a locally compact $G$-space
$X$, and the relative K-theory is defined by
$\K_{G}^{-n}(X,Y)=\widetilde{\K}_{G}^{-n}(X/Y)$ (when the quotient
space makes sense). The advantage of using this exact sequence is that
one may take $Y$ to be the fixed point set of the group action on $X$, such
that the quotient space $X/Y$ has a free $G$-action on it and its equivariant
cohomology can be computed as the ordinary cohomology of its quotient by $G$,
as in \eqn{freeaction}.

{}From now on, we will assume that $G$ is a finite discrete group of
symmetries of the spacetime manifold $X$. Away from any orbifold
singularities of $X/G$, D-brane charge is classified according to
(\ref{trivialaction}) which yields the usual Type II spectrum,
taking into account the mirror images connected by the $G$-action. Therefore,
we want to understand how brane-antibrane pairs behave at the singular
points. For this we need to know how the orbifold projection is realized on the
Chan-Paton factors. From the general theory of D-branes in orbifold
singularities it is known \cite{gimonpol,quivers} that the action of $G$ on
Chan-Paton indices is given by
\beq
g\cdot{\cal X}(\Lambda)={\cal X}(\gamma_g\,\Lambda)\ ,
\eeq
where $\Lambda$ is the Chan-Paton factor of a field $\cal X$ and $\gamma_g$
is the representation of $g\in G$ in the Chan-Paton gauge group. (An example
is the action of the GSO projection that we described at the end of section
3.1). In particular, like a vector potential $A_i(x)$, the tachyon field
transforms in the adjoint representation under the $G$-orbifold
projection, i.e. it is $G$-equivariant:
\beq
g\,:\,~~~~T(x)~~\longmapsto~~\gamma_g~T\left(g^{-1}\cdot x\right)
\,(\gamma_g)^{-1}\ .
\label{equivT}\eeq
In this way, the tachyon field can be thought of as either a
$G$-bundle map $T: E_G\rightarrow F_G$, or
equivalently as a $G$-section of the $G$-bundle $(E\otimes F^*)_G$.

In considering brane charges in terms of $9-\overline{9}$ brane pairs on
orbifold singularities, considerations similar to earlier ones apply, but now
including the mirror images induced under the action of $G$, i.e. at
an orbifold singularity, each brane pair has $|G|$ mirror pairs.
The gauge fields from the vector multiplet of the worldvolume
spectrum in $X/G$ define a connection $A_i(x)$ of the
corresponding Chan-Paton bundle. The GSO projection cancels tachyonic
degrees of freedom leaving only the quiver structure of vector
multiplets and hypermultiplets \cite{quivers}. However, when coincident
branes and antibranes wrap a submanifold $Y/G$ of the
orbifold spacetime, the tachyon field is still preserved by the GSO
projection and the massless vector multiplet is projected out, i.e. the
$G$-action commutes with the GSO projection. The worldvolume field
theory of $N$ branes wrapped on $Y\subset X$ is described via
Chan-Paton bundles $E$ over $Y$ with structure groups $\prod_{I}U(Nn_I)$,
where $n_I$ is the dimension of the $I$-th regular
representation of $G$. The vacuum configuration at infinity is
re-expressed now in a $G$-equivariant way in terms of the characters of $G$.
Then the resulting $G$-invariant
vacuum may be reached by tachyon condensation, provided that
(\ref{equivT}) holds. The bound state construction may now be carried out
just as before. Some explicit constructions of D-branes on orbifolds using
equivariant K-theory may be found in \cite{garcia}.

However, it turns out that for sufficiently ``regular'' orbifolds, equivariant
K-theory does not really provide new information or new states that
are not already described according to ordinary cohomology theory or K-theory
\cite{gukov}. For instance, equivariant Bott periodicity $\wt{\K}_G({\bf
  S}^{p+2})=\wt{\K}_G({\bf S}^p)$  implies that $\wt{\K}_G({\bf
  S}^{2k})=\wt{\K}_G({\bf S}^{0})={\rm R}(G)$, yielding typically $|G|$
copies of the usual RR charge. In the equivariant cases, the Bott
periodicity theorems related different sets of $|G|$ branes to each
other, where $|G|$ is the number of mirror images in the orbifold.
As an illustration, consider the $\zed_3$ AdS-orbifold for Type IIB
supergravity on ${\bf AdS}_5\times {\bf S}^5$ which is dual to the ${\cal
  N}=1$ superconformal field theory on its boundary that is an
$SU(N)^3$ gauge theory on the worldvolume of $N$ parallel D3-branes
placed at an orbifold singularity \cite{lawrence}. The supergravity
horizon is the Lens space
\beq
{\cal H}={\rm L}^2(3)\equiv{\bf S}^5/\zed_3\ .
\eeq
Extended objects in the boundary theory are understood as Type IIB
branes which wrap cycles in ${\cal H}$. The nontrivial homology groups
of the horizon are
\beq
H_1({\cal H})=H_3({\cal H})=\zed_3\ ,
\eeq
which correspond respectively to D3-branes and D5-branes wrapped on a
one-cycle and a three-cycle of ${\cal H}$. However, there are also
wrapped NS5-branes on the three-cycles, corresponding to the discrete
symmetry group $(\zed_3\times\zed_3)\,\semiprod\zed_3$ of the boundary
superconformal field theory. The K-group of the Lens space ${\cal H}$
is \cite{kambe}
\beq
\K({\cal H})=\zed_3\oplus\zed_3\cong H^{\rm even}({\cal H},\zed)\ ,
\eeq
where $H^{\rm even}$ denotes the subring of elements of even degree in the
ordinary cohomology ring. Thus the K-group correctly accounts for the
D3-brane and D5-brane torsion charges, but it is missing the non-commuting
$\zed_3$-valued charge of the NS5-brane. This is not at all surprising,
because topological K-theory always has an underlying {\it commutative} ring
structure, and it does not take into account the Neveu-Schwarz $B$-field
(equivalently the $S$-duality
symmetry of Type IIB superstring theory) \cite{witten}. For this
particular orbifold example, K-theory completely agrees with ordinary
cohomology theory and does not supply us with new objects.
This example lies in a particular class whereby the spacetime manifold is
birationally equivalent to a smooth toric variety for which the K-groups are
torsion-free
and thus the Chern character, to be discussed in section 7.1, yields an
isomorphism
with the corresponding cohomology ring \cite{gukov}.
In light of this feature, we will now turn our attention to orientifolds,
whereby the discrete geometrical action of $G$ on $X$ is further
accompanied by a worldsheet symmetry action on the superstring theory.

\subsection{Real K-Theory}

Real KR-theory \cite{atiyahreal} is a generalized K-theory which merges
complex K-theory, real KO-theory (as well as quaternionic KSp-theory and
self-conjugate KSC-theory) with equivariant K-theory.
We will specialize the orbifold construction above to the
case $G=\zed_2$, so that the space $X$ is equipped with an
involution, i.e. a homeomorphism $\tau: X\rightarrow X$ with
$\tau^2={\rm Id}_{X}$. In addition to the equivariant cohomology, we shall
quotient by the action of the worldsheet parity
transformation $\Omega$. $\Omega$ reverses the orientation of a string, and it
induces an anti-linear involution on gauge bundles $E$ over $X$
that commutes with $\tau$. In making the orientifold projection by
$\Omega$ (in Type IIB string theory on $X$), we need to retain
K-theory classes that are in effect even under the projection by
$\Omega$. Since $\Omega$ acts on 9-brane (and $\overline{9}$-brane)
Chan-Paton bundles by complex conjugation, we consider an induced
anti-linear involution $\tau^*:E_x\rightarrow E_{\tau(x)}$ acting on the
fibers of gauge bundles, with $(\tau^*)^2=1$, that maps $E$ to its complex
conjugate bundle $\overline{E}$. Thus $\tau^*(E)\cong
\overline{E}$ with isomorphism $\psi:\tau^*(E)\rightarrow
\overline{E}$ satisfying $(\psi\tau^*)^2={\rm Id}$. Now we define an
equivalence relation on the category of such vector bundles by
$(E,F)\sim (E\oplus H,F\oplus H)$ for any bundle $H$ that is similarly
mapped by the involution $\tau^*$ to its complex conjugate. The
Grothendieck group of all virtual bundles with involutions on $X$ is
called the Real K-group $\KR(X)$.

As usual, one defines higher groups $\KR^{-m}(X)$ by
\beq
\wt{\KR}^{-m}(X)=\wt{\KR}(X\wedge {\bf S}^m)\ ,
\label{higherKR}\eeq
with the involution $\tau$ on $X$ extended to $X\wedge {\bf S}^m$ by
a trivial action on ${\bf S}^m$. More generally, one can extend the
definition (\ref{higherKR}) to spheres on which $\tau$ acts non-trivially.
Let $\real^{p,q}$ be the $p+q$ dimensional real space in which an involution
acts as a reflection of the last $q$ coordinates, i.e. given
$(x,y)\in \real^p\times\real^q$ we have $\tau:(x,y)\mapsto (x,-y)$. Let
${\bf S}^{p,q}$ be the unit sphere of dimension $p+q-1$ in $\real^{p,q}$ with
respect to the flat Euclidean metric on $\real^p\times\real^q$. With these
definitions, we
may define a two-parameter set of higher degree KR-groups according to
\beq
\wt{\KR}^{p,q}(X)=\wt{\KR}(X\wedge\real_+^{p,q})\ ,
\label{twoparamKR}\eeq
or by using the suspension isomorphism:
\beq
\KR^{p,q}(X)=\KR(X\times \real^{p,q})\ .
\label{KRsuspension}\eeq
With these definitions we have
\beq
\KR^{-n}(X)=\KR^{n,0}(X)\ .
\eeq

Bott periodicity in KR-theory takes the form
\bea
\KR^{p,q}(X)&=&\KR^{p+1,q+1}(X)\ ,\label{KRbottper}\\
\KR^{-m}(X)&=&\KR^{-m-8}(X)\ .\label{KRbott8}
\eea
The relation (\ref{KRbottper}) implies that
$\KR^{p,q}(X)=\KR^{q-p}(X)$ so that $\KR^{p,q}(X)$ only depends on
the difference $p-q$. The relation (\ref{KRbott8}) then states that
$\KR^{p,q}(X)$ depends only on this difference modulo 8.
This implies that one can define negative-dimensional
spheres as those with antipodal involutions in KR-theory, with
${\bf S}^{n,0}$ being identified as ${\bf
  S}^{n-1}$ and ${\bf S}^{0,n}$ as ${\bf S}^{-n-1}$. Note that if
we identify $\real^{1,1}=\complex$ with the involution
$\tau$ acting by complex conjugation, then the $(1,1)$
periodicity theorem (\ref{KRbottper}) takes the particularly nice form
\beq
\KR(X)=\KR(X\times\complex)\ ,
\eeq
for any locally compact space $X$.

KR-theory is a generalization of K-theory and KO-theory because
of the following internal symmetries. If the involution $\tau$ acts
trivially on $X$, then
\bea
\KR^{-m}(X\times {\bf S}^{0,1})&=&\K^{-m}(X)\ ,\label{KRK} \\
\KR^{-m}(X)&=&\KO^{-m}(X)\ .
\label{KRKO}\eea
The relation \eqn{KRK} follows from the fact that the space
$X\times {\bf S}^{0,1}$ can be identified with the double cover
$\tilde{X}=X\amalg X$ of $X$, with $\tau$ acting by exchanging the two
copies of $X$ in $\tilde{X}$. In particular, if $X^{\tau}$ denotes the
set of fixed points of the map $\tau:X\rightarrow X$, then
\beq
\KR^{-n}(X^{\tau})=\KO^{-n}(X^{\tau})\ ,
\eeq
because the involution $\tau$ acts trivially on a fixed point.
There are many further such internal symmetries in Real K-theory, coming from
the usage of negative dimensional spheres. Using the
multiplication maps in the fields $\real$, $\complex$ and $\quater$,
and (1,1) periodicity, one may establish the isomorphisms \cite{atiyahreal}
\beq
\KR(X\times {\bf S}^{0,p})=
\KR^{-2p}(X\times{\bf S}^{0,p})\ ,
\label{KR2p}\eeq
for $p=1,2$ and 4, respectively. This isomorphism for $p=1$ gives the
complex Bott periodicity theorem, while the real periodicity theorem
can be deduced from the case $p=4$. In fact, there is the usual
natural isomorphism
\beq
\KR^{-n}(X\times{\bf S}^{0,p})=
\KR^{-n}(X)\oplus\KR^{p+1-n}(X)\ ,
\eeq
for all $p\geq 3$. The case $p=2$, where there is no splitting into KR-groups
of $X$, is special and will be discussed in
section 6.5. Again, most of the properties discussed in section 2 have
obvious counterparts in the Real case. In particular, the product
formulas derived in section 2 can also be extended to KR-theory (as
they did for KO-theory). For example, by repeating the steps which led
to (\ref{kmxs}) we may obtain, for a trivial action of $\tau$ on $X$, the
product formula
\bea
\widetilde{\KR}^{-1}(X\times{\bf S}^{1,1})&=&\widetilde{\KR}^{-1}
(X\wedge{\bf S}^{1,1})\oplus\widetilde{\KR}^{-1}(X)\oplus
\widetilde{\KR}^{-1}({\bf S}^{1,1})\nonumber\\
&=&\widetilde{\KR}^{1,1}(X)\oplus\widetilde{\KO}^{-1}(X)
\oplus\zed\nonumber\\
&=&\widetilde{\KO}(X)\oplus\widetilde{\KO}^{-1}(X)\oplus\zed\ ,
\label{KRS11}\eea
where we have used $(1,1)$ periodicity, eq. (\ref{KRKO}), and the fact that
\beq
\widetilde{\KR}^{-1}({\bf S}^{1,1})=\KR^{-1+1}({\rm pt})
=\KO({\rm pt})=\zed \ .
\eeq

For $X={\bf S}^n$ the periodicity theorem \eqn{KRbottper} can also be deduced
from the ABS construction for Real K-theory. For this, we define a
two-parameter set of Clifford algebras $\cliff(\real^{n,m})$ of the Real space
$\real^{n,m}$ as the usual algebra associated with $\real^{n+m}$ together with
an involution generated by the action of $\tau$ on $\real^{n,m}$.
A Real module over $\cliff(\real^{n,m})$ is then a finite-dimensional
representation
together with a $\complex$-antilinear involution which preserves the Clifford
multiplication. The corresponding representation ring
${\rm R}[Spin(n,m)]$
is naturally isomorphic to the Grothendieck group generated by the
irreducible $\real$-modules $\Delta_{n,m}$ of the Clifford algebra of the space
$\real^n\oplus\real^m$ with quadratic form of Lorentzian signature
$(n,m)$, as in section 2.8. The
ABS map is now the graded ring isomorphism \cite{lawson,atiyahreal}
\beq
\KR(\real^{n,m})~\cong~{\rm R}\Bigl[Spin(n,m)\Bigr]\,/
\,i^*\,{\rm R}\Bigl[Spin(n+1,m)\Bigr]~=~\KO^{n-m}({\bf S}^0)\ ,
\label{absKR}\eeq
where in the last equality we have used the periodicity relations
(\ref{KRsuspension}), (\ref{KRbottper}) and (\ref{KRbott8}).
This isomorphism relates the groups on the left-hand side of \eqn{absKR} to the
Clifford algebras $\cliff_{n,m}$, so that the topological $(1,1)$ periodicity
\eqn{KRbottper} follows from the algebraic $(1,1)$ periodicity (\ref{cliff11}).

Let us now discuss how Real K-theory can be used to classify D-branes in
$\Omega$-orientifolds. Generally, the fixed point set of a $G$-action on
$X$ is a number of $p+1$ dimensional planes called orientifold $p$-planes, or
O$p$-planes for short. They determine the singular points of the given
orbifold. For the present orientifold group action, these objects are
non-dynamical but they share many of the properties of D-branes themselves.
For instance, they carry RR charge and have light open string states connecting
them and the D-branes, which enhances the gauge symmetry of coincident branes
over an orientifold plane. Having a non-trivial gauge symmetry means that the
supersymmetric vacuum state of these theories must contain 32 D$p$-branes in
order render the vacuum neutral (this is again the requirement of tadpole
anomaly cancellation). We want to determine the charges of stable (but
possibly non-BPS) states localized over an orientifold plane of
$X/\Omega\cdot G$. Note that far away from the orientifold planes, we can
think of the spacetime manifold $X$ as being represented by a double cover
$\tilde{X}\rightarrow X$ with the orientifold group $\Omega\cdot G$ mapping
the two disconnected components of $\tilde{X}$ into each other. Using
(\ref{KRK}), we see that far away from the O$p$-planes
the theory looks just like ordinary Type II superstring
theory. We are therefore interested only in what happens to states
which are localized on the singular O$p$-planes.

{}From the periodicity relations (\ref{KRbottper}) and (\ref{KRbott8}),
we may show quite generally that
\beq
\KR(\real^{d-p,9-d})=\KR(\real^{2d-p-1,0})\ ,
\label{KR9-d}\eeq
where we have identified the spacetime $X$ with the Real space
$\real^{d+1}\times(\real^{9-d}/\Omega\cdot{\cal I}_{9-d})$, with
${\cal I}_{9-d}$ the reflection $\tau$ acting on $9-d$
coordinates of the transverse space. The KR-group (\ref{KR9-d})
classifies D$p$-brane charges localized over an orientifold $d$-plane.
On the right-hand side of (\ref{KR9-d}) we have a
Real space with the KR-involution acting trivially, so that
\beq
\KR(\real^{d-p,9-d})=\KO({\bf S}^{2d-p-1})\ .
\label{KR2d}\eeq
Setting $d=9$ in (\ref{KR2d}) gives the usual group $\KO({\bf
  S}^{9-p})$ that classifies D$p$-brane charge in ordinary Type I
  superstring theory. Setting $d=8$ leads to
\beq
\KR^{-1}(\real^{d-p-1,1})=\KO^{-1}({\bf S}^{15-p})
=\KO({\bf S}^{8-p})\ ,
\label{KO8-p}\eeq
giving a shift by one of the Type I charge spectrum. In the next subsection
it is shown that the $T$-dual of the Type I theory on a spacetime manifold
$X$ is classified by $\KR^{-1}(X)$, in agreement with (\ref{KO8-p}). In
general, for a given dimensionality $d$ of orientifold planes, one may use
\eqn{KR2d} and table \ref{tableKOgroups} with the
appropriate period shift to read off the charges of D-branes located over
the $d$-planes. For example, for $d=5$ we get the classification of stable
D-brane charges localized on an O5-plane. This spectrum resembles that
of Type I string theory in that there is a $\zed$-charged D-string, a
$\zed_2$-charged gauge soliton, and a $\zed_2$-charged gauge
instanton. This spectrum agrees perfectly with the bound state construction
of an orientifold $p$-brane in terms of Type IIB $p$-brane-antibrane pairs,
and the result (\ref{Om2action}) which shows that the tachyonic mode
is removed by the $\Omega$-projection only for $p=-1,7$.
A similar analysis can be carried out for Type IIA orientifolds.

A physical interpretation of the $(1,1)$ periodicity of KR-theory
may also be given \cite{os}. Consider a $p$-brane of codimension $n+m$ in a
Type II orientifold by $\Omega\cdot{\cal I}_m$. The $p$-brane charge is
induced by the tachyon field which is given by Clifford multiplication on the
transverse space $\real^{n,m}$, i.e. $T(x)=\sum_i\Gamma_i\,x^i$ where
$\Gamma_i$ are the generators of the spinor module $\Delta_{n,m}$, and which
generates $\widetilde{\KR}(\real^{n,m})$. Under the ABS isomorphism
(\ref{absKR}), this KR-theory class is multiplied, via the cup product, by
the Hopf generator of $\widetilde{\KR}(\complex P^1)=\zed$ (with its natural
Real structure induced by the antilinear complex conjugation involution), or
equivalently by the spin bundles which carry the spinor representation
$\Delta_{1,1}$. This gives a class with tachyon
field that generates the KR-group of the new transverse space
$\real^{n+1,m+1}$. This class represents a $p-2$-brane of the Type II
orientifold by $\Omega\cdot{\cal I}_{m+1}$. From this mathematical fact one
deduces a new descent relation for Type II orientifold theories, whereby a
$p-2$-brane localized at an O$(8-m)$-plane in a Type II
$\Omega\cdot{\cal I}_{m+1}$ orientifold is constructed as the tachyonic
soliton of a bound state of a $p$-$\overline{p}$ pair located on top of an
O$(9-m)$-plane in a Type II $\Omega\cdot{\cal I}_m$ orientifold. This
realizes the branes of a Type II orientifold as equivariant magnetic
monopoles in the worldvolumes of brane-antibrane pairs of an orientifold with
fixed point planes of one higher dimension. The former orientifold has $2^m$
O$(8-m)$-planes each carrying RR charge $-2^{3-m}$, while the latter one has
$2^{m+1}$ O$(9-m)$-planes of charge $-2^{4-m}$. In the process of tachyon
condensation the number of fixed point planes is doubled while their charges
are lowered by a factor of 2 via a combined operation of charge transfer (via
the equivariant monopole) and dimensional reduction through the orientifold
planes. An example is the non-BPS state consisting of a D5-brane on top of an
orientifold 5-plane in the Type IIB theory \cite{senbps}, which may be
constructed via a tachyon condensate from a pair of D7-D$\overline{7}$
branes on an orientifold 6-plane in the Type IIA theory. The 8 O$6$-planes
each carrying charge $-2$ are transfered to the 16 O$5$-planes of charge $-1$.

\subsection{Type I$\,^\prime$ D-Branes and $\KR^{-1}(X)$}

Type I$\,^\prime$ superstring theory is the $T$-dual of the Type I theory,
which may be obtained as the orientifold of Type IIA string theory of the form
$X/\Omega\cdot {\cal I}_1$. This theory contains unstable spacetime-filling
9-branes, whose configurations up to creation and annihilation of elementary
9-branes classify all D-brane charges. In terms of K-theory this corresponds to
the group $\KR^{-1}(X)$, or the group of equivalence classes
$[(E,\alpha)]$ where $E$ is a Real bundle with
an involution that commutes with the orientifold group and $\alpha$
is an automorphism of $E$ that also preserves the orientifold group
action. In Type I$\,^\prime$ string theory, $E$ is identified with the
Chan-Paton bundle on the worldvolume of the spacetime-filling
9-branes. At the orientifold planes, the gauge symmetry is reduced from
$U(N)$ to $O(N)$. Each individual lower-dimensional brane is
represented as a bound state of a certain number of unstable Type
I$\,^\prime$ 9-branes. The tachyon condensate is required to respect the
$\zed_2$ orientifold symmetry (as in \eqn{equivT}), corresponding to a
$\zed_2$-equivariant monopole. We will discuss this latter property in more
detail in the next subsection.

\subsection{The Bound State Construction for Type II Orientifolds}

A $T$-duality transformation of the Type I theory on an $m$-torus
${\bf T}^m$ gives a Type II orientifold on ${\bf T}^m/\Omega\cdot{\cal I}_m$.
In this section we shall describe some aspects of these orientifold theories
using K-theoretic properties, thereby extending the discussion of the last
subsection. In particular, we will demonstrate how the formalism
allows one to naturally deduce the complete set of vacuum
manifolds for tachyon condensation in the $T$-dual theories of the
Type I theory (see table 7), and hence the worldvolume field contents
of D-branes in these models.
\begin{table}
\begin{center}
\begin{tabular}{|c|c|c|c|c|} \hline
$m$ &  Real Spin Module & Dimension & Vacuum Manifold & Dual Theory\\
\hline\hline{} & {} & {} & {} & {} \\
0 & --- & --- & $O(N)$ & Type I\\
{} & {} & {} & {} & {} \\ \hline
{} & {} & {} & {} & {} \\
1 & $\Delta_1$ & 1 & $\frac{O(2N)}{O(N)\times O(N)}$ & Type I$\,^\prime$\\
{} & {} & {} & {} & {} \\ \hline
{} & {} & {} & {} & {} \\
2 & $\Delta^+_2\oplus\Delta^-_2$ & 2 & $\frac{U(2N)}{O(2N)}$ & IIB on
${\bf T}^2/\Omega\cdot{\cal I}_2$\\
{} & {} & {} & {} & {} \\ \hline
{} & {} & {} & {} & {} \\
3 & $\Delta_3\oplus\Delta_3$ & 4 & $\frac{Sp(2N)}{U(2N)}$ & IIA on
${\bf T}^3/\Omega\cdot{\cal I}_3$\\
{} & {} & {} & {} & {} \\ \hline
{} & {} & {} & {} & {} \\
4 & $\Delta^\pm_4\oplus\Delta^\pm_4$ & 4 & $Sp(2N)$ & IIA on $K3$\\
{} & {} & {} & {} & {} \\ \hline
{} & {} & {} & {} & {} \\
5 & $\Delta_5\oplus\Delta_5$ & 8 & $\frac{Sp(4N)}{Sp(2N)\times Sp(2N)}$ &
IIB on $K3\times{\bf S}^1$\\ {} & {} & {} & {} & {} \\ \hline
{} & {} & {} & {} & {} \\
6 & $\Delta^+_6\oplus\Delta^-_6$ & 8 & $\frac{U(8N)}{Sp(4N)}$ & IIB on
${\bf T}^6/\Omega\cdot{\cal I}_6$\\
{} & {} & {} & {} & {} \\ \hline
{} & {} & {} & {} & {} \\
7 & $\Delta_7$ & 8 & $\frac{O(16N)}{U(8N)}$ & IIA on
${\bf T}^7/\Omega\cdot{\cal I}_7$\\
{} & {} & {} & {} & {} \\ \hline
{} & {} & {} & {} & {} \\
8 & $\Delta^+_8\oplus\Delta^-_8$ & 8 & $O(16N)$ & IIA on ${\bf T}^8/\zed_2$\\
{} & {} & {} & {} & {} \\ \hline
\end{tabular}
\end{center}
\caption{\baselineskip=12pt {\it Type II orientifold theories on spacetimes
$X=Y\times {\bf T}^{1,m}$ whose D-brane charges are classified by the group
$\KR^{-m}(X)$. The general dual orbifold model in each case is listed
(column 5) along with the corresponding vacuum manifold for tachyon
condensation in the worldvolume of $2^{[m/2]+1}N$ spacetime
filling 9-branes (column 4) whose stable homotopy group coincides with
$\KR^{-m}(X)$. The second column lists the
appropriate real spinor module which is used to map each KO-theory
class of the Type I theory into the corresponding KR-theory class of
the orientifold. Their dimensions (column 3) determine the appropriate
increase in the number of 9-branes needed for the bound state
construction as required by K-theoretic stabilization.}}
\label{tablevacuum}\end{table}
(Superstring compactifications will be
discussed in more generality in section 7). The rich structure that now
arises, in contrast to the two unique vacuum manifolds (\ref{IIBvac}) and
(\ref{IIAvac}) for tachyon condensation in the ordinary Type II theories, is
a consequence of the 8-fold periodicity of the KO- and KR-functors. In terms
of iterated loop spaces, $\Omega^nBO(k)$ is of the same homotopy type as
$\Omega^{n+8}BO(k)$, while $\Omega^mO(k)$ for $0\leq m\leq 7$ are of
the same homotopy types as the loop spaces of the Lie groups given in
the fourth column of table 7 \cite{bott}. The vacuum manifolds of the Type II
orientifolds are thereby very natural consequences of the homotopy
properties of KO-theory. Indeed, the identification of these
worldvolume gauge symmetries is a genuinely new prediction made solely
by K-theory. Moreover, the periodicity of 8 is in agreement with the fact that
the cycle of distinguishing properties and dualities of Type II orientifolds
starts over again on the compactification torus ${\bf T}^8$. (A concise
overview of the properties of Type II orientifolds and their moduli spaces
may be found in \cite{mukhi}.) In the rest of this subsection we shall give
some physical interpretations to the appearence of these stable homotopy
properties. More details can be found in \cite{os}.

After a $T$-duality transformation on ${\bf T}^1={\bf S}^1$, the
superstring theory is, as mentioned in the last subsection,
the Type I$\,^\prime$ theory. The mapping between Type I and Type I$\,^\prime$
is similar to the mapping in Type II superstring theory where $T$-duality
maps Type IIB into Type IIA. The induced charge takes values in the
higher KO-group $\widetilde{\KO}^{-1}({\bf S}^{l+1})$,
from which we identify the vacuum manifold in the second line of table
7. Next, consider the toroidal compactification of the Type I theory which is
$T$-dual to Type IIB superstring theory on the ${\bf T}^2/\zed_2$
orientifold. There are $N$ 7-$\overline{7}$ brane pairs that are described in
terms of $2N$ 9-$\overline{9}$ brane pairs which were used in the bound state
construction of D-branes in the original Type I theory. In
the dual orientifold model, lower-dimensional branes may then be
constructed out of the 7-branes using the ``descent'' procedures
described earlier. The appearance of the
unitary group $U(2N)$ in the third line of table 7 is then due to the
following facts. Recall from section 2.8 that the chiral
spinor modules $\Delta_2^\pm$ are complex, so that, in order to preserve the
reality properties of the Type I theory, the desired map which takes
us via the cup product between the K-groups of the two Type I
theories must be taken with respect to the real spinor module
$\Delta_2^+\oplus\Delta_2^-$, as in \eqn{Bottspinmap}. The overall number of
9-branes required for the bound state construction is given by multiplying the
original number of 9-branes by the dimension of the spinor
representation, given in the third column of table 7.
The relevant homotopy is therefore defined with respect to a unitary
symmetric space. Physically, the appearance of a unitary gauge symmetry can
again be understood from (\ref{Om2action}), which leads to an inconsistency
on 7-branes that are therefore quantized using the unprojected unitary gauge
bundles. Thus, while the naive gauge group on the spacetime filling 9-branes
is $O(2N)\times O(2N)\subset O(4N)$, the inconsistent $\Omega$-projection on
IIB 7-branes enhances the symmetry to $U(2N)$. The requisite tachyon field
$T(x)$ is required to be $\zed_2$-equivariant with respect to the orientifold
projection (in order that the resulting lower dimensional brane
configurations be invariant under the $\zed_2$-action), i.e. it transforms
under the orientifold group as in (\ref{equivT}). As shown in \cite{witten},
the tachyon vertex operator for a $p$-$\overline{p}$ brane pair acquires the
phase $(\pm i)^{7-p}$ under the action of $\Omega^2$.
For the 7-branes this operator is even under $\Omega^2$, and so the eigenvalues
of the vacuum expectation value $T_0$ are real. Thus the tachyon
field breaks the $U(2N)$ gauge symmetry down to its orthogonal subgroup
$O(2N)$, and the induced brane charge is labelled by the winding numbers around
the vacuum manifold $U(2N)/O(2N)$ of the IIB orientifold on ${\bf
  T}^2/\zed_2$. Note that, in general, the Type II orientifold on
$X=Y\times {\bf T}^{1,m}$ is described by the KR-group
$\KR^{-m}(X)$. The explicit relation between the KO- and KR-groups
which implements the $T$-duality between the Type I and
orientifold theories will be described in section 7.4.

For $m=3$ we obtain the Type I$\,^\prime$ theory on ${\bf T}^3$ which is
$T$-dual to the ${\bf T}^3/\zed_2$ orientifold of Type IIA superstring theory.
The appearance of a symplectic gauge group in table 7 follows from the
mathematical fact that the complex spinor module $\Delta_3$ is the
restriction of a quaternionic Clifford module, so that the appropriate
augmentation of the spin bundles on the 9-branes is taken with respect
to the rank 4 real representation $\Delta_3\oplus\Delta_3$. This means
that there are now $4N$ unstable 9-branes which have an $Sp(2N)$ worldvolume
gauge symmetry. This enhanced $Sp(2N)$ symmetry comes from the
intermediate representation of a given Type I$\,^\prime$ $p-3$-brane in terms
of 6-$\overline{6}$ brane pairs \cite{os} and is easily understood in terms
of 5-branes, as we discussed in section 5.7. Again by
$\zed_2$-equivariance the tachyon field breaks this gauge group to its complex
subgroup $U(2N)$, so that the vacuum manifold is $Sp(2N)/U(2N)$.
The rest of table 7 can be deduced from similar arguments. Note that the change
of structure of the spinor modules and of the vacuum manifolds after the
$m=4$ compactification is in agreement with the property that the orientifold
planes then begin acquiring fractional RR charges, leading to very
different moduli spaces for these string theories \cite{mukhi}.

In the last column of table 7 we have also indicated the appropriate dual
superstring compactifications to the given toroidal compactification
of the Type I theory (see \cite{mukhi} and references therein). For the cases
$m=4,5$ and 8 we see that the moduli space of the Type I theory (or of the
corresponding Type II orientifold) is actually non-perturbatively dual to a
conventional {\it orbifold} of Type II superstring theory. The corresponding
$\zed_2$-equivariant K-groups have been calculated in \cite{os} using
the product formulas (\ref{kxs}) and (\ref{kmxs}), and the six-term exact
sequence (\ref{sixterm}) (see also the computation at the end of
section 6.6 to follow). This gives a heuristic way to check the given
duality. However, the duality operations involve an intermediate
$S$-duality transformation of the Type II theory, whose description within
the framework of K-theory is not yet known (see again the
discussion in section 6.6 to follow). Thus one does not obtain isomorphisms
of the corresponding K-groups, as would naively be expected. $T$-duality
transformations of K-groups will be described in section 7.4.

Having identified the vacuum manifolds of the Type II
orientifold models, we shall now describe the field content, where one must be
careful about identifying the appropriate homotopy of the relevant vacuum
manifolds. The classifying spaces for Real vector bundles are described in
\cite{krclass}. Consider an orientifold of the Type IIB theory, and a set of
brane-antibrane pairs with worldvolume gauge symmetry $U(N)\times U(N)$. The
$U(N)$ gauge group is endowed with its Hermitian conjugation involution, such
that the fixed point set is the real subgroup $O(N)$. The tachyon field $T$ is
equivariant with respect to the orientifold group, so that
\beq
T(x,-y)=T(x,y)^*
\label{tachyonKR}\eeq
where $(x,y)\in\real^n\oplus\real^m$ are coordinates of the transverse space to
the induced lower dimensional brane configuration. It breaks the worldvolume
gauge symmetry down to $U(N)_{\rm diag}$. The relevant homotopy group
generated by \eqn{tachyonKR} comes from decomposing the one-point
compactification of $\real^{n,m}$ into upper and lower hemispheres as
described in section 2.7, such that the tachyon field is the transition
function on the overlap. The D-brane charges thereby reside in the KR-group
of the transverse space which is given by
\beq
\widetilde{\KR}(\real^{n,m})=\pi_{n,m}\Bigl(U(N)\Bigr)_{\rm R}
\label{KRhomotopy}\eeq
where the homotopy group is defined by the maps ${\bf S}^{n,m}\to U(N)$ which
obey
the Real equivariance condition \eqn{tachyonKR}. The refined Bott
periodicity theorem for stable homotopy in KR-theory then reads
\beq
\pi_{n,m}\Bigl(U(\infty)\Bigr)_{\rm R}=
\pi_{n+1,m+1}\Bigl(U(\infty)\Bigr)_{\rm R}
\label{KRbotthom}\eeq
In a similar way one may relate the Real K-groups
$\widetilde{\KR}^{-1}(\real^{n,m})=\widetilde{\KR}(\real^{n+1,m})$ to the
stable equivariant homotopy of the complex Grassmannian manifold
$U(2N)/[U(N)\times U(N)]$. Note that the gauge fields living on the
brane worldvolumes in these cases must also satisfy an equivariance condition
like \eqn{tachyonKR}. These remarks clarify the meaning of the term
``equivariant soliton'' in the bound state constructions for orbifold and
orientifold theories.

\subsection{Type $\wt{\rm I}$ D-Branes and $\KSC(X)$}

Type I$\,^\prime$ superstring theory has two orientifold O$8^-$ planes which
each carry $-8$ units of RR charge. There is a natural extension
of Type I$\,^\prime$, which involves replacing one of its O$8^-$
planes with an O$8^+$ plane that carries RR charge $+8$ and is
quantized using symplectic gauge bundles (i.e. with $\Omega^2=-1$).
This theory requires no D8-branes to make the supersymmetric vacuum neutral,
so it has no gauge group, yet it still contains interesting stable non-BPS
D-branes in its spectrum. For the classification of D-brane charges,
it is easier to start with the $T$-dual of this theory, which has been
worked out in \cite{wittenvec}. The theory is obtained by gauging a
$\zed_2$-symmetry of Type IIB on a circle, which is realized by the
composition of the worldsheet parity $\Omega$ with a half-circumference shift
along the circle. This theory is usually called Type $\wt{\rm I}$. The
natural K-group of Type $\wt{\rm I}$ D-brane charges is thus
$\KR(X\times {\bf S}^{0,2})$ which, with a trivial involution action on $X$,
is known to be isomorphic to the K-group KSC$(X)$ of self-conjugate bundles
on $X$ \cite{atiyahreal}. This latter group can be defined
as follows. Let $X$ be a compact Real manifold with involution
$\tau$. A {\it self-conjugate} bundle over $X$ is a complex
vector bundle $E$ together with an isomorphism $\alpha:
E\stackrel{\approx}{\longrightarrow}\overline{(\tau^*
  E)}$. Self-conjugate K-theory KSC$(X)$ is then defined as the
Grothendieck group generated by the category of self-conjugate
bundles.

We will first prove that
\beq
\KR(X\times {\bf S}^{0,2})=\KSC(X) \ .
\label{KSCKR}\eeq
Consider the space $X\times{\bf S}^{0,2}$ and decompose the circle
${\bf S}^{0,2}$ into two halves ${\bf S}^{0,2}_{\pm}$ with
${\bf S}^{0,2}_+\cap{\bf S}^{0,2}_-=\{\pm 1\}$. As usual, a Real
vector bundle $E$ over $X\times{\bf S}^{0,2}$ is equivalent to the
specification of a complex vector bundle $E_+$ over $X\times{\bf
  S}^{0,2}_+$ (the corresponding restriction of $E$) together with
an isomorphism
\beq
\psi: E|_{X\times\{ +1\}}\stackrel{\approx}{\longrightarrow}
\tau^*(\overline{E}|_{X\times\{ -1\}})\ .
\eeq
Since $X\times\{ +1\}$ is a deformation retract of $X\times{\bf
  S}^{0,2}_+$, we actually have an isomorphism $E_+|_{X\times\{
  -1\}}\stackrel{\approx}{\longrightarrow}E_+|_{X\times\{ +1\}}$ which
is unique up to homotopy. This means that the specification of $\psi$
is equivalent, up to homotopy, to giving an isomorphism
$\alpha: E\stackrel{\approx}{\longrightarrow}\overline{(\tau^* E)}$. In
other words, isomorphism classes of Real bundles over $X\times{\bf
  S}^{0,2}$ are in one-to-one correspondence with homotopy classes of
self-conjugate bundles over $X$. Taking $\tau$ to be trivial, we
obtain the desired correspondence between $\KR(X\times{\bf S}^{0,2})$
and the K-theory of vector bundles $E$ over a compact manifold $X$
equipped with an antilinear automorphism
$\alpha:E\stackrel{\approx}{\longrightarrow}E$.

Using this equivalence, the D-brane charge spectrum of Type $\wt{\rm I}$
superstring theory can be computed using the results of \cite{green}, and is
summarized in table \ref{tableKSCgroups}.
\begin{table}
\begin{center}
\begin{tabular}{|c|c|c|c|c|c|c|c|c|c|c|} \hline
\ D-brane\ & \ D8\ & \ D7\ & \ D6\ & \ D5\ & \ D4\ & \ D3\
& \ D2\ & \ D1\ & \ D0\ & \ D(--1)\ \\ \hline
Transverse & {} & {} & {} & {} & {} & {} & {} & {} & {} & {} \\
space & \raisebox{1.5ex}[0pt]{${\bf S}^1$} &
\raisebox{1.5ex}[0pt]{${\bf S}^2$} & \raisebox{1.5ex}[0pt]{${\bf S}^3$} &
\raisebox{1.5ex}[0pt]{${\bf S}^4$} & \raisebox{1.5ex}[0pt]{${\bf S}^5$} &
\raisebox{1.5ex}[0pt]{${\bf S}^6$} & \raisebox{1.5ex}[0pt]{${\bf S}^7$} &
\raisebox{1.5ex}[0pt]{${\bf S}^8$} & \raisebox{1.5ex}[0pt]{${\bf S}^9$} &
\raisebox{1.5ex}[0pt]{${\bf S}^{10}$}\\\hline
$\wt{\KSC}({\bf S}^n)$ & $\zed$ & $\zed_2$ & 0 & $\zed$
& $\zed$ & $\zed_2$ & 0 & $\zed$ & $\zed$ & $\zed_2$\\ \hline
\end{tabular}
\end{center}
\caption{\it D-brane spectrum in Type $\wt{\it I}$ superstring theory from
$\wt{\KSC}({\bf S}^n)$.}
\label{tableKSCgroups}\end{table}
This demonstrates that K-theory predicts an interesting spectrum of BPS and
non-BPS states in the Type $\wt{\rm I}$ theory. Upon analyzing the
corresponding groups $\widetilde{\KO}({\bf S}^n)$ and
$\widetilde{\KSp}({\bf S}^n)$ \cite{bgh}, one correctly accounts for the stable
BPS D-branes whose charges are spread out over the two types of
O8-planes. On the other hand, non-BPS $\zed_2$-charged D-branes which
are locally stable near one kind of singular plane can become unstable
due to the other singularities in the complete spacetime
\cite{bgh}. For example, analyzing $\widetilde{\KO}({\bf S}^n)$ shows
that there is a non-BPS D6-brane which is locally stable near the
O8$^-$-plane, because the orientifold projection removes the
tachyonic mode present in the D6-brane mirror D$\overline{6}$-brane
system. However, the orientifold projection is different at the
O8$^+$-plane, so that the tachyon is no longer removed and the non-BPS
D6-brane is no longer stable in the global theory. The $\zed_2$-valued
charges in table \ref{tableKSCgroups} are precisely those non-BPS states
which are globally stable.

The classifying space $BSC(k)$ for self-conjugate vector bundles is described
in \cite{green}, so that KSC-groups are related to homotopy theory by
\beq
\KSC(X)=\Bigl[X\,,\,BSC(\infty)\Bigr]\ .
\eeq
Alternatively, the connection with homotopy theory may be deduced from the
KR-theory representation, from which we can identify the relevant
bound state constructions for D-branes in the Type
$\wt{\rm I}$ theory. From (\ref{KR2p}) it follows that
Bott periodicity of self-conjugate K-theory is 4. Recall that the
group $\KR^{-4}(X\times{\bf S}^{0,2})$ associates a symplectic
projection to $\Omega$. The 4-fold periodicity of KSC-theory is
thereby the indication that the dual of Type $\wt{\rm I}$
has both O8$^-$ and O8$^+$ planes, since it means that
orthogonal and symplectic gauge groups appear on equal footing in this
model. Generally, self-conjugate K-theory is intimately tied to complex, real
and quaternionic K-theories through the following long exact sequences
\cite{anderson}:
\bea
& &\ldots\longrightarrow
\K^{-n-1}(X)\longrightarrow\K^{-n-1}(X)\longrightarrow\KSC^{-n}(X)
\longrightarrow\K^{-n}(X)\longrightarrow\nn
\\& &~~~~~~\longrightarrow\K^{-n}(X)\longrightarrow\ldots
\\& &\ldots\longrightarrow\K^{-n-1}(X)\longrightarrow
\KO^{-n}(X)\oplus\KSp^{-n}(X)\longrightarrow\K^{-n}(X)
\longrightarrow\nonumber\\
& &~~~~~~\longrightarrow\KSC^{-n-2}(X)\longrightarrow\ldots\\
& &\ldots\longrightarrow
\KSC^{-n-1}(X)\longrightarrow\K^{-n}(X)\longrightarrow\KO^{-n}(X)\oplus
\KSp^{-n}(X)\longrightarrow\nonumber\\
& &~~~~~~\longrightarrow\KSC^{-n}(X)\longrightarrow\ldots
\eea
which can be established from the KR version of the Barratt-Puppe exact
sequence \eqn{poppe} and the excision theorem \eqn{excision} applied to the
pairs $(X\times{\bf S}^{0,p},X\times{\bf S}^{0,q})$ for $(p,q)=(2,1)$, $(3,1)$
and $(3,2)$, respectively. These sequences illustrate how the symmetries of
D-brane configurations whose charges
are classified by a given KSC-group are related to webs of gauge
symmetries that appear in the K-theories of Type I and Type II
strings. These interrelationships could prove useful in extending the above
analysis to other Type I models without vector structure \cite{wittenvec}.

\subsection{The Hopkins Groups $\K_{\pm}(X)$}

In this subsection we will study orientifolds of Type IIB superstring theory
obtained via the quotient by the involution $\tau\cdot\klein$, where
$F_{\rm L}$ is the left-moving spacetime fermion number operator. The
operator $\klein$ changes the sign of all spacetime fields in the RR
sector, and therefore the RR charge of a BPS D-brane changes sign and
it gets mapped to its antibrane under $\klein$. In this case, D-brane
configurations on $X/\tau\cdot\klein$ are related to those on $X$ whose
K-theory class is odd under the $\zed_2$ action. This means that $\tau^*$ maps
the pair $(E,F)$ to $(F,E)$, i.e. there are isomorphisms
$\psi:(E,F)\stackrel{\approx}{\longrightarrow}(\tau^*(F),\tau^*(E))$ with
$(\psi\tau^*)^2={\rm Id}$. A trivial pair is $(H,H)$ with
$H\cong\tau^*(H)$. The corresponding Grothendieck group is called the
Hopkins group and is denoted by $\K_{\pm}(X)$ \cite{witten,gukov}.

It can be shown that the group $\K_{\pm}(X)$ may be computed in terms
of conventional equivariant K-theory as
\beq
\wt{\K}_{\pm}(X)=\K^{-1}_{\zeds_2}(X\times\real^{0,1})\ ,
\label{hopkins}\eeq
where the cyclic group $G=\zed_2$ acts on $X\times\real^{0,1}$ as the
product of the action of $\tau$ on $X$ and an orientation-reversing
symmetry of $\real^{0,1}$. The validity of the formula (\ref{hopkins}) may
be argued by defining $\K_{\pm}(X)$ as a (generalized) cohomology theory that
satisfies the exact sequence
\beq
\ldots\longrightarrow\K^{-n}_{\zeds_2}(X)\longrightarrow\K^{-n}(X)
\longrightarrow\K^{-n}_{\pm}(X)\longrightarrow\ldots \ .
\label{exact}\eeq
Comparing \eqn{exact} with the six-term exact sequence (\ref{sixterm}) for
the pair $(M,A)=(X\times\real^{0,1},X\times(\real^{0,1}-{\rm  pt}))$ gives
the pair of exact sequences:
\beq
{\begin{array}{ccccccccc}
\K^{-n}_{\zeds_2}(X)&\rightarrow&\K^{-n}(X)&
\rightarrow&\K^{-n}_{\pm}(X)&\rightarrow&\K^{-n-1}_{\zeds_2}(X)&
\rightarrow&\K^{-n-1}(X)\\ & & & & & & & &\\\parallel& &\parallel& &\downarrow&
&\parallel& &\parallel\\ & & & & & & & &\\
\K^{-n}_{\zeds_2}(A)&\rightarrow&\K^{-n-1}_{\zeds_2}(M,A)&
\rightarrow&\K^{-n-1}_{\zeds_2}(X\times\real^{0,1})&
\rightarrow&\K^{-n-1}_{\zeds_2}(A)&
\rightarrow&\K^{-n}_{\zeds_2}(M,A)\end{array}} \ .
\label{five}\eeq
Applying the five-lemma to (\ref{five}), i.e. that the four isomorphisms
between the two exact sequences in \eqn{five} imply that the remaining middle
vertical mapping is also an isomorphism \cite{spanier}, we arrive at
(\ref{hopkins}). An independent, algebraic argument using automorphism groups
of the corresponding Clifford algebras may also be given \cite{gukov}.

For an orientifold of the type
$X=\real^{d+1}\times(\real^{9-d}/\klein\cdot{\cal I}_{9-d})$, the
corresponding D$p$-brane charge over an orientifold $d$-plane takes values
in $\wt{\K}_{\pm}(\real^{d-p,9-d})$. Since the right-hand side of
(\ref{hopkins}) represents an equivariant functor on the category of complex
vector bundles, we may use the suspension isomorphism with multiplication by
$\complex$ or $\complex/\zed_2$ to derive the periodicities
\beq
\wt{\K}_{\pm}(\real^{p,q})=\wt{\K}_{\pm}(\real^{p,q+2})\ \ \ ,
\ \ \ \wt{\K}_{\pm}(\real^{p,q})=\wt{\K}_{\pm}(\real^{p+2,q})\ .
\label{per}\eeq
This implies that $\wt{\K}_{\pm}(\real^{d-p,9-d})$ depends only on the
parity of $p$ and $d$. Suppose first that $d$ is an even integer. Then
using (\ref{hopkins}) and (\ref{per}) we may compute
\beq
\wt{\K}_{\pm}(\real^{d-p,9-d})=\K^{-1}_{\zeds_2}(\real^{d-p,10-d})
=\K^{p-1}_{\zeds_2}({\rm pt})=
\left\{\begin{array}{cl}{\rm R}[\zed_2]&~~~~,~~p~~{\rm odd}\\0&~~~~,
{}~~p~~{\rm even}\end{array}\right.\ ,
\eeq
where ${\rm R}[\zed_2]=\zed\oplus\zed$ is the representation ring of the
cyclic group $\zed_2$. Thus, when $d$ is even, we obtain the standard
spectrum of BPS D$p$-brane charges for $p$ odd localized over
orientifold planes of odd dimension (the representation ring
${\rm R}[\zed_2]$ accounts for the mirror image brane charges induced by
the given involution). The situation for $d$ odd is a bit more
involved. For this, we apply the six-term exact sequence (\ref{sixterm}) to
the pair $({\bf B}^{d-p,9-d},{\bf S}^{d-p,9-d})$ to get
\beq
\ldots\stackrel{\partial^*}{\longrightarrow}
\K^{-p}_{\zeds_2}({\bf B}^{0,9-d+1},{\bf S}^{0,9-d+1})\longrightarrow
\K^{-p}_{\zeds_2}({\bf B}^{0,9-d+1})\stackrel{i^*}{\longrightarrow}
\K^{-p}(\real P^{9-d})\stackrel{\partial^*}{\longrightarrow}\ldots
\label{BSexact}\eeq
where we have used the suspension isomorphism and $\real P^{9-d}={\bf
  S}^{0,9-d+1}/\zed_2$ is the real projective space. The first K-group
in (\ref{BSexact}) is isomorphic to the Hopkins group $\wt{\K}_{\pm}
(\real^{d-p,9-d})$ that we are interested in. For the second K-group,
we use the fact that the ball ${\bf B}^{0,9-d+1}$ is equivariantly
contractible to get
$\K^{-p}_{\zeds_2}({\bf B}^{0,9-d+1})=\K^{-p}_{\zeds_2}({\rm
  pt})=\delta^{p,{\rm even}}\,{\rm R}[\zed_2]$.
The exact sequence (\ref{BSexact}) thereby relates the K-groups of
interest to the cohomology of the real projective space \cite{karoubi}:
\beq
\K^{-p}(\real P^{9-d})=\delta^{p,{\rm even}}\,\zed\oplus\zed_{2^r}\ ,
\label{Kproj}\eeq
where $r=\left[\frac{9-d}{2}\right]$. A careful analysis of the ring
structure shows that the epimorphism $i^*$ in (\ref{BSexact}) maps both of
the generators of $\K^{-p}_{\zeds_2}({\bf B}^{0,9-d+1})$ into the generator of
$\K^{-p}(\real P^{9-d})$, i.e. $i^*$ is a surjective mapping of the free parts
of the K-groups. The exactness of the sequence \eqn{BSexact} then implies that
\beq
\K^{-p}_{\zeds_2}({\bf B}^{0,9-d+1})=\K^{-p}(\real P^{9-d})\,/\,
\K^{-p}_{\zeds_2}({\bf B}^{0,9-d+1},{\bf S}^{0,9-d+1})\oplus\zed_{2^r} \ ,
\eeq
from which we arrive finally at
\beq
\wt{\K}_{\pm}(\real^{d-p,9-d})=
\left\{\begin{array}{cll}\zed~~~~&,&~~p~~{\rm even}\\0~~~~&,
&~~p~~{\rm odd}\end{array}\right.\ ,
\label{Keven}\eeq
for $d$ an odd integer.

As an example, we see that the D-particle over an O5-plane carries an
integer-valued charge. This configuration is $S$-dual to the stable
non-BPS D-particle on the O5-plane of the corresponding
$\Omega\cdot{\cal I}_4$ orientifold \cite{sentachyon,senparticle}.
The apparent contradiction that arises here owes to the usual fact that
K-theory only classifies the charges of topologically stable objects at weak
string coupling, as mentioned in section 6.4. It is an open problem as of yet
to determine how K-theory correctly incorporates the $S$-duality symmetry of
Type IIB superstring theory. Note that the coincidence
of the brane charges (\ref{Keven}) with those of Type IIA superstring
theory can be traced back to the IIB orientifold boundary
states in the case at hand, which are of the form \cite{sentachyon,senparticle}
\beq
|\widetilde{Dp}\rangle
=\mbox{$\frac{1}{2}$}\,\Bigl(|Up,+\rangle_{\NS}-|Up,-\rangle_{\NS}\Bigr)
+\mbox{$\frac{1}{2}$}\,\Bigl(|Tp,+\rangle_{\RR}+|Tp,-\rangle_{\RR}\Bigr)\ ,
\label{Bptwist}\eeq
where $T$ and $U$ label the twisted and untwisted sectors of the
closed string Hilbert space under the $\klein$ orientifold projection. The
boundary state (\ref{Bptwist}) has precisely the same form as that of the
ordinary Type IIA D$p$-brane.

The relationship with the Type IIA theory can also be seen by taking
$d=0$ in the above construction. In this case we are simply
quotienting the IIB theory by the operator $\klein$, which is known to
map it into Type IIA superstring theory. In general, the operation of
modding out the Type II spectrum $m$ times by $\klein$ determines a
mapping \cite{os}
\beq
\wt{\K}^{-n}(X)\longrightarrow\wt{\K}^{-n-1}_{\zeds_2}(X\times\real^{0,m})\ ,
\label{maps}\eeq
where now the $\zed_2$ acts only as a reflection on $\real^{0,m}$. The
right-hand side of (\ref{maps}) may be evaluated using the six-term exact
sequence (\ref{sixterm}). For example, for $m=1$ we consider in
(\ref{sixterm}) the pair $(X\times\real^{0,1}, X\times\{0\})$. Then
the quotient space $X\times\real^{0,1}/X\times\{0\}$ is homotopic to
two copies of $X\times\real$ which are exchanged by the
involution. Since the $\zed_2$ action on this quotient is free, the
equivariant K-groups may be computed by using the homotopy invariance
of the K-functor and the suspension isomorphism to get
\beq
\K_{\zeds_2}^{-n-1}\Bigl((X\times\real)\amalg(X\times\real)\Bigr)
=\K^{-n-1}(X\times\real)=\K^{-n}(X)\ .
\label{KMA}\eeq
On $X\times\{0\}$ the $\zed_2$ action is trivial, so that
\beq
\K_{\zeds_2}^{-n-1}\Bigl(X\times\{0\}\Bigr)=\K^{-n-1}(X\times{\rm pt})\otimes
{\rm R}[\zed_2]\ .
\label{KA}\eeq
Finally, since in this case $X\times\{0\}$ is an equivariant retract of
$X\times\real^{0,1}$, we have
$\ker\partial^*=\K_{\zeds_2}^{-n-1}(X\times\{0\})$ and so the horizontal
exact sequences in (\ref{sixterm}) split. The general result is then
\beq
\wt{\K}^{-n-1}_{\zeds_2}(X\times\real^{0,m})
=\left(\wt{\K}^{-n-1}(X)\otimes {\rm
R}[\zed_2]\right)\oplus\wt{\K}^{-n-m-1}(X)
\ .\label{Knm}\eeq
The group $\wt{\K}^{-n-1}(X)$ in (\ref{Knm}) comes from the trivial part of
the $\zed_2$ action and as such represents the untwisted brane charges. The
other part $\wt{\K}^{-n-m-1}(X)$ comes from the free part of the $\zed_2$
action and represents the twisted sector.

The case $n=0$, $m=1$ represents the result of quotienting the IIB theory
by $\klein$ \cite{os}. The projection onto the first factor in (\ref{Knm})
thereby represents the condensation of the quotiented IIB brane
configuration onto the corresponding IIA D-brane (along with the
mirror images under the $\klein$ projection). The second direct
summand in (\ref{Knm}) represents the twisted sector of the $\klein$-quotient
which should be properly projected out in the mapping onto the Type IIA
theory. The further quotient by $\klein$ corresponds to
taking $n=1$, $m=2$ in (\ref{Knm}), which maps back into the IIB theory
with the same set of twisted charges projected out.
More details about the explicit construction of these maps in terms of
K-theory classes can be found in \cite{os}. This K-theory construction
agrees with the boundary state description in \cite{daspark} and also the
analysis of the open string spectrum of a Type II $p-\overline{p}$-brane
configuration in \cite{sendescent}. In the former case it was shown that the
result of quotienting the closed superstring Hilbert space by the operator
$\klein$ projects onto the NS-NS part of all IIB $p$-brane boundary
states, with no contributions from the twisted sector. The result is
then a boundary state of the form (\ref{D9IIA}), which, as discussed in
section 4, via tachyon condensation can decay into a stable IIA
D$(p-1)$ configuration. On the other hand, the superposition of a
$p$-brane with a $\overline{p}$-brane can be described by the boundary
state (c.f. eqs. (\ref{Dp}) and (\ref{Dp+-}))
\beq
|Dp\rangle+|D\overline{p}\rangle=|Dp,+\rangle_{\NS}-
|Dp,-\rangle_{\NS}\ ,
\eeq
which thereby produces the same configuration as that obtained above. In these
cases, the K-theory construction shows that the $\klein$ quotient on
the spacetime-filling Type IIB 9-branes leaves an equal number of (identical)
9-branes and $\overline{9}$-branes which are used in the bound state
construction of the $p-\overline{p}$ brane pair \cite{os}. Again this is in
complete agreement with the Type IIA $p-1$-brane configuration that is
eventually reached by tachyon condensation. The naturality of the $\klein$
mapping as a canonical projection on K-theory groups is simply an indication
of the fact that $\klein$ acts as a genuine non-perturbative symmetry of Type
II superstring theory, as discussed in \cite{daspark}.

\newsection{Global Aspects}

This previous section concludes our general analysis of the ways of
classifying D-branes using topological K-theory. There are many
more exotic theories that one would like to study at this stage, for
example orientifolds arising from quotients by the operator
$\Omega\cdot\klein\cdot{\cal I}_m$. However, the corresponding
(equivariant) K-groups for such involutions are not well understood
(see \cite{gukov,os} for some discussion), and such an analysis must await
further developments in the mathematics literature. Let us note that these
latter orientifolds are also important for a more thorough description
of the Type II orientifolds of sections 6.2--6.4 above, in that the
$\Omega\cdot{\cal I}_m$ orientifold projection should strictly
speaking be accompanied by the action of the operator
$(-1)^{\frac{1}{2}(9-p)(8-p)F_{\rm L}}$ on D$p$-brane states in order to
preserve the $\zed_2$-equivariant structures. It is possible that there are
approaches based on {\it algebraic} K-theory which could also be used to
incorporate $S$-duality, and also the construction of M-branes, as has been
recently discussed in \cite{vancea}. We shall not pursue such matters here,
which are still at best at a very preliminary stage. Instead, in this final
section we shall proceed to analyze the interesting D-brane configurations
that arise when one accounts for the global topology of the (possibly
non-trivial) spacetime $X$ and the associated brane worldvolume
embeddings.

\subsection{The Chern Character}

Before proceeding to describe the global aspects of D-branes and their
associated bound state constructions, which we will start in section 7.3, we
shall first need some more mathematical preliminaries. In dealing with global
properties of a space, we shall be forced to consider the cohomology of the
manifolds, in addition to the K-theory of the relevant Chan-Paton bundles.
One of the features of K-theory which makes it so useful in a variety of
applications is the existence of the Chern character homomorphism, which
provides a link between K-theory and ordinary cohomology theory by relating
the ring $\K^\#(X)$ to the usual cohomology ring $H^\#(X)$ (here we shall
deal mostly with \v{C}ech cohomology). In this subsection we will describe the
construction of the Chern character in topological K-theory.

Let $E$ be a complex vector bundle of rank $k$ over a compact topological
space $X$. We can associate to $E$ certain cohomology classes $c_n(E)\in
H^{2n}(X,\zed)$ called the {\it Chern characteristic classes} of
$E$ which measure the twisting of the vector bundle
and which are defined as follows. As in section 2.7, we consider the
universal bundle $Q(k,\infty;\complex)$ over the classifying space $BU(k)$,
whose pullbacks generate vector bundles such as $E$, i.e.
$E=f^*Q(k,\infty;\complex)$ for a certain map $f:\, X\rightarrow BU(k)$.
The cohomology ring $H^\#(BU(k),\zed)$ of the classifying space has
even-degree generators $c_n(Q(k,\infty;\complex))$ whose pullbacks under $f$
are precisely the characteristic classes of $E$:
\beq
c_n(E)\equiv f^*c_n\Bigl(Q(k,\infty;\complex)\Bigr)\in H^{2n}(X,\zed)\ .
\eeq
The basic properties of these characteristic classes are as follows:
\begin{itemize}
\item{{\bf(i)} $c_0(E)=1\in H^0(X,\zed)$.}
\item{{\bf(ii)} For all $l\geq 0$, $c_l(E\oplus F)=\sum_{n+m=l}c_n(E)\wedge
c_m(F)$.}
\item{{\bf(iii)} (Naturality) If $f:Y\rightarrow X$ is a continuous map, then
$c_n(f^*E)=f^*c_n(E)$.}
\end{itemize}
For a rank $k$ bundle $E$, the {\it total Chern class} is defined as
\beq
c(E)=1+c_1(E)+\ldots+c_k(E)\ ,
\eeq
and from property (ii) above it follows that $c(E)$ is multiplicative,
\beq
c(E\oplus F)=c(E)\wedge c(F) \ ,
\eeq
under Whitney sums. In particular, we may invoke the {\it splitting principle}
which states that $E$ is always a Whitney sum of complex line bundles
${\cal L}_n$ (more precisely, $E$ is the pullback of some other vector bundle
which is a sum of line bundles over another space)~\cite{karoubi}, and take
\beq
E={\cal L}_1\oplus{\cal L}_2\oplus\cdots\oplus{\cal L}_k\ .
\eeq
We then have
\beq
c(E)=\prod_{n=1}^kc({\cal L}_n)=\prod_{n=1}^k(1+\lambda_n)\ ,
\eeq
where we have defined $\lambda_n\equiv c_1({\cal L}_n)$. This yields
explicit expressions for the Chern classes of $E$ in terms of
elementary symmetric functions of the two-cocycles $\lambda_n$:
\bea
c_1(E)&=&\sum_n\lambda_n\nn\\c_2(E)&=&\sum_{n<m}\lambda_n\wedge\lambda_m
\nn\\&\cdots&\nonumber\\
c_m(E)&=&\sum_{n_1<n_2<\cdots<n_m}\lambda_{n_1}\wedge \lambda_{n_2}\wedge\cdots
\wedge \lambda_{n_m}\nn\\&\cdots&\nn\\c_k(E)&=&\lambda_1\wedge
\lambda_2\wedge\cdots\wedge \lambda_k\ .
\label{symmfns}\eea

The {\it Chern character} of the vector bundle $E$ is now defined by
\beq
\ch(E)=\sum_{n=1}^k\e^{\lambda_n}\in H^\#(X,\rat)\ ,
\label{chE}\eeq
which can be thought of as a generating function for the characteristic
classes. Note that it takes values in rational cohomology
$H^\#(X,\rat)=H^\#(X,\zed)\otimes_\zeds\rat$, so that $\ch(E)$ cannot detect
any torsion subgroups of the cohomology. Using \eqn{symmfns}, the degree $2m$
part $\ch_m(E)$ of the inhomogeneous cocycle \eqn{chE} can be written in terms
of the characteristic classes of $E$. For example,
\beq
\ch(E)\equiv\sum_{m\geq0}\ch_m(E)=k+c_1(E)+\frac12\Bigl(c_1(E)\wedge c_1(E)
-2c_2(E)\Bigr)+\dots \ .
\label{chEexpansion}\eeq
The definition of the classes $c_m(E)$ (and hence also of the Chern character)
can be generalized to bundles
whose rank is not necessarily constant. For this, one partitions $X$ into open
subsets $X_i$ such that the rank of $E|_{X_i}$ is constant, and then defines
$c_m(E)$ as the unique cohomology class with $c_m(E)|_{X_i}=c_m(E|_{X_i})$.

The Chern character enjoys the following properties:
\begin{itemize}
\item{{\bf(i)} $\ch_0(E)=\rk(E)\in H^0(X,\zed)$.}
\item{{\bf(ii)} $\ch(E\oplus F)=\ch(E)+\ch(F)$.}
\item{{\bf(iii)} $\ch(E\otimes F)=\ch(E)\wedge\ch(F)$.}
\item{{\bf(iv)} (Naturality) $\ch(f^*E)=f^*\,\ch(E)$ for any continuous map
$f:Y\rightarrow X$.}
\end{itemize}
These properties imply that the Chern character respects the semi-ring
structure on the category of vector bundles. Notice that property (i) makes
an explicit connection with the rank function defined in (\ref{rank}), i.e.
the virtual dimension defines a characteristic class in degree 0.
In fact, we can use the Chern character to provide a complete map between
$\K(X)$ and the cohomology ring $H^\#(X)$. Namely, for a virtual bundle
$[(E,F)]\in\K(X)$ we define the homomorphism
\bea
\ch:\, \K(X)&\longrightarrow& H^\#(X,\rat)\nonumber\\
\ch\Bigl(\left[E\right]-\left[F\right]\Bigr)&=& \ch(E)-\ch(F)\ .
\label{chernhomo}\eea
This map is well-defined provided that $[(E,F)]=[(G,H)]$ in $\K(X)$ implies
$\ch(E)-\ch(F)=\ch(G)-\ch(H)$. That this is indeed true is a consequence of
the behaviour (ii) of the Chern character under Whitney sums. For the
particular case where $X={\bf S}^{2n}$, the map ch is an isomorphism onto
$H^\#({\bf S}^{2n},\zed)$. More generally, it can be shown \cite{atiyahhirz}
that the associated map
\beq
\ch:\, \K(X)\otimes_{\zeds}\rat\longrightarrow H^{\rm even}(X,\rat)
\equiv\bigoplus_{n\geq0}H^{2n}(X,\rat)
\label{cheven}\eeq
is an isomorphism, and moreover that this map extends to a ring isomorphism
\beq
\ch:\, \K^\#(X)\otimes_{\zeds}\rat \stackrel{\approx}{\longrightarrow}
H^\#(X,\rat)
\eeq
which maps $\K^{-1}(X)\otimes_{\zeds}\rat$ onto $H^{\rm odd}(X,\rat)$.

In the case where $X$ is a smooth manifold, there is a useful explicit
description of the Chern character. We assume that $E$ is a smooth vector
bundle equipped with a Hermitian connection $\nabla_{E}$, whose curvature is
$\nabla^2_{E}$. The Chern character $\ch(E)\in H^\#(X,\real)$ can then be
represented by the closed inhomogeneous differential form:
\beq
\ch(E)={\rm tr}\,\exp\left(\nabla_{E}^2/2\pi i\right)\ .
\eeq
In this case the $\lambda_n$'s which appear above are the skew-eigenvalues of
the two-form $\nabla_E^2/2\pi i$. To obtain numerical invariants of $X$, we
consider a closed deRham current $\delta_Y$ which is a delta-function supported
representative of the cohomology class of the Poincar\'e dual to an embedded
submanifold $Y\stackrel{i}{\hookrightarrow}X$. Then we can associate to $Y$
a map $I_Y:\,\K^\#(X)\to\complex$ defined by
the natural bilinear pairing on deRham cohomology:
\beq
I_Y(E)=\Bigl\langle\delta_Y\,,\,\ch(E)\Bigr\rangle_{\rm DR}
\equiv\int\limits_X\delta_Y\wedge\ch(E)=\int\limits_Yi^*\,\ch(E)\ .
\label{ICE}\eeq

\subsection{The Thom Isomorphism}

In this subsection we will describe the Thom isomorphism which relates the
K-theory of a manifold $X$ to the K-theory of the total spaces of complex
vector bundles over $X$. In general, this enables one to compute the K-groups
of some relatively complicated spaces in terms of much simpler base spaces.
For example, the K-groups \eqn{Kproj} of real projective spaces may be
determined by the K-theory of a suitable total space over the base
$X={\rm pt}$. In this way the complete set of K-groups for projective spaces
may be determined (see \cite{karoubi} for the details of such calculations).
We shall begin with a description of the map at the level of cohomology, and
then turn to the K-theoretical description. The Thom isomorphism will play an
important role in our discussion of  brane anomalies in section~7.5.

Let $X$ be an oriented manifold of dimension $n$, and let $H^\#(X)$ be its
cohomology ring (it will suffice to consider the cohomology ring with compact
support). A well-known result of differential topology is
{\it Poincar\'e duality}, which gives a canonical isomorphism
\beq
{\cal D}_X:\, H^p(X)\stackrel{\approx}{\longrightarrow} H_{n-p}(X)\ ,
\eeq
for all $p=0,1,\ldots, n$. Now consider another manifold $Y$ of dimension $m$
and let $f:Y\rightarrow X$ be continuous. Then for all $p\geq m-n$ there
is a linear map, called the {\it Gysin homomorphism}:
\beq
f_{!}:\,H^{p}(Y)\longrightarrow H^{p-(m-n)}(X)\ ,
\label{gysin}\eeq
which is defined such that the diagram
\beq
{\begin{array}{ccc}
H^p(Y)&\stackrel{{\cal D}_Y }{\longrightarrow}&H_{m-p}(Y)\\&
&\\{\scriptstyle f_{!}}\downarrow& &\downarrow{\scriptstyle f_*}\\&
&\\H^{p-(m-n)}(X)&\stackrel{{\cal
D}_X^{-1}}{\longleftarrow}&H_{m-p}(X)\end{array}}
\eeq
is commutative, i.e. such that $f_{!}={\cal D}_X^{-1}\,f_*\,{\cal D}_Y$. Here
$f_*$ is the natural push-forward map acting on homology. An
important example to which this construction applies is the case that
$Y$ is an oriented vector bundle $E$ over $X$ of fiber dimension $k$. Then
we consider the canonical projection map $\pi:E\rightarrow X$
and the inclusion $i:X\rightarrow E$ of the zero section. They induce
maps on homology with $\pi_*i_*={\rm Id}$, so that
\bea
\pi_{!}:\, H^{p+k}(E)&\stackrel{\approx}{\longrightarrow}&
H^p(X)\ ,\label{invthommap}\\
i_{!}:\, H^p(X)&\stackrel{\approx}{\longrightarrow}&
H^{p+k}(E)\ ,\label{thommap}
\eea
are isomorphisms for all $p$. The Gysin map $\pi_{!}$ can be thought of as
integration over the fibers of $E\to X$. It is easy to see that
$\pi_{!}i_{!}={\rm Id}$, so that $\pi_{!}=(i_{!})^{-1}$. The map \eqn{thommap}
 is called the {\it Thom isomorphism} of the oriented vector bundle $E$.

An important special instance of the Thom isomorphism (\ref{thommap}) is the
case $p=0$. This defines a map $H^0(X)\rightarrow H^k(E)$, and the image of
$1\in H^0(X)$ thereby determines a cohomology class
\beq
\Phi[E]=i_!(1)\in H^k(E)\ ,
\eeq
which is called the {\it Thom class} of $E$. The Thom isomorphism
\eqn{thommap} is then generated by taking the cup product with this class:
\beq
i_!(\omega)=\pi^*(\omega)\wedge\Phi(E)\ .
\label{thomclassmap}\eeq
This cohomology class will play
a central role in section 7.5. It is related to the {\it Euler class}
$\chi(E)$ of the (even dimensional) real vector bundle $E\rightarrow X$ of
rank $k=2m$, which is a characteristic class of the bundle taking values in
$H^{2m}(X)$. It can be defined as the pullback of the Thom class by the zero
section:
\beq
\chi(E)=i^*\,\Phi[E]\ .
\eeq
When $E$ is a {\it complex} vector bundle of rank $k$, then the Euler
class of $E$ is defined as the Euler class of its underlying real
bundle $E_r$ (of real rank $2k$): $\chi(E)\equiv \chi(E_r)$. Moreover, in
this case the Euler class of $E$ can be shown to coincide with the top
Chern class:
\beq
\chi(E)=c_k(E)=\prod_{n=1}^k\lambda_n\ .
\eeq
If the (real) rank of the vector bundle $E$ coincides with the
dimension of $X$, then one can also introduce the {\it Euler number} $e(E)$,
which is defined as the Euler class evaluated on the homology cycle $[X]$:
\beq
e(X)=\chi(E)[X]=\int\limits_X\chi(E)\ .
\label{eulernumber}\eeq
Furthermore, if $X$ is compact, then for all $\phi\in H^\#(X)$ we have the
identity \cite{lawson}
\beq
i^*\,i_!(\phi)=\chi(E)\wedge\phi\ ,
\label{thomisochi}\eeq
which follows from the fact that the Euler class is given as
$\chi(E)=i^*\,i_!(1)$. Another important
property of these cohomology classes is that if
$s:X\to E$ is ${\it any}$ section of $E$, then $s^*\Phi[E]$ is a closed form
whose cohomology class coincides with the Euler class. From this fact one may
also deduce that $s^*\Phi[E]=\delta_{{\cal Z}(s)}$, where
${\cal Z}(s)\stackrel{i}{\hookrightarrow}X$ is the
zero locus of the section $s$, so that
\beq
\int\limits_{{\cal
    Z}(s)}i^*\,\omega=\int\limits_Xs^*\,\Phi[E]\wedge\omega\ .
\label{localization}\eeq

Let us now describe the Thom isomorphism in K-theory which, using the
Chern character, can be related to the cohomological Thom isomorphism above.
Let $E\rightarrow X$ be a complex vector bundle over $X$. Then $\K^\#(E)$ is
naturally a $\K^\#(X)$-module, with an associative and distributive
module multiplication,
\beq
\K^\#(X)\otimes_\zeds\K^\#(E)\longrightarrow \K^\#(E)\ ,
\eeq
defined according to the sequence of homomorphisms
\beq
\K^\#(X)\otimes_\zeds\K^\#(E)\longrightarrow\K^\#(X\times E)\longrightarrow
\K^\#(E)\ .
\eeq
Here the first map is induced by the cup product and the
second map is the pullback on K-theory of the map $\pi\times {\rm Id}$.
An important example is the case when $E=I^m=X\times\complex^m$ is
the trivial complex vector bundle over $X$. Define $\omega\in\K(E)$ to
be the class
\beq
\omega=
\Bigl[\pi^*\Lambda^{\rm even}E\,,\,\pi^*\Lambda^{\rm odd}E\,;\,\mu\Bigr]\ ,
\eeq
where $\pi^*\Lambda E$ is the trivial $m$-plane bundle over $E$ and
$\Lambda^{\rm even,odd}E$ denote the even and odd degree exterior product
bundles corresponding to $E$. The isomorphism $\mu$ is defined by
\beq
\mu_{x,v}(\phi)=v\wedge\phi-v^\dagger\,\neg\,\phi\ ,
\label{muxv}\eeq
for $(x,v)\in X\times\complex^m$ and $\phi\in\pi^*\Lambda^{\rm even}E$.
Using the identification $\real^{2m}\cong\complex^m$ and choosing the
canonical orientation, this element can be written as
\beq
\omega=[{\cal S}^+,{\cal S}^-;\mu]\ ,
\eeq
where ${\cal S}={\cal S}^+\oplus{\cal S}^-$ is the irreducible complex
graded $\cliff_{2m}$-module (extended trivially over $X$), so that, according
to \eqn{cliffmul}, $\mu_{x,v}(\phi)=v\cdot\phi$ coincides with the usual
Clifford multiplication. The fundamental Bott periodicity theorem then
implies that $\K^\#(E)$ is a free $\K^\#(X)$-module of rank 1
with generator $\omega$,
so that $\omega$ gives a K-theory orientation for the bundle $E$.

Now consider a general (possibly non-trivial) complex vector bundle over $X$.
We say that $\omega\in \K(E)$ is a {\it Bott class}
if $\omega$ determines a K-theory orientation in any local trivialization of
$E$ over a closed subset $C\subset X$, i.e. $\K^\#(E|_C)$ is a free
$\K^\#(C)$-module generated by $\omega$ whenever $E|_C$ is trivial. It can
be shown \cite{karoubi,lawson} that any Bott class
is a K-theory orientation for $E$. In particular, if
$E\rightarrow X$ is a complex Hermitian vector bundle over a compact space
$X$, then the class
\beq
\Lambda_{-1}(E)=\Bigl[\pi^*\Lambda^{\rm even}E\,,\,
\pi^*\Lambda^{\rm odd}E\,;\,\mu\Bigr]\in
\K(E)\ ,
\label{Lambda-1}\eeq
with $\mu_v(\phi)=v\wedge\phi-v^\dagger\,\neg\,\phi$, defines a K-theory
orientation for $E$. This follows from the example above which showed
that $\Lambda_{-1}(E)$ is a Bott class. The K-group
element \eqn{Lambda-1} is the K-theoretic Thom class of the vector bundle
$E$, which is natural and multiplicative:
\beq
\Lambda_{-1}(E\oplus F)=\Lambda_{-1}(E)\otimes\Lambda_{-1}(F) \ .
\label{thommult}\eeq
By taking cup products with it, it follows that the map
$i_!:\K(X)\rightarrow \K(E)$ defined by
\beq
i_{!}(\alpha)=\pi^*(\alpha)\otimes\Lambda_{-1}(E)\ \ , \ \ \alpha\in\K(X)\ ,
\eeq
is an isomorphism. This is the Thom isomorphism in complex K-theory.
When $X={\rm  pt}$ is the space consisting of a single point, and
$E=\complex^n$ is the trivial bundle over $X$, then the Thom isomorphism is
just the statement of Bott periodicity in the form
$\widetilde{\K}({\bf S}^{2n})=\zed$. This follows from the fact that
$\K(X)=\K({\rm pt})=\zed$ and
$\K(E)=\K(\complex^n)=\widetilde{\K}({\bf S}^{2n})$.
More generally, taking $E=I^m=X\times\complex^m$ and using $\real^{2m}
\cong\complex^m$, the Thom isomorphism is just the statement of Bott
periodicity in the form of the suspension isomorphism \eqn{suspension}.
For some more examples and applications, as well as the description of the
Thom isomorphism in KO-theory and KR-theory, see \cite{karoubi,lawson}.

The relationship between the K-theoretic and cohomological Thom
isomorphisms may be described as follows. Let $E\to X$ be a complex vector
bundle of rank $m$, and let $i_!^\K:\K(X)\to\K(E)$ and
$i_!^{H}:H^\#(X,\rat)\to H^\#(E,\rat)$ be the Thom isomorphisms in
K-theory and cohomology, respectively. We introduce the natural,
multiplicative {\it Todd class} ${\rm Td}(E)\in H^{\rm even}(X,\rat)$ by
\bea
{\rm Td}(E)&=&\prod_{n=1}^m\frac{\lambda_n}{1-\e^{-\lambda_n}}\nonumber\\
&=&1+\mbox{$\frac12$}\,c_1(E)+\mbox{$\frac1{12}$}\,\Bigl(c_1(E)\wedge
c_1(E)+c_2(E)\Bigr)+\ldots \ .
\label{todd}\eea
Then for each class $\omega\in\K(X)$, we have the formula:
\beq
\ch\Bigl(i_!^\K(\omega)\Bigr)=i_!^{H}\Bigl(\ch(\omega)\wedge
{\rm Td}(E)\Bigr) \ .
\label{toddKH}\eeq

The Thom isomorphism also enables the construction of a K-theoretic Gysin
map which will be a crucial ingredient in the global bound state construction
that will be presented in the next subsection. Consider an embedding
$f:Y\hookrightarrow X$ of a submanifold $Y$ of even codimension $2k$ in $X$.
(The restriction to embeddings is not necessary but is assumed for
simplicity.) The normal bundle $N(Y,X)$ of $Y$ in $X$ can be defined through
the exact sequence of vector bundles:
\beq
0\longrightarrow TY\stackrel{f_*}{\longrightarrow} TX\longrightarrow
N(Y,X)\longrightarrow0
\label{normalseq}\eeq
which decomposes the tangent bundle $TX$ of $X$ as $TX=TY\oplus N(Y,X)$.
This identifies the normal bundle with a tubular neighbourhood of $Y$ in $X$
(this means that one chooses a suitable metric on $X$ and defines $N(Y,X)$ to
be the set of all points of distance $<\epsilon$ from $Y$ in $X$, for some
small $\epsilon$), and also with the bundle $f^*(TX)/TY$ over $Y$. The vector
bundle $N(Y,X)$ has structure group $SO(2k)$, which we assume is extendable
globally to $Spin(2k)$, i.e. $N(Y,X)$ admits a spin structure. (Again this
requirement can be relaxed, but we will defer this discussion to the next
subsection). Given the Thom isomorphism $i_!:\K(Y)\to\K(N(Y,X))$, we then
define the Gysin homomorphism by
\beq
f_*=j_*\circ i_!:\, \K(Y)\longrightarrow\K(X)
\label{gysinK}\eeq
where $j_*$ is induced by the morphism $N(Y,X)\stackrel{j}{\hookrightarrow}X$
of locally compact spaces. The map $f_*$ is independent of the choice of
tubular neighbourhood and it depends only on the homotopy class of $f$. Its
basic properties are as follows. First of all, if $f:Y\to X$ and
$g:Z\to Y$ are two embeddings, then $(f\circ g)_*=f_*\circ g_*$.
Furthermore, there are the identities
\beq
f_*(\omega\otimes f^*\alpha)=f_*(\omega)\otimes\alpha \ ,\ \ \
\forall\omega\in\K(Y),\alpha\in\K(X)
\label{fomid}\eeq
and, if $X$ is compact,
\beq
f^*\circ f_*(\omega)=\chi\Bigl(N(Y,X)\Bigr)\otimes\omega
\label{fchiid}\eeq
where the K-theoretic Euler class is defined as the restriction of the
corresponding K-theoretic Thom class to the zero section. In the same way,
one may construct the Gysin homomorphism for KO-theory, with the further
requirements that $\dim X-\dim Y\equiv0~{\rm mod}\,8$ and that $N(Y,X)$
admits a spin structure.

\subsection{Global Version of the Bound State Construction}

The bound state constructions that we have described thus far only apply
locally in the
spacetime $X$. In this subsection we will discuss the features that arise when
global
topology is taken into account. We shall describe the details only for Type IIB
superstring theory, as then the generalization to other string theories will be
evident.
For this, we must be careful about the topology of the (non-trivial)
normal bundle of the D-brane worldvolume in $X$, which must thereby be treated
more carefully
using the Thom isomorphism and the Gysin map discussed in the previous
subsection.
Actually, the mapping \eqn{absmapK} is a local version of the Thom isomorphism,
with the
transverse space ${\bf S}^{2k}$ identified with the normal bundle of $Y$ in $X$
and
$Y\times{\bf B}^{2k}$ with a small neighbourhood of $Y$ in spacetime. Globally
then, the
Thom isomorphism $f_!:\K(Y)\stackrel{\approx}{\to}\K(X)$ applied to the normal
bundle
$N(Y,X)\stackrel{\pi}{\to}Y$ yields the mapping
\beq
[E]\longmapsto f_![E]=\pi^*\Bigl([E]\Bigr)\otimes\Lambda_{-1}\Bigl(N(Y,X)\Bigr)
\ .
\label{thomnormal}\eeq
A representative of the Thom class of the normal bundle is then given by the
ABS
construction, as described above. However, to achieve the map \eqn{thomnormal}
one needs to extend the bundle $\pi^*E$ to the whole of $X$, which requires
some special
care and treatment of the normal bundle topology that we shall now discuss. The
main idea
is that the global obstructions which prevent the ABS class $[{\cal S}^+,{\cal
S}^-;\mu]$
from producing a K-theory class of $\K(X)$
can be typically eliminated by nucleating extra 9-branes and
$\overline{9}$-branes. In certain cases (to be described below) one has to
stabalize (in
the K-theory sense) the original configuration of 9-branes and
$\overline{9}$-branes by
pair creating extra $9-\overline{9}$-brane configurations and thus yield a
configuration
of $9-\overline{9}$-brane pairs with K-theory class $[{\cal S}^+\oplus H,{\cal
S}^-\oplus H;
\mu\oplus{\rm Id}]$. This construction will then demonstrate that,
globally, brane charges in a spacetime $X$ can always
be described by a configuration of 9-branes and $\overline{9}$-branes
\cite{witten} and are therefore classified by $\K(X)$.

Let us start with the case of codimension 2. Recall that in the case of flat
brane worldvolumes, in order to build a $p$-brane we need a
$p+2$-brane-antibrane pair
wrapping a submanifold $\real^{p+3}$ of the spacetime $X$ which gives
rise to a $U(1)\times U(1)$ gauge field and a tachyon field $T$ of
charges $(1,-1)$. $T$ vanishes on a codimension 2 subspace that is
identified with the worldvolume of the $p$-brane and breaks the gauge
symmetry from $U(1)\times U(1)$ to $U(1)$. For the global construction, let
$Y\subset Z$ be the worldvolume manifold of the $p$-brane
embedded in the $p+3$ dimensional submanifold $Z$ of spacetime.
To build such a $p$-brane we consider a $p+2$-brane-antibrane pair on
$Z$. Let ${\cal L}$ be a complex line bundle over $Z$ and $\mu$
a section of ${\cal L}$ that vanishes on $Y$. By placing a
$U(1)$ gauge field on the $p+2$-brane, with the same $p$-brane charge
as that of a $p$-brane on $Y$, and a trivial $U(1)$ gauge
field on the $\overline{p+2}$-brane, the system can be interpreted as a
$p$-brane wrapping $Y$.

However, a $p$-brane wrapping $Y$ also has in general
lower-dimensional brane charges $p-2, p-4,\dots$ which depend on the
choice of a line bundle ${\cal K}$ on $Y$. If the line bundle
${\cal K}$ extends over $Z$ then a $p$-brane wrapping $Y$ is
described by taking the bundle ${\cal L}\otimes{\cal K}$ on the $p+2$
brane and the bundle ${\cal K}$ on the $\overline{p+2}$ brane.
If ${\cal K}$
does not extend over $Y$, then one uses the following classic K-theory
construction
\cite{abs}.
Let $Y'$ be a tubular neighborhood of $Y$ in $Z$, whose closure we
denote by $\overline{Y}$ and whose boundary is $\partial Y$.
If $E$ and $F$ are bundles over $Y$ of the same rank then they determine an
element of $\wt{\K}(Y)$. The inclusion
$i:\,Y\hookrightarrow\overline{Y}$
then induces a map on K-theory such that $(E,F)$ also defines a unique
element of $\wt{\K}(\overline{Y})$. The tachyon field is a map
$T:E\longrightarrow F$,
which is an isomorphism of vector bundles outside an open set
$U\subset X$ whose closure $\overline{U}$ is compact. Now suppose that
$T$ is also a tachyon field on $\overline{Y}$ which is an
isomorphism on $\partial{Y}$. In that case one can construct a
natural map $\wt{\K}(\overline{Y})\hookrightarrow\wt{\K}(Z)$,
showing that D-branes wrapping $Y$ are classified by $\K(Z)$, as
desired. This map can be described as follows.
Let $Z'=Z-Y'$. If we can extend the bundle $F$ from
$\partial Y$ to all of $Z'$ then $F$ would be defined over all
of $Z$. Since $E$ and $F$ are isomorphic (under the map $T$) on $\partial Y$,
in that case $E$ can also be extended over $Z'$, so that
$(E,F)$ would define an element of $\wt{\K}(Z)$. If $F$ does
not extend over $Z$, then we may use Swan's theorem to construct a bundle $H$
over $Y$ such that $F\oplus H$ is trivial (assuming $Y$ is
compact) and therefore also trivial on $\overline{Y}$. Now we
replace $E\rightarrow E\oplus H$, $F\rightarrow F\oplus H$ and
$T\rightarrow T\oplus{\rm Id}$. Then we can extend $F\oplus H$ over $Z$ and
also extend $E\oplus H$ by setting it equal to $F\oplus H$ over $Z'$,
so that $(E\oplus H, F\oplus H)$ defines an element of $\wt{\K}(Z)$.
In summary, if ${\cal K}$ does not extend over $Z$, then one instead
finds a bundle $H$ over $Y$ such that ${\cal K}\oplus H$ is
trivial. Then the bundle ${\cal L}\otimes{\cal K}\oplus H$ can be
extended over $Z$. If we now consider a collection of $p+2$-branes on
$Z$ with gauge bundle ${\cal L}\otimes{\cal K}\oplus H$ and a
collection of $\overline{p+2}$-branes on $Z$ with bundle
${\cal K}\oplus H$, along with a tachyon field which equals $T\oplus{\rm Id}$
near
$Y$ and is in the gauge orbit of the vacuum outside $Y'$, then this
system describes a $p$-brane on $Y$ with gauge bundle ${\cal K}$.

In the case that $Y$ is of codimension greater than 2 in $X$, one proceeds as
follows.
Let $Y$ be of codimension $2k$ in $X$. Its normal bundle
$N(Y,X)$ in $X$ then has structure group $SO(2k)$. Suppose
first that $N(Y,X)$ is a spin manifold, so that its second Stiefel-Whitney
class vanishes in $H^2(N(Y,X),\zed_2)$, $w_2(N(Y,X))=0$.
Then associated with the $2^{k-1}$
$9-\overline{9}$-brane pairs we get a pair of spinor bundles ${\cal
  S}^{\pm}$ which are identified with the gauge bundles on the
9-branes. As usual, the tachyon field is a map $T:\, {\cal
  S}^-\rightarrow {\cal S}^+$ with
\beq
T(x)=\sum_{i=1}^{2k}\Gamma_i\, x^i\ \ \ \ , \ \ x\in Y'
\eeq
and the system describes a $p$-brane wrapped on $Y$. This
configuration can be extended over $X$ if ${\cal S}^-$
extends. Otherwise one can find a bundle $H$ such that
${\cal S}^-\oplus H$ extends and then replace $({\cal S}^+,{\cal
  S}^-)\rightarrow({\cal S}^+\oplus H,{\cal S}^-\oplus H)$ and also
$T\rightarrow T\oplus{\rm Id}$. Similarly, for a $p$-brane with line bundle
${\cal K}$, we start from the pair of bundles ${\cal K}\otimes{\cal
  S}^{\pm}$ and use the same construction just presented.

Let us now relax the requirement that $N(Y,X)$ be a spin manifold. According to
the analysis of \cite{sharpe2}, for Type II compactifications with vanishing
cosmological constant, the normal bundle to a D-brane wrapping a supersymmetric
cycle
always admits a spin$^c$ structure. This means that instead of being extendable
to a
principal $Spin(2k)$ bundle over $Y$, the structure group of the normal bundle
extends to
$Spin^c(2k)$, where $Spin^c(n)=Spin(n)\times_{\zeds_2}U(1)$ is the quotient of
the product
group $Spin(n)\times U(1)$ by the equivalence relation $(p,z)\sim(-p,-z)$ and
it
covers the rotation group $SO(n)$ according to the split exact sequence
\beq
1\longrightarrow U(1)\longrightarrow Spin^c(n)\longrightarrow SO(n)
\longrightarrow1 \ .
\eeq
The criteria for the existence of a spin$^c$ structure can be
formulated as follows. Consider the split exact sequence
\beq
0\longrightarrow\zed\stackrel{\times2}{\longrightarrow}\zed\longrightarrow
\zed_2\longrightarrow 0\ ,
\eeq
where the third map is reduction modulo 2. This sequence gives rise to a
long exact sequence in cohomology
\beq
\ldots\longrightarrow H^n(X,\zed)\stackrel{\times2}{\longrightarrow}
H^n(X,\zed)\longrightarrow H^n(X,\zed_2)\stackrel{\beta}{\longrightarrow}
H^{n+1}(X,\zed)\longrightarrow\ldots
\eeq
where the map $\beta$ is called the {\it Bockstein homomorphism}. The kernel
of $\beta$ is the set of classes in $H^\#(X,\zed_2)$ which are modulo 2
reductions of integral cohomology classes. If $w_n\in H^n(X,\zed_2)$
denotes the $n$-th
Stiefel-Whitney class of $X$, then $W_n\equiv\beta(w_{n-1})$ measures whether
or not the $(n-1)$-th Stiefel-Whitney class is the modulo 2 reduction of an
integral class. The normal bundle $N(Y,X)$ admits a spin$^c$ structure if and
only if $W_3(N(Y,X))=0$ (so that in particular any spin manifold is canonically
a
spin$^c$ manifold). Since $X$ is a spin manifold, $w_1(X)=w_2(X)=0$, and
$Y$ is orientable, $w_1(Y)=0$, one can easily show
using multiplicativity of the total Stiefel-Whitney class \cite{witten} that
$w_2(N(Y,X))=w_2(Y)$ and therefore also that
$W_3(N(Y,X))=W_3(Y)$. Thus $N(Y,X)$ admits a spin$^c$ structure only if
the $p$-brane worldvolume manifold $Y$ does.

The existence of a spin$^c$ structure on $N(Y,X)$ implies the following
features for the
bound state construction.
Let $U_i$ be an open covering of $X$. The transition
functions $g_{ij}$ of ${\cal S}^+$ on $U_i\cap U_j$ are then maps
$g_{ij}:\, U_i\cap U_j\rightarrow Spin(2k)$. The existence of a spin
structure is equivalent to the vanishing of the two-cocycle
\beq
\varphi_{ijk}\equiv g_{ij}g_{jk}g_{ki}:\, U_i\cap U_j\cap U_k\longrightarrow
\zed_2\ ,
\eeq
in $H^2(X,\zed_2)$. This defines a cohomology class $[\varphi]\in
H^2(X,\zed_2)$,
which vanishes precisely when $N(Y,X)$ is a spin manifold and
$N(Y,X)$ admits a spin$^c$ structure if $[\varphi]$ is the
modulo 2 reduction of an integral class in $H^2(X,\zed)$. Let
${\cal L}$ be the complex line bundle corresponding to this integral class
(i.e. $c_1({\cal L})$
is equal to this element in $H^2(X,\zed)$), and
let $\gamma_{ij}:\, U_i\cap U_j\rightarrow {\bf S}^1$ be the transition
functions for ${\cal L}$.
Suppose we want to find a square root of ${\cal L}$, i.e. a
line bundle ${\cal L}^{1/2}$ with ${\cal L}^{1/2}\otimes{\cal L}^{1/2}={\cal
L}$.
Then since $U_i\cap U_j$ is contractible we can
define a square root $\tilde{\gamma}_{ij}\equiv\pm\sqrt{\gamma_{ij}}:\,
U_i\cap U_j\rightarrow {\bf S}^1$. The obstruction to the existence of
a consistent set of transition functions $\tilde{\gamma}_{ij}$ is
the two-cocycle
\beq
\varphi'_{ijk}\equiv\tilde{\gamma}_{ij}\tilde{\gamma}_{jk}
\tilde{\gamma}_{ki}:\, U_i\cap U_j\cap U_k\longrightarrow \zed_2
=\ker\sigma
\eeq
where $\sigma$ is the map $\sigma(z)\equiv z^2$
corresponding to the split exact sequence
\beq
0\longrightarrow\zed_2\longrightarrow{\bf
  S}^1\stackrel{\sigma}{\longrightarrow}
{\bf S}^1\longrightarrow 0\ .
\eeq
The class $[\varphi']\in H^2(X,\zed_2)$ is
the coboundary of $[\tilde\gamma]\in H^1(X,{\bf S}^1)$ under the
associated long exact sequence in cohomology. In fact, consider the
following commutative diagram:
\beq
{\begin{array}{ccccc}
H^1(X,{\bf S}^1)&\stackrel{\sigma}{\longrightarrow}&H^1(X,{\bf S}^1)
&\stackrel{\varphi'}{\longrightarrow}&H^2(X,\zed_2)\\&
&\\\downarrow\approx& &\downarrow\approx& &\parallel \\&
&\\H^2(X,\zed)&\stackrel{\times2}{\longrightarrow}&H^2(X,\zed)
&\stackrel{\rho}{\longrightarrow}&H^2(X,\zed_2)\end{array}}
\eeq
It follows that $[\varphi']=\rho(c_1({\cal L}))=[\varphi]$ and therefore
$[\varphi']+[\varphi]=0$, or equivalently
\beq
c_1({\cal L})\equiv w_2\Bigl(N(Y,X)\Bigr)~~{\rm mod}~2 \ .
\eeq
This means that while we cannot construct
the spinor bundles and we cannot construct the complex line bundle ${\cal
L}^{1/2}$
globally, we can
construct their tensor product. Thus, the existence of a spin$^c$ structure
means that ${\cal L}^{1/2}\otimes{\cal S}^{\pm}$ exist as vector
bundles even though ${\cal L}^{1/2}$ and ${\cal S}^{\pm}$ do not.
This in turn means that if $N(Y,X)$ is a spin$^c$ bundle then we can proceed as
in the
case of spin bundles with the pair $({\cal L}^{1/2}\otimes{\cal S}^+,
{\cal L}^{1/2}\otimes{\cal S}^-)$ determining an element of $\K(X)$
and representing a D-brane wrapped on $Y$.

\subsection{Compactifications and $T$-Duality}

A $T$-duality transformation maps Type IIA superstring theory to
Type IIB superstring theory, under which a D$p$-brane is mapped to a
D$(p-1)$-brane
if the transformation is done in a direction transverse to the
brane worldvolume. Since Type IIB branes are classified by $\K(X)$ and
Type IIA branes by $\K^{-1}(X)$, it is natural to study the action of
$T$-duality at the level of K-groups \cite{bgh,hori,os}.
For this, we shall need to understand how to measure D-brane charge on
spacetime
compactifications in terms of K-theory and how to achieve natural isomorphisms
of the
corresponding K-groups.

We first need to explain an intimate connection between
the index theory of Fredholm operators and topological
K-theory, which will also be used in the next subsection.
A {\it Fredholm operator} ${\cal T}$, acting on a separable
Hilbert space ${\cal H}$, is a bounded linear operator whose kernel and
cokernel are
finite dimensional subspaces of $\cal H$.
Such operators therefore have a well-defined {\it index}:
\beq
\ind\,{\cal T}=\dim\ker{\cal T}-\dim{\rm coker}\,{\cal T}\ ,
\label{index}\eeq
which is invariant under perturbations by any compact operator
${\cal A}$,
\beq
\ind({\cal T}+{\cal A})=\ind\,{\cal T}\ .
\eeq
Moreover, if ${\cal S}$ is a bounded operator that is sufficiently close in the
operator
norm to
${\cal T}$, then ${\cal S}$ is also a Fredholm operator and $\ind\,{\cal
  T}=\ind\,{\cal S}$.

The importance of these properties stems from the fact that one can also
describe the
group $\K(X)$
in terms of Fredholm operators. For this, let ${\cal F}$ be the space of
Fredholm operators on ${\cal H}$ with the operator norm topology. Then
\eqn{index}
defines a continuous map
\beq
\ind:\, {\cal F}\longrightarrow\zed\ ,
\eeq
which can be shown to induce a bijection
\beq
\pi_0({\cal F})\longrightarrow\zed
\label{indexbij}\eeq
between the set of connected components of ${\cal F}$
and the integers. More generally, let $X$ be a compact topological
space and consider the set $[X,{\cal F}]$ of homotopy classes of maps from $X$
to
${\cal F}$. Since the product
of two Fredholm operators is again a Fredholm operator, $\left[X,{\cal
    F}\right]$ is a monoid. It can be shown that there is
an isomorphism:
\beq
\left[X,{\cal F}\right]\stackrel{\approx}{\longrightarrow}\K(X)\ ,
\label{FredholmK}\eeq
which may be described as follows. Let ${\cal T}_x$ be a continuous
family of Fredholm operators labelled by the parameter $x\in X$.
Then the family of vector
spaces $\ker{\cal T}_x$ forms a vector bundle $\ker{\cal T}$ over $X$. This
statement is also true for the cokernel of ${\cal
  T}$, so that we can define the index of a family of operators
${\cal T}_x$ as the class
\beq
\Ind\,{\cal T}\equiv\Bigl[(\ker{\cal T}\,,\,
{\rm coker}\,{\cal T})\Bigr]\in \K(X)\ .
\label{Index}\eeq
Note that this is similar to the correspondence that was made in
\eqn{grothrep}.
With this correspondence, the composition of operators in ${\cal F}$
corresponds to the addition in $\K(X)$, while adjoints correspond to inversion.
In particular, in the case where $X$ is
a point (so that $\K(X)=\zed$) the isomorphism (\ref{FredholmK}) is
just the index map (\ref{indexbij}). In other words, the virtual dimension of
the K-theory
class \eqn{Index} coincides with the index defined in \eqn{index}:
\beq
\ch_0(\Ind\,{\cal T})
=\ind\,{\cal T} \ .
\eeq
Moreover, the set of
homotopy classes of Fredholm operators defines the K-homology group $\K_0(X)$.
The duality
with K-theory is provided by the natural bilinear pairing
\beq
\Bigl([E]\,,\,[{\cal F}]\Bigr)\longmapsto\ind\,{\cal F}_{E}\in\zed
\label{bipairing}\eeq
where $[E]\in\K(X)$ and ${\cal F}_E={\cal F}_{E,E}$
denotes the action of the Fredholm operator ${\cal F}$
on the Hilbert space ${\cal H}=L^2(\Gamma(X,E))$ of square-integrable sections
of the vector
bundle $E\to X$ as ${\cal F}:\Gamma(X,E)\to\Gamma(X,E)$.

For the present purposes we shall be interested in applying these ideas
to a special class of operators, namely the Dirac operators associated to
vector bundles
over a spin manifold $X$. Dirac operators are examples of pseudo-differential
elliptic
operators, which are Fredholm operators when viewed as operators on a Hilbert
space.
To this end, we consider the case ${\cal F}=i\dirac:\Gamma(X,{\cal S}
_E^+)\to\Gamma(X,{\cal S}_E^-)$, where $E\to X$ is a real spin bundle (of even
rank)
and ${\cal S}_E^\pm$
are the corresponding twisted chiral spinor bundles lifted from $E$.
The Chern character \eqn{chernhomo} (along with a version of the Gysin map
introduced
at the end of section 7.2) then allows
one to map the analytical
index of $i\dirac$ defined in terms of K-theory classes into a topological
index which
can be expressed in terms of cohomological characteristic classes. The result
is the
celebrated
Atiyah-Singer index theorem \cite{atiyahsing}:
\beq
\ind\,i\dirac=-\int\limits_X\ch(E)\wedge\widehat{A}(TX)
\label{atiyahsingthm}\eeq
where the Dirac genus of the vector bundle $E$ is defined by
\bea
\widehat{A}(E)&=&\prod_n\frac{\lambda_n/2}{\sinh(\lambda_n/2)}\nonumber\\
&=&1-\frac1{24}\,p_1(E)+\frac1{5760}
\,\Bigl(7p_1(E)\wedge p_1(E)-4p_2(E)\Bigr)+\dots
\label{diracgenus}\eea
and $p_n(E)=(-1)^nc_{2n}(E\otimes_\reals\complex)$ is the $n$-th Pontryagin
class of $E$.
An important special
instance of this index formula is obtained by taking $E=TX$ to be the tangent
bundle of the manifold $X$. Then the Euler number \eqn{eulernumber} can be
expressed in
terms of the Euler-Poincar\'e characteristic of $X$:
\beq
e(X)=\dim\K(X)\otimes_\zeds\rat-\dim\K^{-1}(X)\otimes_\zeds\rat \ .
\label{eXK}\eeq

We can apply these ideas to give an
index-theoretical interpretation of $T$-duality acting on K-theory
classes in various superstring theories \cite{hori}. The basic motivation for
this
analysis is the expression for the transformation of RR tensor fields under
$T$-duality \cite{bho}. It can be shown that the RR fields on spacetimes
of the form ${\bf T}^n\times M$ and those of the $T$-dual theory on
$\widehat{\bf T}^n\times M$ (where $\widehat{\bf T}^n$ is the dual
torus of ${\bf T}^n$) are related according to (in the absence of a
Neveu-Schwarz
$B$-field)
\beq
\widehat{H}=\int\limits_{{\bf T}^n}\ch({\cal P})\wedge H\ ,
\label{RRdual}\eeq
where $H=\sum_pH^{(p+2)}$ is the gauge-invariant,
total RR form field strength. Here
\beq
\ch({\cal P})=\exp\left(\sum_{i=1}^nd\widehat{y}_i\wedge dy^i\right)
\eeq
is the Chern character of the Poincar\'e (complex line) bundle $\cal P$ over
$\widehat{\bf T}^n
\times{\bf T}^n$,
with $y^i$ and $\widehat{y}_i$ dual coordinates on ${\bf T}^n$ and
$\widehat{\bf T}^n$. The Poincar\'e bundle is defined as the quotient of the
trivial
bundle ${\bf T}^n\times(\real^n)^*\times\complex$ by the action of the rank $n$
lattice $2\pi\Lambda
^*$ (where ${\bf T}^n=\real^n/2\pi\Lambda$) defined by
$(x,x^*,z)\mapsto(x,x^*+m^*,\e^{im_i^*
x^i}z)$.
The relationship (\ref{RRdual}) is reminescent of a formula that arises
in the family index theory \cite{atiyahsing} for a family of Dirac operators on
${\bf T}^n$
parametrized by $\widehat{\bf T}^n$ which is carried by the bundle $\cal P$
over
$\widehat{\bf T}^n
\times{\bf T}^n$.
This motivates the search for a relatively simple explanation of the
transformation
property \eqn{RRdual} in terms of K-theory which provides the analogous
transformation
rule for D-branes (which are sources for the RR fields).

For illustration, let us consider the case of D-branes in Type IIB superstring
theory compactified on a circle ${\bf S}^1$. Spacetime is then ${\bf
S}^1\times M$, where $M$ is a nine-dimensional manifold, and the dual
geometry is $\widehat{\bf S}^1\times M$. As usual, a Type IIB D-brane is
constructed as a bound state of $9-\overline{9}$-branes with Chan-Paton
bundles ${\cal S}^\pm$, gauge connections $A^\pm$ and a tachyon field $T:\,
{\cal S}^+\rightarrow{\cal S}^-$. We probe this system with a D1-brane wrapped
on ${\bf
S}^1$, so that the dual system is a D0-brane moving in $\widehat{\bf S}^1\times
M$. The mass matrix of the
fermionic modes coming from the $1-9$ and $1-\overline{9}$ strings is
given by the Dirac operator
\beq
i\dirac=\left( \begin{array}{cc}D_+& -T^\dagger\\
T&-D_-\end{array}\right)=
\left( \begin{array}{cc}\partial_y-ia&0\\ 0&-\partial_y+ia
\end{array}\right)+{\cal A}\ .
\label{massmatrix}\eeq
Here $D_\pm=\partial_y+A^\pm_y-ia$ is the Dirac operator on ${\bf
S}^1$ coupled to the connection $A^\pm_y-ia$ on the bundle
${\cal S}^\pm\otimes{\cal P}$, where ${\cal P}$ is the Poincare bundle over
${\bf S}^1\times
\widehat{\bf S}^1$ with curvature $-ida\wedge dy$.
The operator \eqn{massmatrix} can be interpreted as the usual Dirac operator
twisted by the
superconnection \eqn{superconn} on the $9-\overline{9}$-branes coupled to the
probe
D-strings. It can
also be interpreted as the tachyon field of the unstable Type IIA 9-branes of
the
$T$-dualized system \cite{hori}, whereby the Wilson line on a D$p$-brane is
mapped onto
the position of a D$(p-1)$-brane (c.f. eq. \eqn{IIAIIBmap}).

Since $i\dirac$ is a skew-adjoint operator its index vanishes
identically as an element of $\K({\bf S}^1\times M)$. Rather, it can be
shown \cite{atiyahsing} that the index takes values in the higher
K-group $\K^{-1}(\widehat{\bf S}^1\times M)$ of the parameter space for the
family through the following
construction. Given the family $i\dirac(x)$ of skew-adjoint Fredholm operators
labelled by $x\in W=\widehat{\bf S}^1\times M$, one can define a family over
$[-\frac{\pi}{2},\frac{\pi}{2}]\times W$ by
\beq
i\widetilde{\dirac}(t,x)=-\sin t+i\dirac(x)\cos t\ .
\eeq
This is no longer a skew-adjoint operator and therefore it can have distinct
kernel
and cokernel. Furthermore, since $i\wt{\dirac}
(-\frac{\pi}{2},x)=-i\wt{\dirac}(\frac{\pi}{2},x)=1$, its kernel and
cokernel are isomorphic at $t=\pm\frac{\pi}{2}$ and therefore
$\Ind\,i\wt{\dirac}\in\K([-\frac{\pi}{2},\frac{\pi}{2}]\times W,
\partial[-\frac{\pi}{2},\frac{\pi}{2}]\times W)=\K^{-1}(W)$.
It follows then that $T$-duality determines a map:
\beq
\K({\bf S}^1\times M)\longrightarrow\K^{-1}(\widehat{\bf S}^1\times M)
\eeq
which can be identified as the sequence of homomorphisms:
\beq
\K({\bf S}^1\times M)\stackrel{\otimes[{\cal P}]}{\longrightarrow}
\K({\bf S}^1\times\widehat{\bf S}^1\times M)
\stackrel{\Ind\,i\dirac}{\longrightarrow}\K^{-1}(\widehat{\bf S}^1\times
M) \ ,
\label{indsequence}\eeq
where the last map is defined by $[(E,F)]\mapsto\Ind\,i\dirac_{E,F}$. In an
analogous way
one may construct the inverse map, so that the transformation \eqn{indsequence}
is actually
an isomorphism of K-groups. To compare the transformation \eqn{indsequence}
with
\eqn{RRdual}, we compute the index using the family index theorem to get
\beq
\ch(\Ind\,i\dirac_{E\otimes{\cal P}})=\int\limits_{{\bf S}^1}\ch(E\otimes{\cal
P})\wedge
\widehat{A}(T{\bf S}^1)
\label{famindex}\eeq
with $\widehat{A}(T{\bf S}^1)=1$ and $\ch(E\otimes{\cal
P})=\ch(E)\wedge\ch({\cal P})$.
Since the K-groups of ${\bf S}^1$ are torsion free, the Chern character
\eqn{chernhomo}
is an isomorphism onto the subring $H^{\rm even}({\bf S}^1,\zed)$ of $H^{\rm
even}
({\bf S}^1,\rat)$, which makes the connection with the formula \eqn{RRdual}.
This construction can
be generalized to compactifications on the $n$-torus ${\bf T}^n$, thus
defining maps
$\K^{-m}({\bf T}^n\times M)\longrightarrow
\K^{-m-1}(\widehat{\bf T}^n\times M)$.
Similar arguments can be applied to Real vector bundles
\cite{atiyahsing}, yielding the corresponding maps on KR-groups appropriate for
the
Type I and Type II orientifold theories.

There is another way to see the $T$-duality isomorphism in terms of {\it
relative}
K-theory \cite{bgh}.
Consider a general compactification manifold $Z$ of dimension $d$.
We want to determine all D-brane charges of codimension $m$ in the non-compact
space $\real^{9-d}$. These charges arise from D-branes which wrap
non-trivial cycles of $Z$ and from D-branes located at particular
points in $Z$. As usual, one considers configurations of finite energy
and therefore only those which are equivalent to the vacuum
asymptotically in the transverse space $\real^m$. So $\real^m$ is
replaced by its one-point compactification ${\bf S}^m$ by
the addition of a copy of the compactification manifold $Z$ at infinity.
This corresponds precisely to considering charges which take values in
the relative K-group (\ref{Krelative}) (with $Y=Z$ and $X={\bf S}^m\times
Z$). Thus, for example, for compactifications of Type IIB
superstring theory on a submanifold $Z$, D-brane charges are classified by
$\K({\bf S}^m\times Z,Z)$, and by $\K^{-1}({\bf S}^m\times Z,Z)$ for
Type IIA compactifications. For instance, consider the compactification of Type
II on an
$n$-torus ${\bf T}^n$. By iterating the relations \eqn{kxs}, \eqn{kmxs} and
\eqn{Kretract},
one may easily derive the natural isomorphisms
\bea
\K(M\times{\bf T}^n,{\bf
T}^n)&=&\bigoplus_{k=0}^n\wt{\K}^{-k}(M)^{\oplus{n\choose k}}
{}~=~\wt{\K}(M)^{\oplus2^{n-1}}\oplus\K^{-1}(M)^{\oplus2^{n-1}}\nonumber\\
&\cong&\K^{-1}(M\times{\bf T}^n,{\bf T}^n)\ .
\label{KTisos}\eea
{}From this point of view, Type II $T$-duality is then a consequence of the
periodicity
of 2 of complex K-theory. Furthermore, from \eqn{KXYcup} we see that under the
isomorphism \eqn{KTisos} of K-groups for $n=1$, $\wt\K(M)\otimes_\zeds\K({\bf
S}^1)$
maps to $\wt\K(M)\otimes_\zeds\K^{-1}({\bf S}^1)$ with the summands $\K({\bf
S}^1)$ and
$\K^{-1}({\bf S}^1)$ interchanged. From this it follows that $T$-duality
exchanges
wrapped and unwrapped D-brane configurations. For $n>1$, the decomposition
\eqn{KTisos} gives the anticipated degeneracies $2^{n-1}$ of brane charges
arising from the
higher supersymmetric branes wrapped on various cycles of the torus ${\bf
T}^n$. This
may be attributed to the fact that the $T$-duality mapping generates the spinor
representation of the target space duality group $O(n,n,\zed)$, in agreement
with the
fact that $O(n,n,\zed)$ acts on the IIA and IIB RR potentials in the positive
and
negative chirality spinor representations, respectively. The complete agreement
with
the predictions of cohomology theory is once again a consequence of the Chern
isomorphism
of the integer K-groups of ${\bf T}^n$ with the corresponding integer
cohomology ring.

This analysis generalizes to other string theories as well. For instance, we
can
write down the explicit $T$-duality isomorphism between D-brane charges of Type
I
compactified on a torus and those of the corresponding Type II orientifold
compactification. Using the analog of the decomposition \eqn{Kretract} for
KO-theory
and \eqn{KOXS1}, we may iteratively compute the relevant group for the
compactification
of the Type I theory,
\beq
\KO(M\times{\bf T}^n,{\bf
T}^n)=\bigoplus_{k=0}^n\wt\KO^{-k}(M)^{\oplus{n\choose k}}
\label{KOTdecomp}\eeq
whereas for the corresponding $T$-dual orientifold theory we may use
\eqn{KRS11} to get
\bea
\KR^{-n}(M\times{\bf T}^{1,n},{\bf T}^{1,n})&=&
\bigoplus_{k=0}^n\wt\KR^{n,k}(M)^{\oplus{n\choose
k}}~=~\bigoplus_{k=0}^n\wt\KO^{k-n}(M)^{\oplus{n\choose k}}\nonumber\\
&\cong&\KO(M\times{\bf T}^n,{\bf T}^n) \ .
\label{KORTisos}\eea
where we have used the fact that the KR-involution acts trivially on $M$.
The corresponding spectrum of BPS and $\zed_2$ non-BPS D-branes agrees again
with the
degeneracies of the various wrapped branes. The complexity of the decomposition
\eqn{KORTisos} as compared to the Type II case owes to the periodicity of 8 of
the KO
and KR-groups, as discussed in section 6.4. In these cases, the precise
bookkeeping of
D-brane charges requires the concept of ``D-brane transfer", whereby a D-brane
which
is located over an orientifold plane is ``transfered" via a wrapped D-brane of
one
higher dimension to another orientifold plane. This is required to compensate
for the
apparently absent $\zed_2$ charges in the K-theory spectrum \eqn{KOTdecomp}
(see
\cite{bgh} for more details).

Other K-theoretic interpretations of the $T$-duality
isomorphism may also be given. In \cite{sharpe} it was discussed how to
describe Type II
D-branes wrapped on complex submanifolds of complex varieties using a
holomorphic
version of K-theory (more precisely, the Grothendieck groups of coherent and
locally
free sheaves), which further
encodes a choice of connection on the brane worldvolume, and how the action of
$T$-duality
can be understood in terms of Fourier-Mukai transformations (see also
\cite{hori}).
In \cite{os}, $T$-duality was interpreted as being a consequence of the weak
Bott
periodicity sequence for
the stable homotopy groups of the finite-dimensional vacuum manifolds
for the Type II and Type I theories (c.f. section 6.4).

\subsection{D-Brane Anomalies}

In the final part of this review we will derive an explicit
formula for the charge of a D$p$-brane when
it wraps a submanifold $Y$ of the spacetime $X$. Locally, this formula has its
origin
in ordinary cohomology theory, but as we shall demonstrate, when global
topology is
taken into account the expression involves quantities which are most naturally
understood in terms of K-theory \cite{minmoore} in exactly the same spirit as
our
previous discussions of D-brane charge. The basic idea comes from the fact that
a Weyl fermion on an even-dimensional manifold always yields an anomalous
variation of its
action given by the well-known descent formula \cite{atiyahdescent}.
This formula determines the anomaly in terms of the
representation of the gauge group carried by the fermions, and the
corresponding Yang-Mills
and gravitational connections. The same phenomenon occurs whenever a D-brane
wraps around a
non-trivial supersymmetric cycle of a curved manifold, because the twisting of
its normal
bundle can induce chiral asymmetry in its worldvolume field theory. The form of
these chiral
anomalies can be deduced by considering the field content on the intersection
of two branes,
which contains chiral fermions. The anomaly term then comes from the tensor
product of the
spinor bundles with the Chan-Paton vector bundles over the two D-branes. The
anomalous zero
modes on the intersection of the branes come from the massless excitation
spectrum of the
worldvolume field theory which consists of Weyl fermions in the mixed sector
${\bf N}_1\otimes
\overline{\bf N}_2$ and $\overline{\bf N}_1\otimes{\bf N}_2$
representations of the gauge group
$U(N_1)\times U(N_2)$ on the intersecting brane worldvolume. To render the
theory
anomaly-free thereby requires the addition of Wess-Zumino terms to the D-brane
action. These
induced terms imply that topological defects (such as instantons or monopoles)
on the
D-branes carry their own RR charge determined by their topological quantum
numbers
\cite{semz,witteninst}.

Let $f:Y\hookrightarrow X$ be the embedding of a $p+1$ dimensional brane
worldvolume
$Y$ into the spacetime manifold $X$ of Type IIB superstring theory. The
anomalous
D-brane coupling takes the form of a Wess-Zumino type action,
\beq
S_Y=\int\limits_Yf^*\,C\wedge {\cal Y}(\nabla_E^2,g)\ ,
\label{IM}\eeq
where $C=\sum_pC^{(p+1)}$ is the total RR form potential and
${\cal Y}(\nabla_E^2,g)$ is the D-brane source field which is an invariant
polynomial of the Yang-Mills field strength and gravitational curvature
on $Y$. Here
$\nabla_E$ is the Hermitian curvature of a $U(N)$ gauge bundle $E$ on the
brane,
while $g$
is the restriction of the spacetime metric to $Y$. The anomalies on the D-brane
result
from the chiral asymmetry of their massless fermionic modes
which are in one-to-one correspondence
with the ground states of the relevant open string Ramond sectors. Open string
quantization
requires the Ramond ground states to be sections of the spinor bundle lifted
from the
spacetime tangent bundle $TX$ tensored with a vector bundle in the adjoint
${\bf N}\otimes\overline{\bf N}$ representation of the brane gauge group
$U(N)$,
as dictated by the
incorporation of the usual Chan-Paton factors. The GSO projection restricts the
fermions to
have a definite $SO(9,1)$ chirality. When the normal bundle of $Y$ in $X$ is
trivial,
so that $TX=TY$, a standard index-theoretical calculation gives
\beq
{\cal Y}_0(\nabla_E^{2},g)=\ch(E)\wedge f^*\sqrt{\widehat{A}(TX)}\ .
\label{sqhat}\eeq
However, the cohomology class (\ref{sqhat}) needs to be refined in the case
that the normal
bundle is non-trivial and, as we will demonstrate,
this refinement leads to a formula for the
D-brane RR charge which is most naturally understood
in terms of K-theory classes, rather than cohomology classes.
Assuming that $N(Y,X)$ admits a spin
structure, one can determine the fermion
quantum numbers of the spinor bundle associated with
$N(Y,X)$. When $N(Y,X)\neq \emptyset$, the fermions have quantum numbers
$(+,+)\oplus(-,-)$ under the worldvolume Lorentz group $Spin(1,p)$ and the
spacetime Lorentz
group $Spin(9-p)$ restricted to $N(Y,X)$. If the normal bundle is flat, then
left-
and right-moving fermions in the worldvolume field theory are treated equally
and the theory
is non-chiral. However, when $N(Y,X)$ has a non-vanishing curvature, chiral
asymmetry
is induced on the brane worldvolume and a distinction arises between the
$(+,+)$ and $(-,-)$
quantum numbers.

It is well-known that the index of the Dirac operator on an even-dimensional
manifold $X$
gives the perturbative chiral gauge anomaly of a Dirac spinor on $X$. The
positive and
negative chirality spinor bundles ${\cal S}_{TX}^{\pm}$ corresponding to the
tangent bundle
of $X$ can be decomposed in terms of the positive and negative chirality spin
bundles
${\cal S}_{TY}^{\pm}$ and ${\cal S}_{N(Y,X)}^{\pm}$ lifted from the tangent
and normal bundles to $Y$ in $X$:
\beq
{\cal S}_{TX}^{\pm}=\left[{\cal S}_{TY}^{\pm}\otimes
{\cal S}_{N(Y,X)}^{+}
\right]\oplus\left[{\cal S}_{TY}^{\mp}\otimes
{\cal S}_{N(Y,X)}^{-}\right]
\label{S+-}\eeq
The Dirac operator for the charged and reduced fermions acts on sections of the
bundles
(\ref{S+-}) via the two-term complex
\beq
i\dirac:\, \Gamma(Y,E^+)\longrightarrow \Gamma(Y,E^-)\ ,
\label{dirac2term}\eeq
where
\beq
E^{\pm}=\left(\left[{\cal S}_{TY}^{\pm}\otimes
{\cal S}_{N(Y,X)}^{+}\right]\oplus\left[{\cal S}_{TY}^{\mp}\otimes
{\cal S}_{N(Y,X)}^{-}\right]\right)\otimes E\ .
\label{2complex}\eeq
The standard index theorem applied to the two-term complex
(\ref{dirac2term},\ref{2complex}) yields
\bea
\ind\,i\dirac&=&(-1)^{\frac{(p+1)(p+2)}{2}}\int\limits_Y\ch(E)\wedge
\left[\ch(S_{TY}^+)-\ch(S_{TY}^-)\right]\nonumber\\
&& \wedge\left[\ch(S_{N(Y,X)}^+)-\ch(S_{N(Y,X)}^-)\right]
\wedge\frac{{\rm Td}(TY)}{\chi(TY)}
\eea
with $\sqrt{{\rm Td}(TY\otimes_\reals\complex)}=\widehat{A}(TY)$ and
$\ch(S_E^\pm)=\prod_n\e^{\pm\lambda_n/2}$.
Using the identity
\beq
\ch(S_{E}^+)-\ch(S_{E}^-)=\frac{\chi(E)}{\widehat{A}(E)}
\eeq
which holds for any orientable, real spin bundle $E$, we see that the
appropriate
modification of (\ref{sqhat}) due to the normal bundle topology is
\beq
{\cal Y}(\nabla_E^2,g)={\cal Y}_0(\nabla_E^2,g)\wedge
\left[\widehat{A}\Bigl(N(Y,X)\Bigr)\right]^{-1} \ .
\label{modi}\eeq
In arriving at (\ref{modi}) we have re-written (\ref{IM}) as an integral over
$X$ using the
appropriate deRham current $\delta_Y$, and used the discussion of section 7.2
(c.f. eq.
\eqn{localization}) to write
\beq
\delta_Y\wedge\chi\Bigl(N(Y,X)\Bigr)
=\delta_Y\wedge\delta_Y\ .
\eeq
Finally, as with the total Chern class, the Dirac genus is a multiplicative
characteristic
class, so that $\widehat{A}(TX)=\widehat{A}(TY)\wedge\widehat{A}(N(Y,X))$ and
eq. (\ref{modi}) can be written as
\beq
{\cal Y}(\nabla^2_E,g)=\ch(E)\wedge\sqrt{\frac{\widehat{A}(TY)}
{\widehat{A}\Bigl(N(Y,X)\Bigr)}}\ .
\label{final}\eeq

Now we will describe how the anomalous coupling affects brane charges in the
language of
K-theory. To obtain the D-brane charge, we study the RR equations of motion and
Bianchi
identity coming from the complete action for the RR tensor fields:
\beq
S=-\frac{1}{4}\int\limits_{X}H(C)\wedge\,^*H(C)
-\frac{\mu_{(p)}}{2}\int\limits_X\delta_Y\wedge f^*C\wedge {\cal
Y}(\nabla_E^{2},g)\ ,
\eeq
where $H(C)$ is the curvature of $C$.
Then the equations of motion and Bianchi identity for a given $(p+1-m)$-form
potential are
\bea
d\,^*H(C)&=&\mu_{(p)}\,\delta_Y\wedge {\cal Y}(\nabla_E^{2},g)\nonumber\\
dH(C)&=&-\mu_{(p)}\,\delta_Y\wedge \overline{{\cal Y}}(\nabla_E^{2},g)\ ,
\label{eqbianchi}\eea
where $\overline{{\cal Y}}$ is obtained from ${\cal Y}$ by complex conjugation
of the
Chan-Paton gauge group representation (note that
$c_n(\overline{E})=(-1)^nc_n(E)$, so that
the Chern classes are torsion cohomology classes in the case that
$E\cong\overline{E}$
and $n$ is odd). From eq. (\ref{final}) it follows that the formula
for the charge vector $Q\in H^\#(X)$ defined on the
right-hand side of (\ref{eqbianchi}) is
\beq
Q=f_{!}\left(\ch(E)\wedge\widehat{A}(TY)
\wedge\frac{1}{f^*\sqrt{\widehat{A}(TX)}}\right)\ ,
\label{Qind}\eeq
where $f_!: H^{n}(Y,\zed)\rightarrow H^{n+9-p}(X,\zed)$ is the (push-forward)
Gysin
map acting on cohomology as defined in (\ref{gysin}).
{}From the point of view of the worldvolume field theory on
the D-brane, the characteristic class ${\cal Y}(\nabla^2_{E},g)$ on the
right-hand side of (\ref{Qind}) measures the topological charge of a
gravitational/Yang-Mills ``instanton". From eq. (\ref{eqbianchi}) we see
that $\delta_Y\wedge {\cal Y}$ can be thought of as the brane
current for a ``fat" D$(p-m)$-brane bound to and spread out over the
D$p$-brane. When the instanton shrinks to zero size, ${\cal Y}$ acquires
a delta-function singularity, so that the quantity $\delta_Y\wedge
{\cal Y}$ behaves just like a brane current. For some specific examples
wherein the twisting of the normal bundle $N(Y,X)$ modifies the
induced charge, see \cite{cheung}.

To write the class (\ref{Qind}) in a more suggestive form,
we make use of the Thom isomorphism for cohomology in
the form of eq. \eqn{thomisochi} and the identity
\beq
f_!f^*\phi={\cal D}_0\wedge\phi\ ,
\eeq
where $\phi\in H^\#(X,\zed)$ and ${\cal D}_0$ is the Poincar\'e
dual of the zero section. Then we have
\beq
Q=f_!\left(\ch(E)\wedge\widehat{A}(TY)\right)
\wedge\frac{1}{\sqrt{\widehat{A}(TX)}}\ .
\eeq
Now we apply the Atiyah-Hirzebruch version of the Riemann-Roch theorem
\cite{karoubi} which gives (see eq. \eqn{toddKH})
\beq
f_!\left(\ch(E)\wedge\widehat{A}(TY)\right)=
\ch(f_!E)\wedge\widehat{A}(TX)\ ,
\label{atiyahhirz}\eeq
where $f_![E]\in \K(X)$ is defined using the Thom isomorphism \eqn{thomnormal}.
{}From (\ref{atiyahhirz}) it
follows that, as an element of $H^\#(X)$, the RR charge associated to a D-brane
wrapping a
supersymmetric cycle in spacetime $f:Y\hookrightarrow X$ with Chan-Paton bundle
$E\rightarrow Y$ is given by
\beq
Q=\ch(f_!E)\wedge\sqrt{\widehat{A}(TX)}\ .
\label{Qfinal}\eeq
The result (\ref{Qfinal}) has a very natural K-theory interpretation using the
Chern
isomorphism \eqn{cheven}.
The cohomology rings $\K(X)\otimes_\zeds\rat$ and $H^{\rm even}(X,\rat)$
both have natural inner products
defined on them. On $H^{\rm even}(X,\rat)$, the bilinear form is given as in
eq. (\ref{ICE}),
while the pairing on
$\K(X)$ is given by the index of the Dirac operator (c.f. eq. \eqn{bipairing}):
\beq
\Bigl\langle [E]\,,\,[F]\Bigr\rangle_{\K}=\ind\,i\dirac_{E\otimes F}
\eeq
which using the Atiyah-Singer index theorem \eqn{atiyahsingthm}
may be written in terms of the deRham inner
product as
\beq
\biggl\langle [E]\,,\,[F]\biggr
\rangle_{\K}=\left\langle\ch(E)\wedge\sqrt{\widehat{A}(TX)}\,,
\,\ch(F)\wedge\sqrt{\widehat{A}(TX)}
\,\right\rangle_{\rm DR}\ .
\eeq
This implies that the modified Chern isomorphism
\beq
[E]\longmapsto\ch(E)\wedge\sqrt{\widehat{A}(TX)}
\eeq
is an isometry with respect to the natural inner products on $\K(X)$ and
$H^\#(X)$. Thus, the result (\ref{Qfinal}) is in complete agreement with
the fact that D-brane charge is
given by $f_![E]\in \K(X)$, and it moreover gives an explicit formula for
the brane charges in terms of the Chern character homomorphism on K-theory.
Integrating \eqn{Qfinal} over suitable cycles of the spacetime manifold $X$, as
in
\eqn{ICE}, one thereby
obtains the various $p'$-brane charges of the D$p$-brane.

\subsection*{Acknowledgements}

We thank J. Correia, P. Di Vecchia, T. Harmark, G. Landi, A. Liccardo, R. Marotta, N.
Obers,
J.L. Petersen, B. Pioline and P. Townsend for questions and comments about the
topics
discussed in this review which have prompted us to clarify various aspects.
R.J.S.
would like to thank G. Semenoff for hospitality at the University of British of
Columbia, where this work was completed. R.J.S. would also like to thank the
organisors
and participants of the PIms/APCTP/CRM workshop ``Particles, Fields and Strings
'99",
which was held at the University of British Columbia in Summer 1999, for having
provided a stimulating environment in which to work. The work of R.J.S. was
supported in part by the Natural Sciences and Engineering Research Council of
Canada.

\end{document}